\documentclass[twocolumn]{aastex62}

\usepackage{mathrsfs}

\makeatletter
\def\cleardoublepage{\clearpage\if@twoside \ifodd\c@page\else
\hbox{}
\thispagestyle{empty}
\newpage
\if@twocolumn\hbox{}\newpage\fi\fi\fi}
\makeatother 
\newcommand{\BAYMAX}{{\tt BAYMAX}}

\shortauthors{Foord et al.}

\usepackage{relsize}

\usepackage{relsize}
\makeatletter
\providecommand{\ion}[2]{#1$\;$\textsmaller{\@Roman{#2}}}
\makeatother

\usepackage{amsmath,amssymb} 
\usepackage{natbib} 

\newcommand{\beq}{
\begin{equation}
}
\newcommand{\eeq}{
\end{equation}
}

\newcommand{\beqa}{
\begin{eqnarray}
}
\newcommand{\eeqa}{
\end{eqnarray}
}

\begin{document}

\title{AGN Triality of Triple Mergers: Detection of Faint X-ray Point Sources}

\author[0000-0002-1616-1701]{Adi Foord}
\affil{Department of Astronomy and Astrophysics, University of Michigan, Ann Arbor, MI 48109}
\affil{Kavli Institute of Particle Astrophysics and Cosmology, Stanford University, Stanford, CA 94305}

\author[0000-0002-1146-0198]{Kayhan G\"{u}ltekin}
\affil{Department of Astronomy and Astrophysics, University of Michigan, Ann Arbor, MI 48109}

\author{Jessie C. Runnoe}
\affil{Department of Physics and Astronomy, Vanderbilt University, Nashville, TN 37235}

\author{Michael J. Koss}
\affil{Eureka Scientific Inc, Oakland, CA, 94602}



\begin{abstract}
We present results from our X-ray analysis of the first systematic search for triple AGN in nearby ($z<0.077$) triple galaxy mergers. We analyze archival \emph{Chandra} observations of 7 triple galaxy mergers with \BAYMAX{} (Bayesian Analysis of Multiple AGN in X-rays), fitting each observation with single, dual, and triple X-ray point source models. In doing so, we conclude that 1 triple merger has one X-ray point source (SDSS J0858+1822, although it's unlikely to be an AGN); 5 triple mergers are likely composed of two X-ray point sources (NGC 3341, SDSS J1027+1749, SDSS J1631+2352, SDSS J1708+2153, and SDSS J2356$-$1016); and one system is composed of three X-ray point sources (SDSS J0849+1114). By fitting the individual X-ray spectra of each point source, we analyze the $2-7$ keV luminosities as well as the levels of obscuration associated with each potential AGN. We find that 4/5 dual X-ray point source systems have primary and secondary point sources with bright X-ray luminosities ($L_{2-7\mathrm{kev}}>10^{40}$erg s$^{-1}$), possibly associated with 4 new undetected dual AGN. The dual and triple point source systems are found to have physical separations between 3$-$9 kpc and flux ratios between 2$\times10^{-3}-$0.84. A multi-wavelength analysis to determine the origin of the X-ray point sources discovered in this work is presented in our companion paper (Foord et al. 2020c).

\end{abstract}
\keywords{galaxies: active --- galaxies – X-rays --- galaxies: interactions}
\section{Introduction}
\label{chap6:intro}
Systems with multiple supermassive black holes (SMBHs) are expected as a result of hierarchical galaxy formation (e.g., \citealt{WhiteandReese1978}). If two massive galaxies are in the process of merging, it is expected that dynamical friction will drag their respective nuclear SMBHs toward the center of the gravitational potential well (see, e.g., \citealt{Begelman1980}). Through the process of merging, significant quantities of gas can be efficiently funneled down to physical scales at which the SMBHs can accrete -- and the multiple SMBH systems can become multiple active galactic nuclei (AGN; \citealt{Barnes1991, DiMatteo2008, Angles2017}). Thus pairs of AGN, such as dual AGN (two AGN that are in the process of merging, but are not yet gravitationally bound) and binary AGN (two interacting AGN that are gravitationally bound), are signposts of ongoing galaxy formation, and represent unique systems where the link between environment and black hole growth (or, lack thereof) can be probed.
\par An important subset of the multiple SMBH population are triple SMBHs, and in particular, \emph{nearby} triple SMBH systems.  Nearby triple SMBHs ($z<0.3$) are theorized to play important roles in the coalescence of SMBHs and the stochastic gravitational wave background (GWB). For example, without additional interactions from a tertiary SMBH, two interacting SMBHs may not merge within the Hubble time due to an empty loss cone (\citealt{Milosavljevic2003}), producing a stochastic GWB undetectable by pulsar timing arrays (PTAs; ``the nightmare scenario", \citealt{Dvorkin2017}).  Although triaxial potentials have been shown to prevent SMBH stalling \citep{Vasiliev2015}, it is not clear that triaxiality will sufficiently refill the loss cone in time \citep{Merrit1996}. More importantly, the process of two SMBHs becoming bound via dynamical friction may be far slower than initially thought \citep{Tremmel2015}, and it has recently been found that gas dynamics may cause the pair of SMBHs to stall at $\sim$1 kpc \citep{Munoz2019, Duffell2019}. Interactions with a third SMBH can enhance the loss cone refilling rate by disturbing stellar orbits \citep{Perets2008}, shrinking the binary semimajor axis and increasing the eccentricity via the Kozai–Lidov mechanism \citep{Blaes2002}. All these effects can dramatically reduce the merger time of binaries by more than a factor of 10 \citep{Blaes2002}. On top of this, it is predicted that the SMBH binary population driving the detectable GWB signal comes from redshifts close to $z=0.3$ \citep{Kelley2017a}, and thus the rate of nearby triple AGN has significant implications for the GWB measurable with PTAs.  Lastly, triple AGN are expected to be common. A recent study on Illustris SMBH binaries found that $>$30\% of AGN pairs have subsequent merger events, over a large range of redshift \citep{Kelley2017a}, while analysis of the Millennium simulation found that 42\% of massive ($10^{11}<M_{\star}<10^{12}$) galaxies undergo more than one significant merger \citep{Ryu2018}. 
\par A handful of multiple AGN candidates have been identified using multiwavelength diagnostics \citep{Liu2011b, Schawinski2012, DeRosa2015, Kalfountzou2017}; however to date, only one serendipitously discovered X-ray triple AGN has been identified \citep{Pfeifle2019b}. This is may be due to observational constraints and the lack of systematic surveys searching for triple AGN. 
\par Regarding the lack of observational constrains, confirmation of multiple AGN systems requires high spatial resolution X-ray observations.  Although other wavelengths and techniques are useful for finding candidate multiple AGN systems, they are not always reliable \citep{Koss2011}.  For example, double-peaked optical emission lines are insufficient to confirm dual AGN due to ambiguity in interpretation, as single AGN spectra are known to also have double-peaked characteristics  \citep{Comerford2015, Nevin2016}.  On top of this, it is difficult to identify multiple AGN using optical line diagnostics as a result of optical extinction and contamination from star formation \citep[e.g.,][]{Koss2012}.  This is especially problematic for multiple AGN systems, as late-stage mergers tend to be highly obscured (e.g., \citealt{Kocevski2015, Koss2016, Ricci2017, Blecha2018, DeRosa2018, Koss2018, Lanzuisi2018, TorresAlba2018}; however see \citealt{Gross2019} for results that argue X-ray deficits in merging AGN may be due to heightened star formation rather than elevated X-ray obscuration).
\par X-rays are thus crucial to detecting closely separated triple AGN. Nuclear point sources with 2--10 keV X-ray luminosities $L_{X}> 10^{40}\,\mathrm{erg\,s^{-1}}$ are almost always SMBHs \citep{Foord2017a, Lehmer2019}, and \emph{Chandra} can uniquely resolve them. Although various X-ray surveys targeting dual AGN exist, all X-ray confirmed dual AGN have large physical separations ($>$ 1 kpc, see \citealt{Deane2014, DeRosa2019, Hou2020}) and count ratios ($f\sim1$, where $f$ represents the ratio of the X-ray counts associated with the secondary to that of the primary AGN). Specifically, because 1$\arcsec$ corresponds to roughly 1 kpc at $z=0.05$, many \emph{Chandra} investigations are limited to dual AGN with large separations. On top of this, past analyses have been limited to detecting sources with similar brightness. We have developed \BAYMAX{}, a code that quantitatively and rigorously analyzes whether a given \emph{Chandra} source is more likely composed of one or multiple point sources \citep{Foord2019, Foord2020a}.  \texttt{BAYMAX} (i) takes calibrated \emph{Chandra} events and compares them to the expected distribution of counts for single/multiple point source models; (ii) calculates a Bayes factor to determine which model is preferred, automatically taking into account model complexity; (iii) calculates likely values for separations ($r$) and count ratios ($f$); and (iv) fits spectra to each component.  Analyses without \texttt{BAYMAX} are likely to lead to false negatives/positives \citep[e.g.,][]{Koss2015,
Foord2020a}, especially systems with small separations ($r < 1$\arcsec) and fainter secondaries. 
\par Regarding the lack of systematic surveys searching for triple AGN, in this paper we present results from our systematic search for triple AGN in nearby triple galaxy mergers. In the following paper, we analyze a sample of 7 optically-identified, nearby ($z<0.1$), triple galaxy-mergers, that have existing archival \emph{Chandra} data. We analyze the \emph{Chandra} observations using \BAYMAX{}, with the goal of identifying faint multiple point source systems that may otherwise go undetected. The \emph{Chandra} observations of 3 of these triple merger systems were previously analyzed in studies focusing on multiple AGN \citep{Pfeifle2019b, Bianchi2013}, which we now re-analyze via our robust statistical analysis. The remaining 4 have no existing analyses of their \emph{Chandra} observations.
\par The remainder of the paper is organized into 5 sections. In section~\ref{chap6:sample} we introduce the sample and the existing multi-wavelength coverage. In section~\ref{chap6:methods} we review the Bayesian framework behind \BAYMAX{} and the models used to fit each observation. In section~\ref{chap6:results} we present our results from running \BAYMAX{} on the \emph{Chandra} observations and quantify the strength of each result. In section~\ref{chap6:spectral} we analyze the X-ray spectra of each multiple AGN candidate in order to measure their 2$-$keV luminosities and levels of surrounding gas. Lastly, we summarize our findings in section~\ref{chap6:conclusions}. Throughout the paper we assume a $\Lambda$CDM universe, where $H_{0}=69.6$, $\Omega_{M}=0.286$, and $\Omega_{\Lambda}=0.714$.
%
\section{Sample}
\label{chap6:sample}
Our sample was created by cross-matching the AllWISE AGN catalog \citep{Secrest2015} with the SDSS Data Release 16 (SDSS DR16) catalog \citep{SDSSDR16}, for all AGN within $z<0.3$. We then visually identify the systems composed of three interacting galaxies via the SDSS DR16 data, and we further filter the sample by enforcing that (i) a photometric or spectroscopic redshift measurement is available for each galaxy in a triple merger system and (ii) that the respective redshifts of each galaxy in a triple merger system are consistent with one another at the 3$\sigma$ confidence level. While spectroscopic measurements of the redshift are generally well constrained (with fractional errors on the order of $\sim10^{-3}$), photometric measurements of the redshift tend to have larger error bars and are estimated by SDSS via the kd-tree nearest neighbor fitting procedure (see \citealt{Csabai2007} for explicit details). Because at least one galaxy member in each triple merger has a spectroscopic redshift measurement, we assume this value for all members (see Table~\ref{chap6:tabTripleGalInfo}). From this larger sample of 12 triple galaxy mergers, 4 systems have existing, on-axis, archival \emph{Chandra} observations: SDSS J1708+2153, SDSS J2356$-$1015, SDSS J1631+2252, SDSS J0849+1114. To this list, we add 3 triple galaxy mergers from the literature with archival \emph{Chandra} and SDSS DR16 observations that meet our redshift criteria as described above: NGC 3341 \citep{Bianchi2013}, SDSS J0858+1822 \citep{Liu2011} and SDSS J1027+1749 \citep{Liu2011b}. Thus, while all of our triple mergers have \emph{Chandra} and SDSS DR16 coverage, only 4 are included in the AllWISE AGN catalog. Because we visually identify each triple galaxy merger using the SDSS catalog snapshots, we are most sensitive to triple galaxy mergers with angular separations between galactic nuclei that are larger than the SDSS optical fiber ($\sim3\arcsec$), corresponding to physical separations larger than 3 kpc and smaller than 15 kpc. Given these larger angular separations, the ability for BAYMAX to probe fainter sources is of particular importance to our study. Specifically, in past systematic X-ray studies searching for dual AGN, flux ratios between the primary and secondary AGN are large ($\sim$ 1 in \citealt{Hou2020}); and in the largest X-ray study to date \citep{Koss2012}, the typical flux ratio was approximately 0.08. However, BAYMAX is capable of detecting dual AGN systems down to flux ratios $\lessapprox10^{-2}$. 
\begin{table*}[t]
\begin{center}
\caption{Triple Galaxy Merger Sample Properties}
\label{chap6:tabTripleGalInfo}
\begin{tabular*}{0.85\textwidth
}{lcccccc}
	\hline
	\hline
	\multicolumn{1}{c}{Galaxy Name} & \multicolumn{1}{c}{$\alpha$} & \multicolumn{1}{c}{$\delta$} & \multicolumn{1}{c}{Redshift} &
	\multicolumn{1}{c}{$D_{A}$ (Mpc)} & \multicolumn{1}{c}{$\Delta \theta$ ($\arcsec$)} & \multicolumn{1}{c}{$r$ (kpc)}\\
	\multicolumn{1}{c}{(1)} & \multicolumn{1}{c}{(2)} & \multicolumn{1}{c}{(3)} & \multicolumn{1}{c}{(4)} & \multicolumn{1}{c}{(5)} & \multicolumn{1}{c}{(6)} &  \multicolumn{1}{c}{(7)}\\
	\hline
	SDSS J084905.51+111447.2 & 08:49:05.51 & +11:14:47.26 & 0.077 & 306.4 & \dots & \dots \\ 
	SDSS J084905.51+111447.2 SW & 08:49:05.41 & +11:14:45.94 & \dots   & \dots& 2.3 & 3.4 \\
	SDSS J084905.51+111447.2 NW & 08:49:05.43 & +11:14:50.97 & \dots   & \dots & 3.6 & 5.3 \\
	\hline
	SDSS J085837.67+182223.3 & 08:58:37.67 & +18:22:23.35 & 0.059 & 236.9 & \dots & \dots \\
	SDSS J085837.67+182223.3 SW & 08:58:37.52 & +18:22:21.56 & \dots & \dots & 2.8 & 3.2 \\
	SDSS J085837.67+182223.3 SE & 08:58:37.85 & +18:22:22.43 & \dots & \dots & 2.8 & 3.2 \\
	\hline
	SDSS J102700.40+174900.8   & 10:27:00.56 & +17:49:00.38 & 0.066 & 262.9 & \dots & \dots \\	SDSS J102700.40+174900.8 N & 10:27:00.38 & +17:49:02.89 & \dots & \dots& 3.6 & 4.6 \\
	SDSS J102700.40+174900.8 W & 10:27:00.39 & +17:49:00.95 &  \dots & \dots & 2.4 & 3.0\\ 
    \hline
    NGC 3341  & 10:42:31.75 & +05:02:52.82 & 0.027 & 112.6 & \dots & \dots   \\
    NGC 3341 SW & 10:42:31.47 & +05:02:37.80 & \dots & \dots & 15.6 & 8.5 \\ 
    NGC 3341 NW & 10:42:32.05 & +05:02:41.95 & \dots & \dots & 9.6 & 5.2 \\
    \hline
    SDSS J163115.52+235257.5 &  16:31:15.52 & +23:52:57.51 & 0.059 & 236.9 & \dots & \dots \\
    SDSS J163115.52+235257.5 NE & 16:31:15.62 & +23:52:59.56 & \dots & \dots & 2.5 & 2.9  \\
    SDSS J163115.52+235257.5 NW & 16:31:15.41 & +23:53:08.44 & \dots & \dots & 11 & 12.6   \\
    \hline
    SDSS J170859.12+215308.0 & 17:08:59.12 & +21:53:08.08 & 0.072 & 284.8 & \dots & \dots \\
    SDSS J170859.12+215308.0 NE & 17:08:59.42 & +21:53:13.51 & \dots & \dots & 6.6 & 9.1 \\
    SDSS J170859.12+215308.0 SW & 17:08:58.40 & +21:53:05.12	 & \dots & \dots & 10.5 & 14.5 \\
    \hline
    SDSS J235654.30-101605.3 & 23:56:54.30 & -10:16:05.31 & 0.074 & 292.0 & \dots & \dots \\
    SDSS J235654.30-101605.3 SE & 23:56:54.49 & -10:16:07.40 & \dots & \dots & 3.5 & 4.9 \\
    SDSS J235654.30-101605.3 NE & 23:56:54.78 & -10:16:01.06	 & \dots & \dots & 8.2 & 11.6 \\
    \hline
	\hline 
\end{tabular*}
\end{center}
Note. -- Columns: (1) Galaxy name; (2) R.A. and (3) Dec. (J2000) from SDSS DR16; (4) spectroscopic redshift from SDSS DR16; (5) angular diameter distance; (6) angular separation from primary galaxy; (7) projected physical separation from primary galaxy.
\end{table*}

%
\begin{table*}
\begin{center}
\caption{\emph{Chandra} Observation Information}
\label{chap6:tabTripleGalXray}
\begin{tabular*}{0.6\textwidth
}{lcc}
	\hline
	\hline
	\multicolumn{1}{c}{Galaxy Name} &  \multicolumn{1}{c}{\emph{Chandra} Obs. ID} & \multicolumn{1}{c}{\emph{Chandra} Exp.~Time (s)}\\
	\multicolumn{1}{c}{(1)} & \multicolumn{1}{c}{(2)} & \multicolumn{1}{c}{(3)} \\
	\hline
    SDSS J084905.51+111447.2 & 14969 & 19800 \\
	\dots & 18196 & 20980 \\
	SDSS J085837.67+182223.3 & 14970 & 19800 \\
	SDSS J102700.40+174900.8 & 14971 & 49410 \\
    NGC 3341 & 13871 & 49330 \\ 
    SDSS J163115.52+235257.5 & 13901 & 18150 \\
    SDSS J170859.12+215308.0 & 13903 & 18200 \\
    SDSS J235654.30-101605.3 & 18195 & 8620 \\
	\hline
\end{tabular*}
\end{center}
Note. -- Columns: (1) Galaxy name; (2) \emph{Chandra} Observation ID; (3) exposure time of \emph{Chandra} observation.
\end{table*}

\begin{figure*}
\centering
    \begin{minipage}{0.9\linewidth}
    \includegraphics[width=\linewidth]{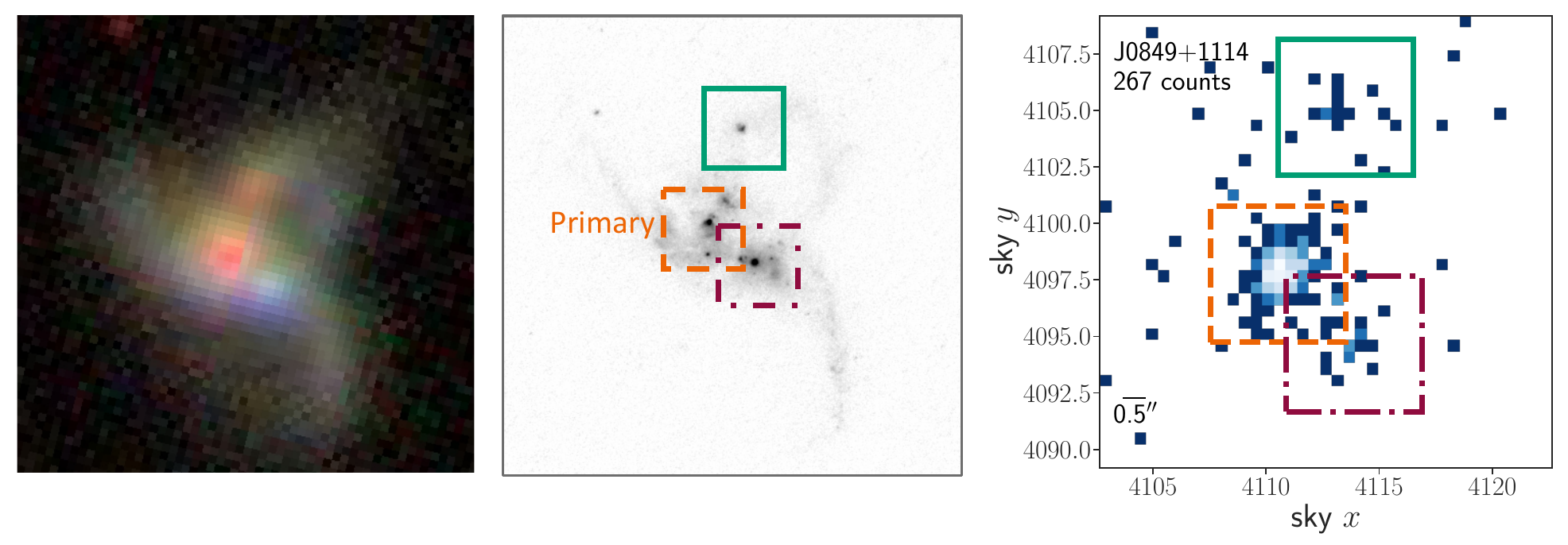}
    \end{minipage}
    \vspace{-0.1cm}
 
    \begin{minipage}{0.9\linewidth}
    \includegraphics[width=\linewidth]{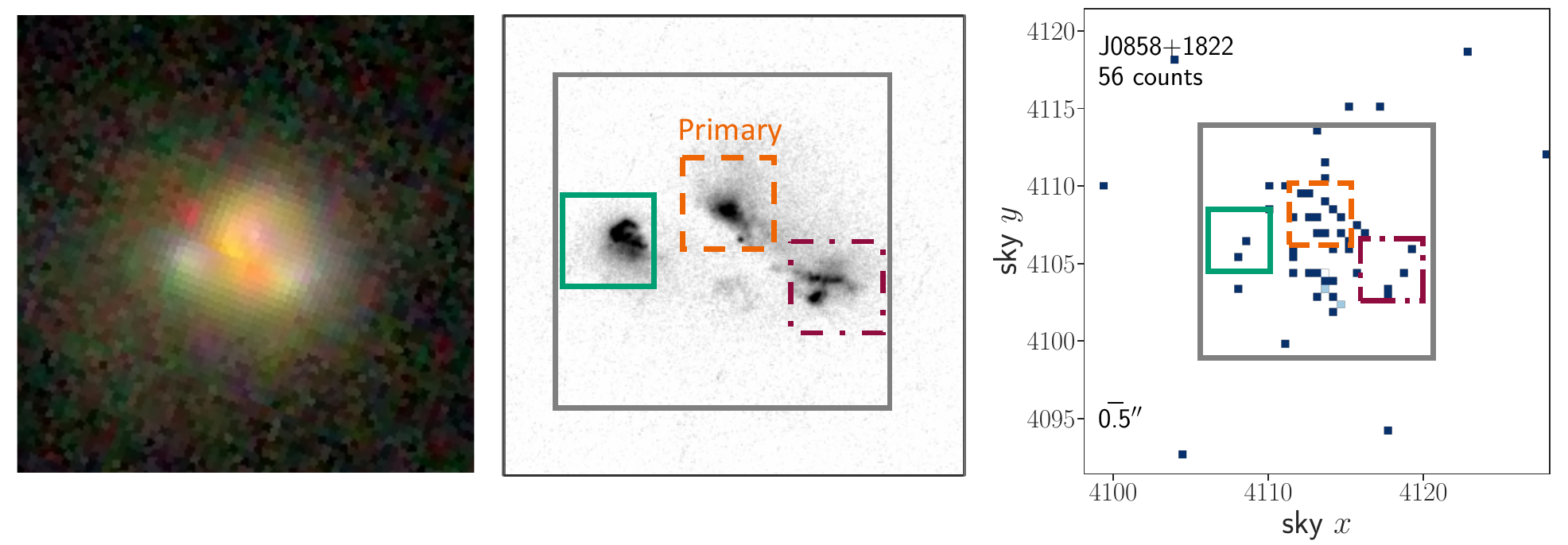}
    \end{minipage}
    \vspace{-0.1cm}
 
    \begin{minipage}{0.9\linewidth}
    \includegraphics[width=\linewidth]{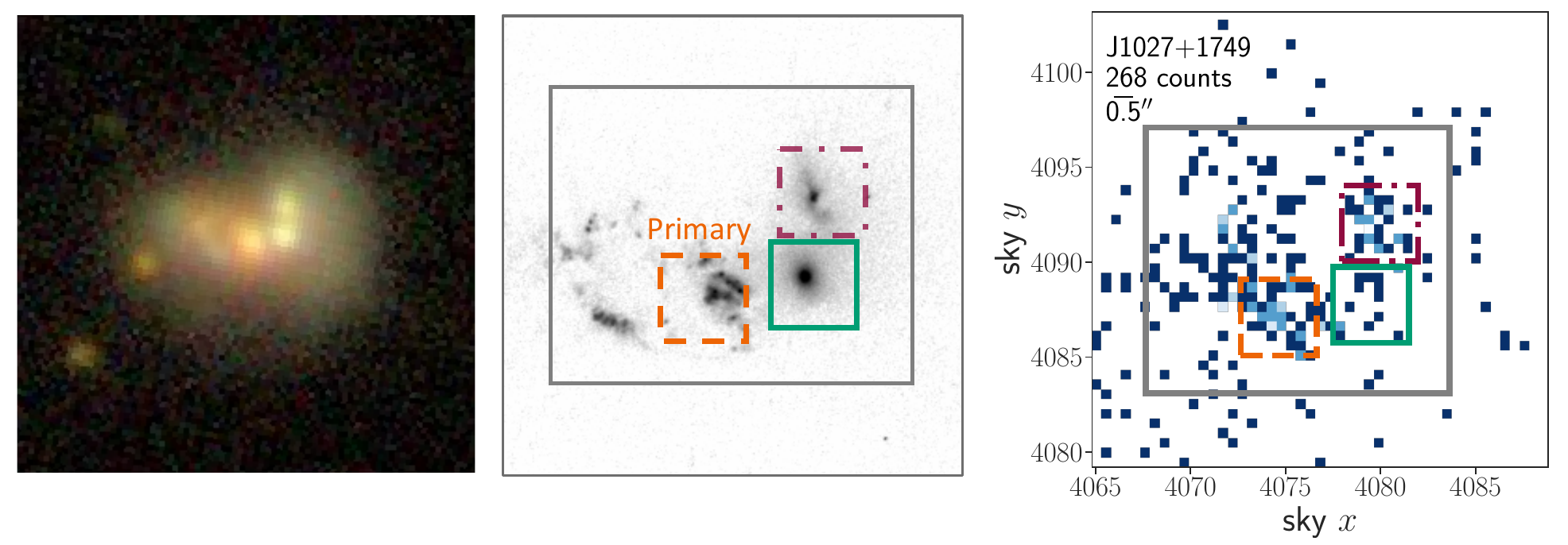}
    \end{minipage}
\caption{SDSS gri color composite observations (\emph{left}), \emph{HST} F366W observations (\emph{center}), and \emph{Chandra} $0.5$--$8$ keV observations (\emph{right}) of the triple mergers in our sample with \emph{HST} observations.  In the \emph{HST} and \emph{Chandra} datasets, we show the sky $x$, sky $y$ region, within which the informative priors for $\mu$ (sky x, y, position) are constrained to in orange (dashed), purples (dot-dashed), and green boxes.  When using non-informative priors, the central locations for the primary and secondary are allowed to be anywhere within the FOV shown in the X-ray image. For SDSS J0858+1822 and SDSS J1027+1729 we denote the region within which the diffuse emission background component is restricted to with a gray box. Additionally, for SDSS J0849+1114 we show the combined X-ray emission for all \emph{Chandra} observations, where we use the best-fit astrometric shift values as found by \BAYMAX{}. The X-ray images have been binned to \emph{Chandra}'s native pixel resolution. In all panels, north is up and east is to the left, and a 0\farcs{5} bar is shown to scale.}
\label{chap6:figTripleGalaxyImages}
\end{figure*}

\begin{figure*}
\centering
    \begin{minipage}{0.68\linewidth}
    \includegraphics[width=\linewidth]{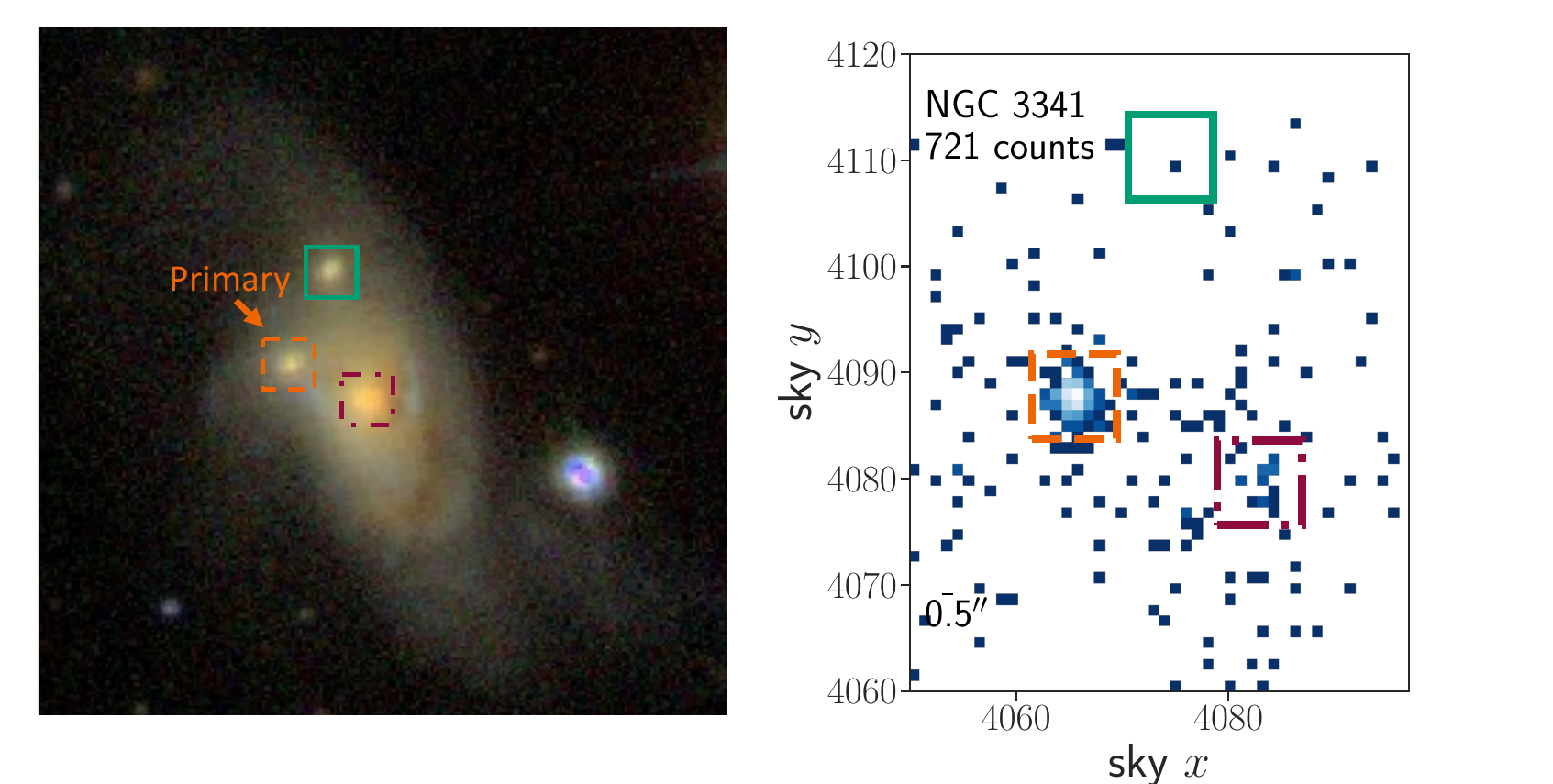}
    \end{minipage}
 
    \begin{minipage}{0.68\linewidth}
    \includegraphics[width=\linewidth]{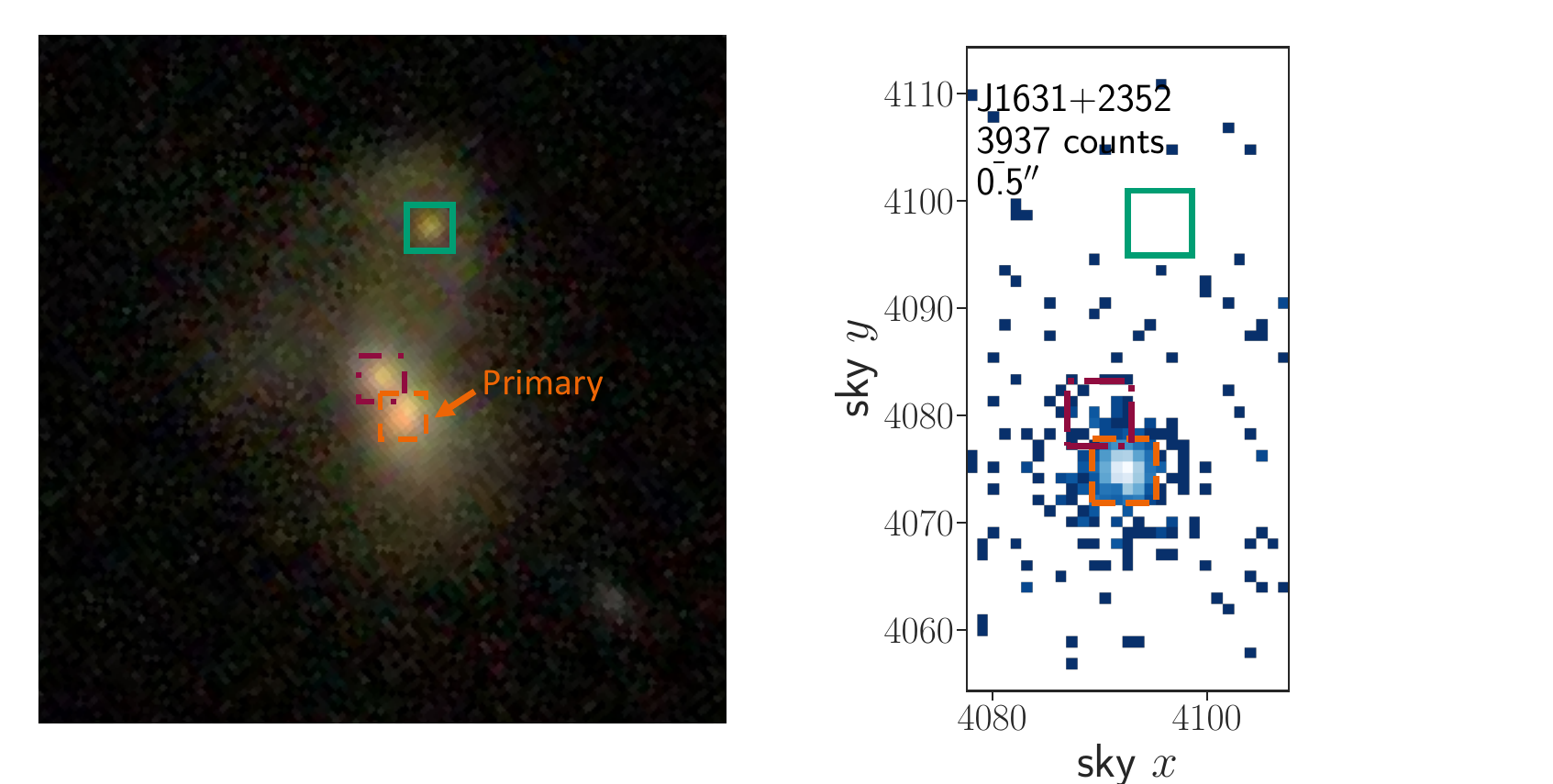}
    \end{minipage}
 
    \begin{minipage}{0.68\linewidth}
    \includegraphics[width=\linewidth]{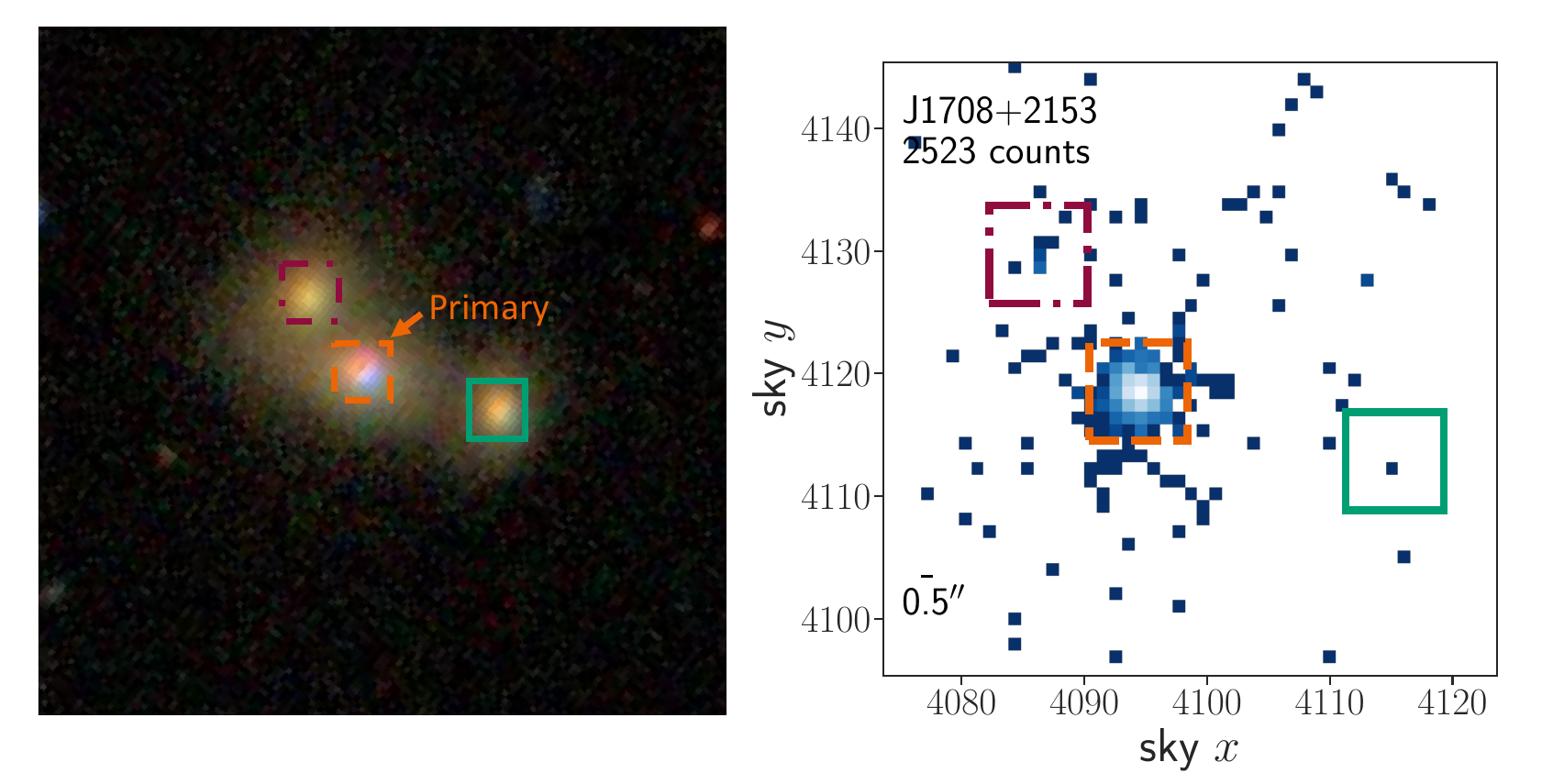}
    \end{minipage}
    \vspace{-0.05cm}
    
    \begin{minipage}{0.68\linewidth}
    \includegraphics[width=\linewidth]{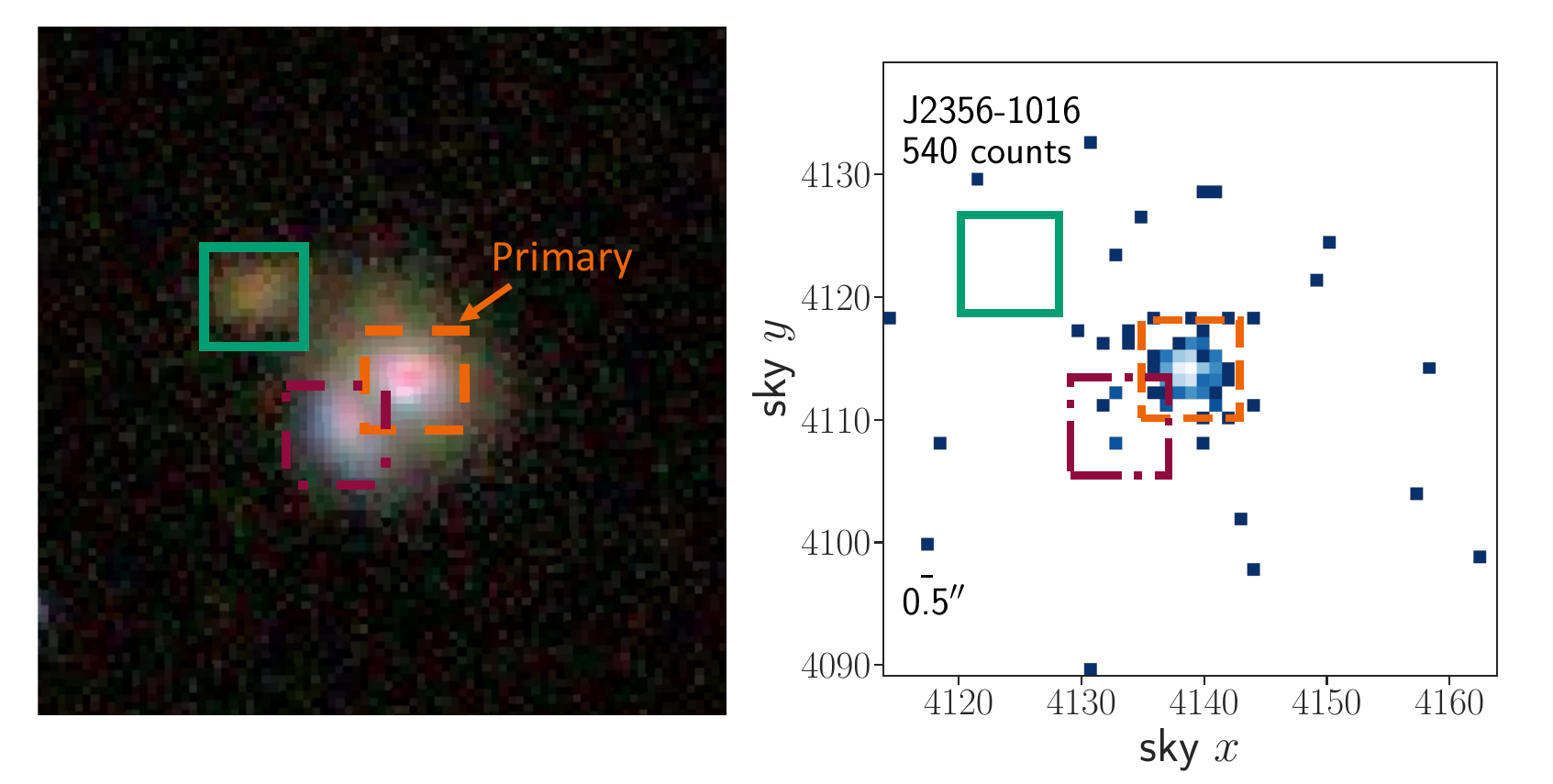}
    \end{minipage}
 \vspace{-0.25cm} 
\caption{SDSS gri color composite observations (\emph{left}), and \emph{Chandra} $0.5$--$8$ keV observations (\emph{right}) of the triple mergers in our sample with no \emph{HST} observations.}
\label{chap6:figTripleGalaxyImages2}
\end{figure*}

The \emph{Chandra} observations of 3/7 systems in our sample have been previously analyzed for the presence of multiple AGN: SDSS J0849+1114 (classified as a triple AGN system in \citealt{Liu2019, Pfeifle2019b}), NGC 3341 (classified as a single AGN system in \citealt{Bianchi2013}), and SDSS J2356$-$1016 (classified as a single AGN system in \citealt{Pfeifle2019a}). The remaining four systems have no existing analyses on the possible multiplicity of AGN in the \emph{Chandra} observations, although SDSS J1027+1749 was claimed as triple AGN in \cite{Liu2011b} based on optical emission diagnostics. The archival \emph{Chandra} observations for all of these systems were part of different programs studying merging systems: SDSS J0849+1114, SDSS J0858+1822, and SDSS J1027+1749 were observed by \emph{Chandra} as part of a program analyzing triple-AGN candidates; SDSS J1708+2153 and SDSS J1631+2252 were observed as part of a program studying mergers in BAT AGN (e.g. \citealt{Koss2010}); SDSS J2356$-$1016 was observed as part of a program studying interacting galaxies with extreme red mid-IR colors; while the \emph{Chandra} observations NGC 3341 was not part of a larger program, but the system was similarly observed to search for the presence of 3 AGN. Analyzing these galaxies using \BAYMAX{}, we aim to identify new X-ray point sources, as well as re-evaluate the true nature of the previously claimed multiple AGN systems. The galaxies are located at redshifts $0.059 < z < 0.077$, and three (SDSS J0849+1114, SDSS J0858+1822, SDSS J1027+1749) have additional multiband HST/WFC3 (F506W and F5336W) imaging (PI: Liu, Proposal ID: 13112). In Table~\ref{chap6:tabTripleGalInfo} we list the properties of each triple merger system, while in Table~\ref{chap6:tabTripleGalXray} we list the \emph{Chandra} observation information.
\subsection{X-ray Data Analysis}
\par Each \emph{Chandra} observation was on-axis and placed on the back-illuminated S3 chip of the ACIS detector.  We follow a similar data reduction as described in previous \emph{Chandra} analyses searching for AGN (e.g., \citealt{Foord2017a, Foord2017b, Foord2019, Foord2020a}), using \emph{Chandra} Interactive Analysis of Observations software ({\tt CIAO}) v4.12 \citep{Fruscione2006}. 
\par We first correct for astrometry, cross-matching the \emph{Chandra}-detected point-like sources with the SDSS Data Release 9 (SDSS DR9) catalog. The \emph{Chandra} sources used for cross-matching are detected by running {\tt wavdetect} on the reprocessed level-2 event file. We require each observation to have a minimum of 3 matches with the SDSS DR9, and each matched pair to be less than 2\arcsec~from one another.  Five of the seven systems meet the criterion for astrometrical corrections, while the remaining 2 observations were taken in sub-arrays and do not have enough X-ray point sources to match with the SDSS DR9 catalog. However, we note that the lack of astronomical corrections have no effect on our X-ray data analysis, as the locations of each putative AGN are assumed to be relative to the primary X-ray point source, and our prior distributions for locations of each AGN are wide enough to account for the relative astrometric shifts between the \emph{Chandra} and optical observations (see section~\ref{chap6:methods} for more details). Lastly, for each observation we find the background flaring contribution to be negligible, with no time interval containing a background rate 3$\sigma$ above the mean level.
\par Both SDSS J1708+2153 and SDSS J1631+2352 were observed in 1/8 sub-array mode to reduce pile-up. We quantify the levels of pile-up by running the {\tt CIAO} tool {\tt pileup\textunderscore map} on both observations, which generates an output image with the number of counts per ACIS frame per pixel (these values can be converted to pile-up fractions; see \citealt{Davis2001}). We find that both SDSS J1708+2153 and SDSS J1631+2352 reach pile-up fractions $<5\%$ in the brightest pixel. Although both observations have visible read-out streaks, neither spatially intercept the locations of the other galaxies in each merger. The minimal pile-up and read-out streaks should have a negligible affect on our X-ray analysis.

\section{Methods}
\label{chap6:methods}
To analyze each \emph{Chandra} observation for the presence of multiple X-ray point sources, we use \BAYMAX{} (Bayesian AnalYsis of Multiple AGN in X-rays). \BAYMAX{} is a tool designed to quantitatively evaluate whether a given \emph{Chandra} observation is a single or multiple point source via a Bayesian framework (\citealt{Foord2019, Foord2020a}). \BAYMAX{} is especially powerful for systems with low count ratios between the AGN ($0.1 < f< 1.0$), and angular separations ($r$) below $0\farcs{5}$. For example, analyses on multiple AGN systems in the low-$f$ and/or low-$r$ regime without \texttt{BAYMAX} are likely to lead to false negatives/positives \citep[e.g.,][]{Koss2015, Foord2020a}. With {\tt BAYMAX}, we've analyzed many low count ratio dual AGN candidates, and are methodically expanding the small group of X-ray confirmed dual AGN \citep{Foord2020a}.
\par 
In order to determine the likelihood of a given triple merger hosting multiple AGN, \BAYMAX{} calculates the Bayes factor (hereafter denoted by $\mathcal{B}$). The Bayes factor represents the ratio of the marginal likelihood of two competing hypotheses. The value can be interpreted as a measure of the strength of evidence in favor of one hypothesis over the other. The Bayes factor can be defined mathematically as:
\begin{equation}
\mathcal{B}_{j/i} = \frac{\int P(D\mid\theta_{j},M_{j}) P(\theta_{j}\mid M_{j}) d\theta_{j}}{\int P(D\mid\theta_{i},M_{i}) P(\theta_{i}\mid M_{i}) d\theta_{i}}
\end{equation}
Here, the marginal probability density of the observed data $D$ under one model ($M$) is represented by $P(D\mid\theta,M)$, while each model is parameterized by a parameter vector, $\theta$, and thus the prior densities are represented by $P(\theta\mid M)$. Regarding our X-ray analysis on the triple galaxy mergers, we calculate the evidence for three different models: a single point source model (i.e., the data are consistent with the X-ray emission of a single AGN; $M_{1}$), a dual point source model ($M_{2}$), and a triple point source model ($M_{3}$). Thus, we calculate the Bayes factor twice, comparing the evidence of the dual point source model to that of the single point source model ($\mathcal{B}_{2/1}$), and the evidence of the triple point source model to the dual point source model ($\mathcal{B}_{3/2}$).
\par The main components of \BAYMAX{} are as follows: (i) take calibrated \emph{Chandra} events and compare them to the expected distribution of counts for single/multiple point sources models; (ii) calculate a Bayes factor to determine which model is preferred; (iii) calculate likely values for separations ($r$) and count ratios ($f$); and (iv) fit spectra to each model component. Regarding step (i), the probability densities of each model are estimated by comparing the sky coordinates ($x$, $y$) and energies ($E$) of each detected X-ray event to simulations based on single and multiple point source models. 
\par The properties of the \emph{Chandra} PSF are characterized by simulating the PSF of the optics with {\tt MARX} \citep{Davis2012}. Testing has shown that the MARX PSF is more narrow than the \emph{Chandra} PSF in real observations of point-like sources (testing was carried out using observations of TYC 8241 2652 1, a young star that was observed in 1/8 sub-array mode to reduce pile-up), where the effect is most pronounced at a radius of 1 sky pixel\footnote{https://space.mit.edu/cxc/marx/tests/PSF.html}. At 1 pixel, the difference in the enclosed count fractions between the MARX and \emph{Chandra} PSF is $\sim10\%$. However, we've found that both in the low-count and wide-separation regime, the effect of these PSF differences is expected to be negligible. On top of this, all multiple AGN detected by \BAYMAX{} in this study have separations larger than 1 sky pixel, such that differences in the MARX and \emph{Chandra} PSF should not effect results. 
\par For a source with multiple observations, \BAYMAX{} first models the PSF of each observation and calculates the likelihoods for each observation individually (which is expected to depend on the detector position and start time of the observation), and then fits for astrometric shifts between different observations of the same source. Regarding step (ii) and (iii), \BAYMAX{} uses nested sampling \citep{Skilling2004} to estimate the evidence of each model (via the {\tt PYTHON} package {\tt nestle}\footnote{https://github.com/kbarbary/nestle}) and uses {\tt PyMC3} \citep{Salvatier2016} for parameter estimation. Below, we review the details of the prior densities used when running \BAYMAX{} on observations of our triple galaxy merger sample.  We refer the reader to \cite{Foord2019, Foord2020a} for an in-depth review of the statistical techniques used to estimate likelihoods and posterior densities.

\subsection{Prior Densities}
\par For all models, each photon is assumed to originate from either a point source component or the background. Thus, $M_{\mathrm{1}}$ is parameterized by vector $\theta_{1}=[\mu, \log{f_{bkg}}$], while $M_{\mathrm{2}}$ and $M_{\mathrm{3}}$ are parameterized by vector $\theta_{2,3}=[\mu_{{N}}, \log{f_{n}}, \log{f_{bkg}}]$. Here, $\mu_{N}$ represents the location for $N$ point sources ($N=[1, \dots,  N]$); $f_{bkg}$ represents the ratio of counts between the background and the combined counts from all point source components; $f_{n}$ represents the ratio of the total counts between a given point source and the primary point source ($n=[2, \dots, N]$), and in Fig.~\ref{chap6:figTripleGalaxyImages} we show which X-ray point source we classify as the primary. For the one system with multiple observations (SDSS J0849+1114, see Table~\ref{chap6:tabTripleGalXray}), the parameter vectors additionally include $\Delta x_{k}$ and $\Delta y_{k}$, which account for the translational components of the relative astrometric registration for the $k=[1, ..., K-1]$ observation (where $K=2$ for SDSS J0849+1114).  For a single point source, the probability that a photon observed at sky coordinate ($x$, $y$) with energy $E$ is described by the PSF centered at $\mu$ is $P(x,y \mid \mu, E)$, while for multiple point sources the total probability is $P(x, y \mid \mu_{N}, E, f_{n}, f_{\mathrm{bkg}})$. 
\par We assume that events associated with the background are uniformly distributed across a given region, such that the probability that a photon observed at location $x$,$y$ on the sky with energy $E$ is associated with a background component is $P(x,y \mid f_{bkg},E)$. Because we assume that each background component is uniformly distributed, $P(x_{i},y_{i}\mid f_{bkg},E)$ is always constant over a given region of interest. With this current model, it is possible that a spatially uniform background with a higher count-rate sitting among a spatially uniform background with a lower count-rate can be mistaken for a resolved point source.  Specifically, 2 of the 7 triple merger systems (SDSS J0858+1822 and SDSS J1027+1749), show evidence in their \emph{Chandra} observations for a high-count, diffuse, emission surrounding the galactic nuclei (see Fig.~\ref{chap6:figTripleGalaxyImages}). Extended, hot, gas is frequently detected in both simulations and observations of merging systems (see, e.g., \citealt{Cox2006, Brassington2007, Sinha2009, Hopkins2013, Smith2018, Foord2020a}). For example, multiple analyses on dual AGN candidate SDSS J0841+0101 have found that the emission can be well-fit by two point sources (\citealt{Comerford2015, Pfeifle2019a, Foord2020a}), but in \cite{Foord2020a} we found that a uniform, high-count, distribution better explains the X-ray emission associated with secondary galactic nucleus, instead of point-like emission from an AGN. Thus, for SDSS J0858+1822 and SDSS J1027+1749 we add an additional background component to our model. In Figure~\ref{chap6:figTripleGalaxyImages} we show these additional regions of background components in gray regions, where the position and size of these regions are chosen such that they cover the majority of the optical emission of the triple galaxy merger, as determined from the SDSS DR16 observations. Within these regions, \BAYMAX{} fits for a different background fraction, $f_{diff}$ than for outside these regions.
\par All prior distributions of $\mu$ for all models are described by continuous uniform distributions. When using non-informative priors, the coordinates of each $\mu$ are allowed to be anywhere within a given region centered on the X-ray source centroid position; when using informative priors, the coordinates of each $\mu$ are defined by the locations of the galaxy nuclei as determined in the SDSS DR16 observations shown in Figure \ref{chap6:figTripleGalaxyImages}. We note that our informative prior distributions for $\mu$ are wide enough to account for the relative astrometric shifts between the \emph{Chandra} and optical observations ($>1\arcsec$). The prior distribution of $\log{f_{bkg}}$ is described by a truncated Gaussian distribution, $N$($\mu_{bkg}$, $\sigma^{2}_{bkg}$), where $\mu_{bkg}$ is estimated by evaluating the count-rate in 10 random and source-free regions within a 20\arcsec\ radius of the X-ray source centroid position.  We set $\sigma_{bkg}$ to 0.5, allowing for \BAYMAX{} to easily move in parameter space, and we truncate $\log{f_{bkg}}$ at $-3$ and $0$.  For $M_{2}$ and $M_{3}$, the prior distributions of $\log{f_{n}}$ are described by uniform distributions and are constrained between $-4$ and $4$, accounting for a large range of possible count fractions between the X-ray emission of the AGN (and allowing for instances where the secondary or tertiary point source has more counts than the primary). For SDSS J0849+1114, the prior distributions of $\Delta x_{1}$ and $\Delta y_{1}$ are described by a uniform distribution constrained between $\delta \mu_{\mathrm{obs}}-3$ and $\delta \mu_{\mathrm{obs}}+3$, where $\delta \mu$ represents the difference between the observed central X-ray coordinates of two given observations. Here, we define $\Delta x_{k}$ and $\Delta y_{k}$ relative to the longest observation (ObsID: 18196). Lastly, for SDSS J0858+1822 and J1027+1749, the prior distribution of $\log{f_{diff}}$ is described by a uniform distribution constrained between $-2$ and $0$.
%

\section{Bayes Factor Results}
\label{chap6:results}
For each observation, we restrict our analysis to photons with energies between $0.5$--$8$ keV.  We analyze the photons contained within rectangular regions that are centered on the midpoint of the nominal X-ray coordinates of the three point sources, $\mu_{\mathrm{obs}}$. The sky $x$- and $y$-lengths of each rectangle are defined as $l_{x}$ and $l_{y}$, where $l_{x}$ and $l_{y}$ vary between 20 and 60 sky-pixels for each observation (9.9\arcsec and 29.7\arcsec, respectively; see Figure~\ref{chap6:figTripleGalaxyImages}). Most analyses use square regions with $l_{x}$=$l_{y}$, however we use a rectangular region for our analysis of NGC 3341 to avoid the inclusion of a nearby, bright, X-ray source (not associated with the triple merger) and our analysis of SDSS J1631+2532 due to the large ($\sim$4000) number of 0.5$-$8 keV counts in the observation. In the latter case, this helps to lower the computational time (as the probability densities are calculated by summing the log probability of each individual X-ray event, and thus the computational time increases as a function of X-ray events analyzed). The known asymmetric \emph{Chandra} PSF feature is within this extraction region \citep{Juda&Karovska2010}, and sits approximately 0\farcs7 from the center of the AGN.  Because our PSF model does not take into account this asymmetry, we mask the feature in all exposures before running \BAYMAX{}.
\par For each galaxy, we run \BAYMAX{} using non-informative priors --- where the prior distributions for $\mu$ are uniform distributions bound between ($\mu_{\mathrm{obs}}-\frac{l_{x}}{2}$, $\mu_{\mathrm{obs}}+\frac{l_{x}}{2}$) and  ($\mu_{\mathrm{obs}}-\frac{l_{y}}{2}$, $\mu_{\mathrm{obs}}+\frac{l_{y}}{2}$) --- and informative priors --- where the distributions for $\mu$ are constrained by and centered on each galaxy in the triple merger system (see Figure~\ref{chap6:figTripleGalaxyImages}).  Here, the sky $x$ and sky $y$ limits of each prior distribution were determined by visually identifying the possible extent of a galactic nucleus via the optical observations. 
%
\begin{table*}
\begin{center}
\caption{Bayes Factor Results}
\label{chap6:tabBF}
\begin{tabular*}{0.65\textwidth
}{lrrrr}
	\hline
	\hline
	\multicolumn{1}{c}{Galaxy Name} &  \multicolumn{1}{c}{$\ln{\mathcal{B}_{2/1}}$} & \multicolumn{1}{c}{$\ln{\mathcal{B}_{3/2}}$} &  \multicolumn{1}{c}{$\ln{\mathcal{B}_{2/1,\mathrm{inform}}}$} & \multicolumn{1}{c}{$\ln{\mathcal{B}_{3/2,\mathrm{inform}}}$}\\
	\multicolumn{1}{c}{(1)} & \multicolumn{1}{c}{(2)} & \multicolumn{1}{c}{(3)} & \multicolumn{1}{c}{(4)} &
	\multicolumn{1}{c}{(5)}\\
	\hline
    SDSS J0849+1114 & $23.4 \pm 1.9$ & $22.4 \pm 2.1$ & $25.6 \pm 1.4$ & $19.4 \pm 1.9$ \\
	SDSS J0858+1822 & $-2.3\pm 3.6$ & $2.1\pm 3.8\mathlarger{\dagger}$ & $0.6\pm 1.0$ & $-0.8 \pm 1.1$ \\
	SDSS J1027+1749 & $28.5 \pm 4.4$ & $-4.5 \pm 5.3$ & 34.2 $\pm$ 1.8 & $-$0.1 $\pm$ 1.9 \\
    NGC 3341 & $22.5 \pm 1.8$ & $5.8\pm 4.7\mathlarger{\dagger}$ & $2766.9 \pm 1.6$ & $-0.1 \pm 1.4$\\ 
    SDSS J1631+2352 & 2.7 $\pm$ 1.9 & 2.3$\pm$3.8$\mathlarger{\dagger}$ & 6.2 $\pm$ 1.6  & $-$0.8 $\pm$ 1.9 \\
    SDSS J1708+2153 & 16.6 $\pm$ 1.9 &  0.5 $\pm$ 3.9 & 18.4 $\pm$ 1.6 & $-$1.9 $\pm$ 1.8\\
    SDSS J2356$-$1016 & 0.7 $\pm$ 1.6 & $-$0.9 $\pm$ 3.2 &  3.1 $\pm$ 1.3 & $-$2.3 $\pm$ 1.4\\
	\hline
\end{tabular*}
\end{center}
Note. -- Columns: (1) Galaxy name; (2) and (3) Bayes factor in favor of dual point source model and triple point source model, using non-informative priors on the locations ($\mu$) of the point sources; (4) and (5) Bayes factor in favor of dual point source model and triple point source model, using informative priors on the locations ($\mu$) of the point sources. Error bars represent the 99.7\% confidence intervals. 
$\mathlarger{\dagger}$ -- Although $\ln{\mathcal{B}_{3/2}}$ favors the triple point source model for these systems, we emphasize that for SDSS J1027+1749 and SDSS J1631+2352 $\ln{\mathcal{B}_{3/2}}$ is consistent with 0 at the 99.7\% C.L. For SDSS J1027+1749, the triple point source model is \emph{not} favored over the single point source model i.e, $\ln{\mathcal{B}_{3/1}} \approx -0.15$. These results are likely due to clumpy, diffuse, X-ray emission, see section~\ref{chap6:spectral} for more details. For NGC 3341, $\ln{\mathcal{B}_{3/2}}$ weakly favors the triple point source model (when taking into account to the large error bars), again likely due to the presence of hot clumpy gas (see section~\ref{chap6:results}).
\end{table*}
%
\par Of the seven triple mergers, using informative priors, we find 1 that favors the triple X-ray point source model (SDSS J0849+1114), 5 that favor the dual X-ray point source model (SDSS J1027+1749, NGC 3341, SDSS J1631+2352, SDSS J1708+2153, SDSS J2356$-$1016) and 1 that favors the single point source model (SDSS J0858+1822). Here, if both $\ln{\mathcal{B}_{1/2,\mathrm{inform}}}$ and $\ln{\mathcal{B}_{3/2,\mathrm{inform}}}$ are consistent with 0, we classify the system as favoring the single point source model. In most cases where the Bayes factor favors a multi-point source model: (i) the Bayes factor favors the same model when using informative and non-informative priors, and (ii) the evidence for a given model is stronger when using informative priors. The one exception to this is SDSS J2356$-$1016, where the Bayes factor only favors the dual point source model when using informative priors. In Table~\ref{chap6:tabBF} we list the various $\ln{\mathcal{B}_{2/1}}$ and $\ln{\mathcal{B}_{3/2}}$ values for each of the 7 triple merger systems.  
\par The error bars on the Bayes factor represent the 99.7\% confidence intervals, as determined by {\tt nestle}. The errors provided by {\tt nestle} are calculated for each model, at each iteration, and are assumed to be proportional to the ratio of the ``information" (denoted \emph{H}, and represents the logarithm of the fraction of prior mass that contains the bulk of the posterior mass) to the number of live points (see \citealt{Skilling2004} for explicit details). We have found these errors to be consistent with the 3$\sigma$ spread of the $\ln{\mathcal{B}}$ values when running \BAYMAX{} multiple times on a single dataset.
\subsection{Multiple X-ray Point Sources: Strength of the Bayes factor}
\label{chap6:strength}
For each triple merger that has a Bayes factor that favors either the triple or dual point source model, we run false-positive tests to better, and uniquely, define a ``strong" Bayes factor. Past analyses have defined arbitrary Bayes factor thresholds, above which values are deemed strong \citep{Jeffreys1935, Jeffreys1961, Kass&Raftery1995}. However, the interpretation of a strong Bayes factor value depends on the details of the dataset, and for our particular observations it depends on parameters such as the number of counts, count ratio, and separation. 
\par Following the procedure outlined in \cite{Foord2019, Foord2020a}, for each system that has a Bayes factor favoring a model with multiple point sources, we simulate a suite of single point source simulations that are based on the \emph{Chandra} observations. The simulations are created via {\tt MARX} and use the same detector position, pointing, and exposure time of the \emph{Chandra} observations (such that, on average, the point source will have the same number of counts as the primary point source). Additionally, the point source has the same spectrum as that of the primary point source. We only analyze the counts contained within the same sky coordinates and energy range as the observations, we use the same informative priors, and we add a uniform background contribution with a similar background fraction as each observation. For SDSS J1027+1749, we also add a synthetic diffuse emission component, constrained within the same region as shown in Fig.~\ref{chap6:figTripleGalaxyImages}. For each system, we run \BAYMAX{} on 100 such simulations and calculate what fraction have $\ln{\mathcal{B}_{2/1}} > 0$, or $\ln{\mathcal{B}_{3/2}} > 0$ (for the one triple point source system, SDSS J0849+1114). 
\par For the false-positive runs based on SDSS J1027+1749, NGC 3341, SDSS J1631+2352, and SDSS J1708+2153, 0/100 of the $\ln{\mathcal{B}_{2/1}}$ values are larger than the measured values of 34.20, 2766.90, 6.20, and 18.40; for SDSS J2356$-$1016, only 2/100 of the $\ln{\mathcal{B}_{2/1}}$ values are larger than the measured value $3.10$; while for SDSS J0849+1114, 0/100 of the $\ln{\mathcal{B}_{3/2}}$ values are larger than the measured value of 19.40. We interpret these results to mean that there is $\le$99\% (or, $<98\%$ for SDSS J2356$-$1016) chance that a single point source with a comparable number of counts would return a Bayes factor value, in favor of a multiple point source model, greater than what we measure. Thus, we classify each Bayes factor value as ``strong" in favor of the dual, or triple, point source model as indicated in Table~\ref{chap6:tabBF}.

\section{X-ray Spectral Analysis}
\label{chap6:spectral}
In the following section we analyze the X-ray spectra of each point source individually, as well as analyze the posterior distributions returned from \BAYMAX{}. For each system, we determine the best-fit values of each model parameter using the median values of their posterior distributions, which is appropriate given their unimodal nature. In Table~\ref{chap6:tabpymc3} we list the best-fit values for $r$, $\theta_{\mathrm{PA}}$, $\log{f}$ and $\log{f_{bkg}}$. Here, $\theta_{\mathrm{PA}}$ is the position angle between the primary and secondary (or, tertiary, for SDSS J0849+1114). All errors bars reported in this section are evaluated at the 99.7\% confidence level, unless otherwise stated.
\subsection{XSPEC Modeling}
\par From our analysis in section~\ref{chap6:results}, we find 5 triple merger systems with Bayes factors that strongly favor the dual point source model: SDSS J1027+1749, NGC 3341, SDSS J1631+2352, SDSS J1708+2153, SDSS J2356$-$1016; while we find one triple merger system with a Bayes factor that strongly favors the triple point source model: SDSS J0849+1114. As outlined in \cite{Foord2020a}, we use \BAYMAX{} to carry out a spectral analysis of individual point source components in each observation. The spectral fits are determined via XSPEC, version 12.9.0 \citep{Arnaud1996}. Each system has either 2 or 3 point sources, hereafter the ``primary" (as defined in Fig.~\ref{chap6:figTripleGalaxyImages}),
``secondary", or ``tertiary". We create 100 spectral realizations of each point source component by probabilistically sampling from the full distribution of counts.  Each spectral realization uses $\theta_{2}$ or $\theta_{3}$ values that are drawn from the posterior distributions as determined by \BAYMAX{}.  For each iteration, \BAYMAX{} assigns each count to a specific model component, based on the relative probabilities of being associated with each component. 
\par By fitting the 100 spectra of each point source component, we create distributions of the best-fit values for various spectral parameters, as well as the 0.5$-$8 keV flux and unabsorbed 2$-$7 keV luminosity. This additional analysis expands on the posterior distributions returned by \BAYMAX{}, as fitting the spectral realizations for each point source results in estimates of the flux ratio, instead of the count ratio. 
\subsubsection{Fitting a Phenomenological Model}
\par
\begin{figure*}
\centering
    \includegraphics[width=0.48\linewidth]{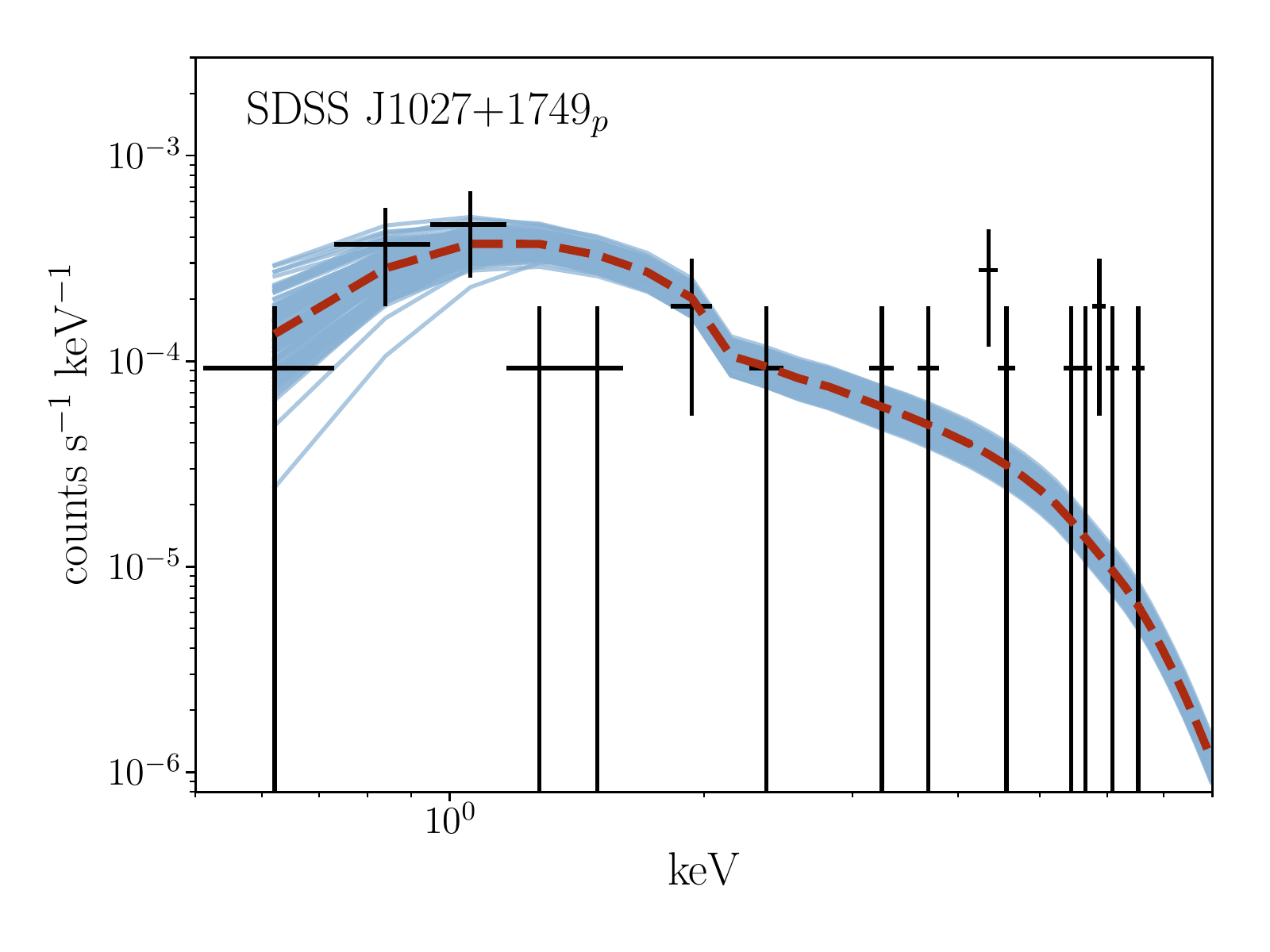}{\hskip 0pt plus 0.3fil minus 0pt}
    \includegraphics[width=0.48\linewidth]{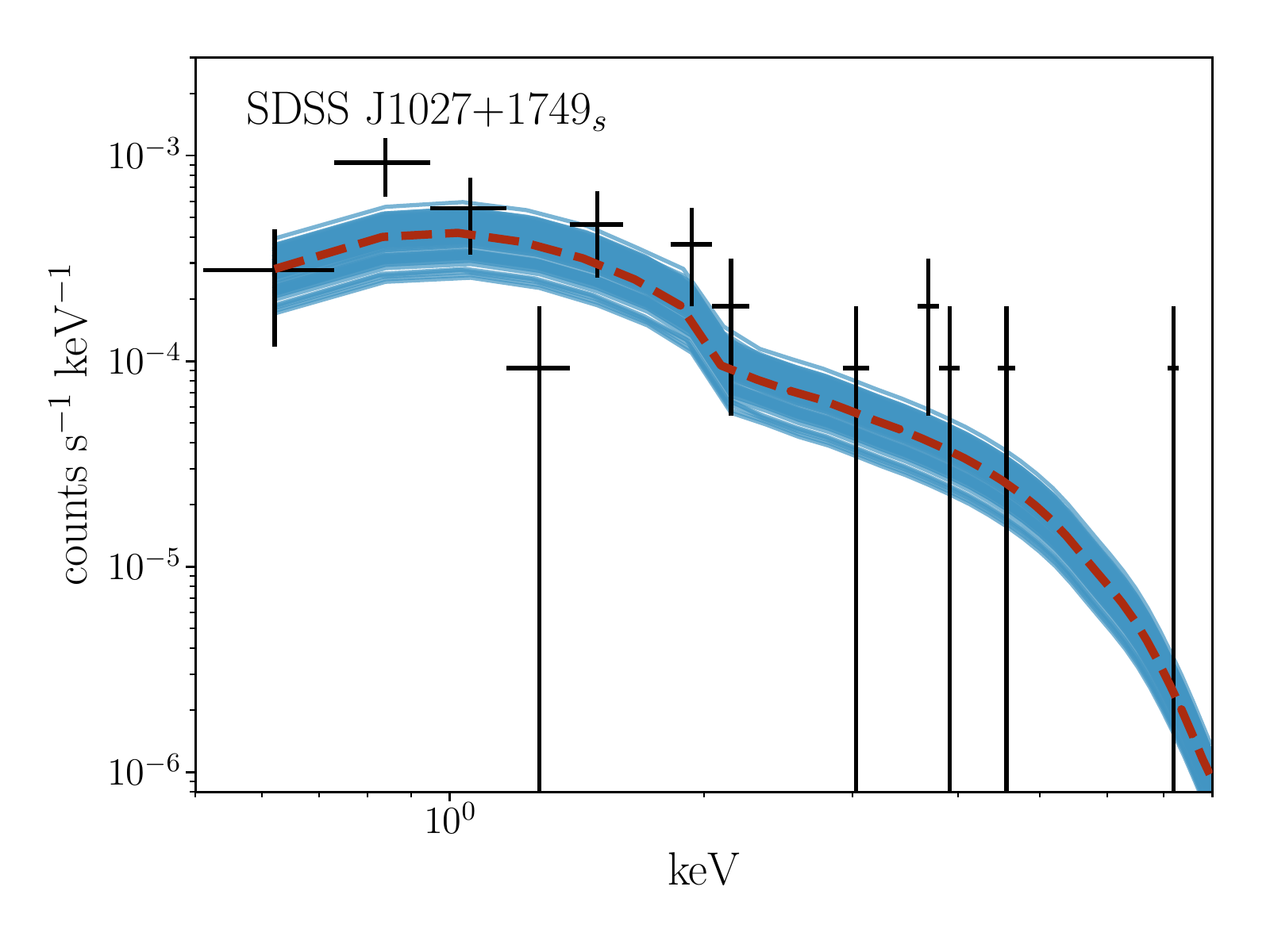}{\vskip -0.2cm plus 1fill}
     \includegraphics[width=0.48\linewidth]{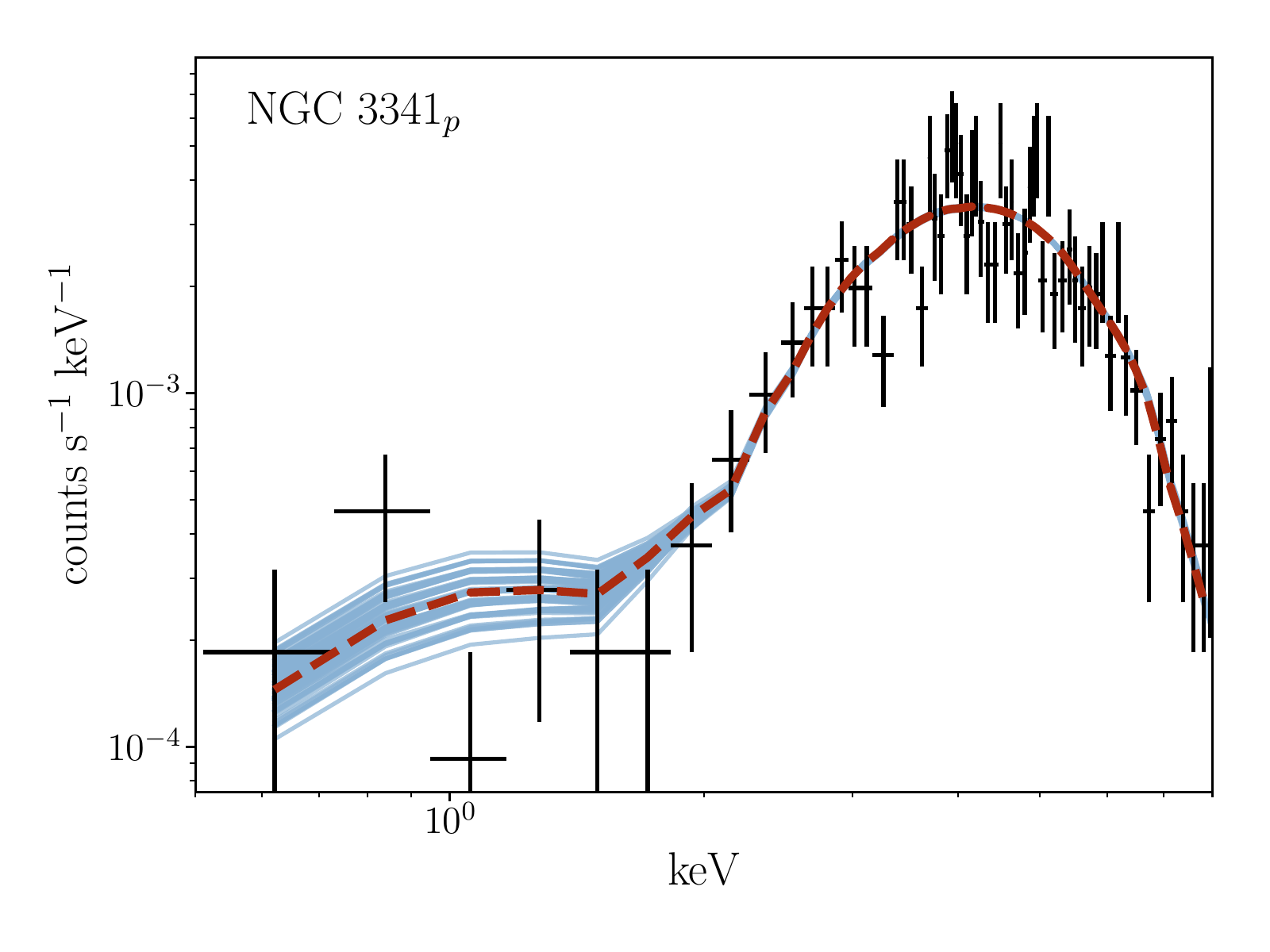}{\hskip 0pt plus 0.3fil minus 0pt}
    \includegraphics[width=0.48\linewidth]{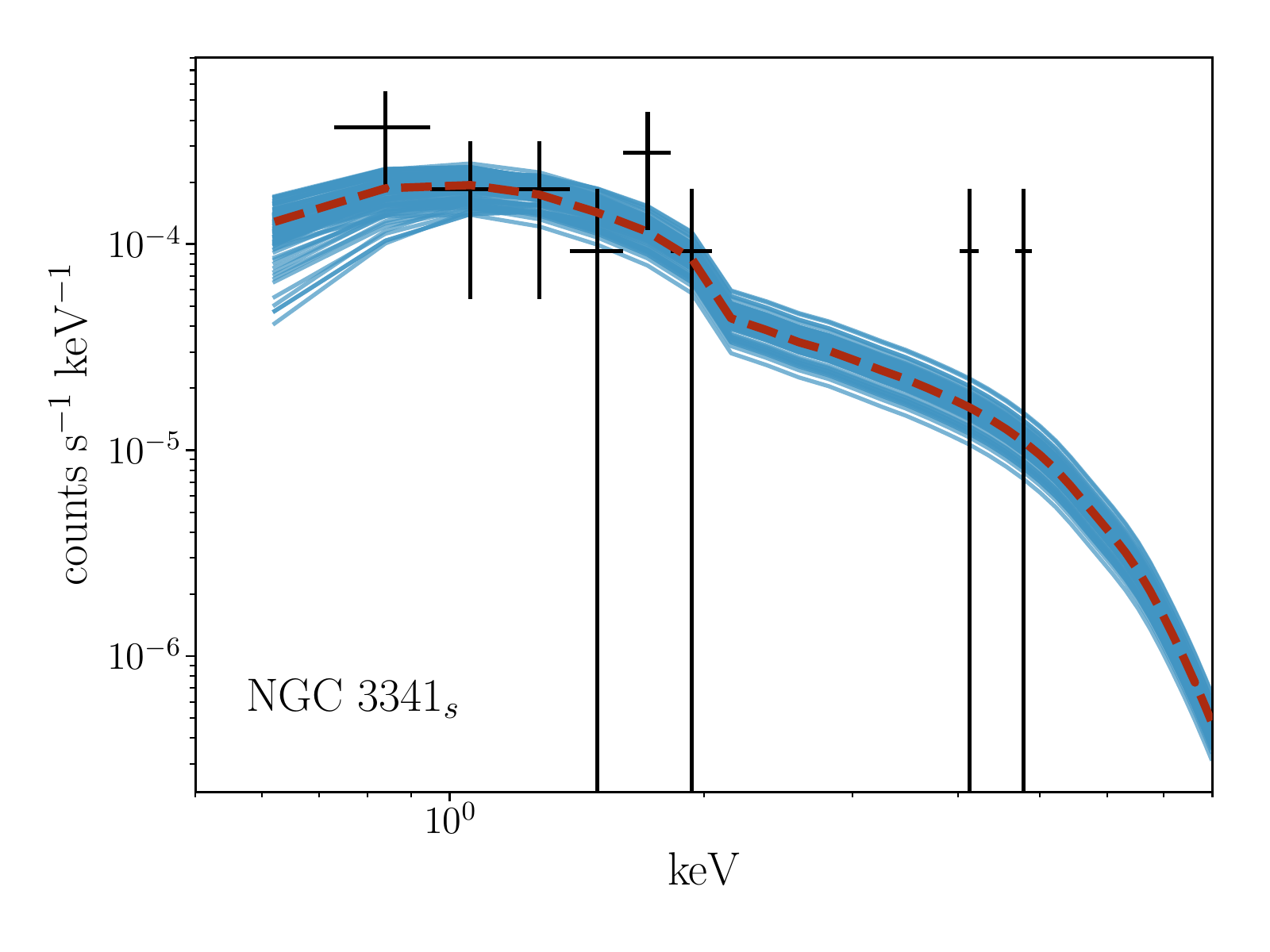}{\vskip -0.2cm plus 1fill}
    \includegraphics[width=0.48\linewidth]{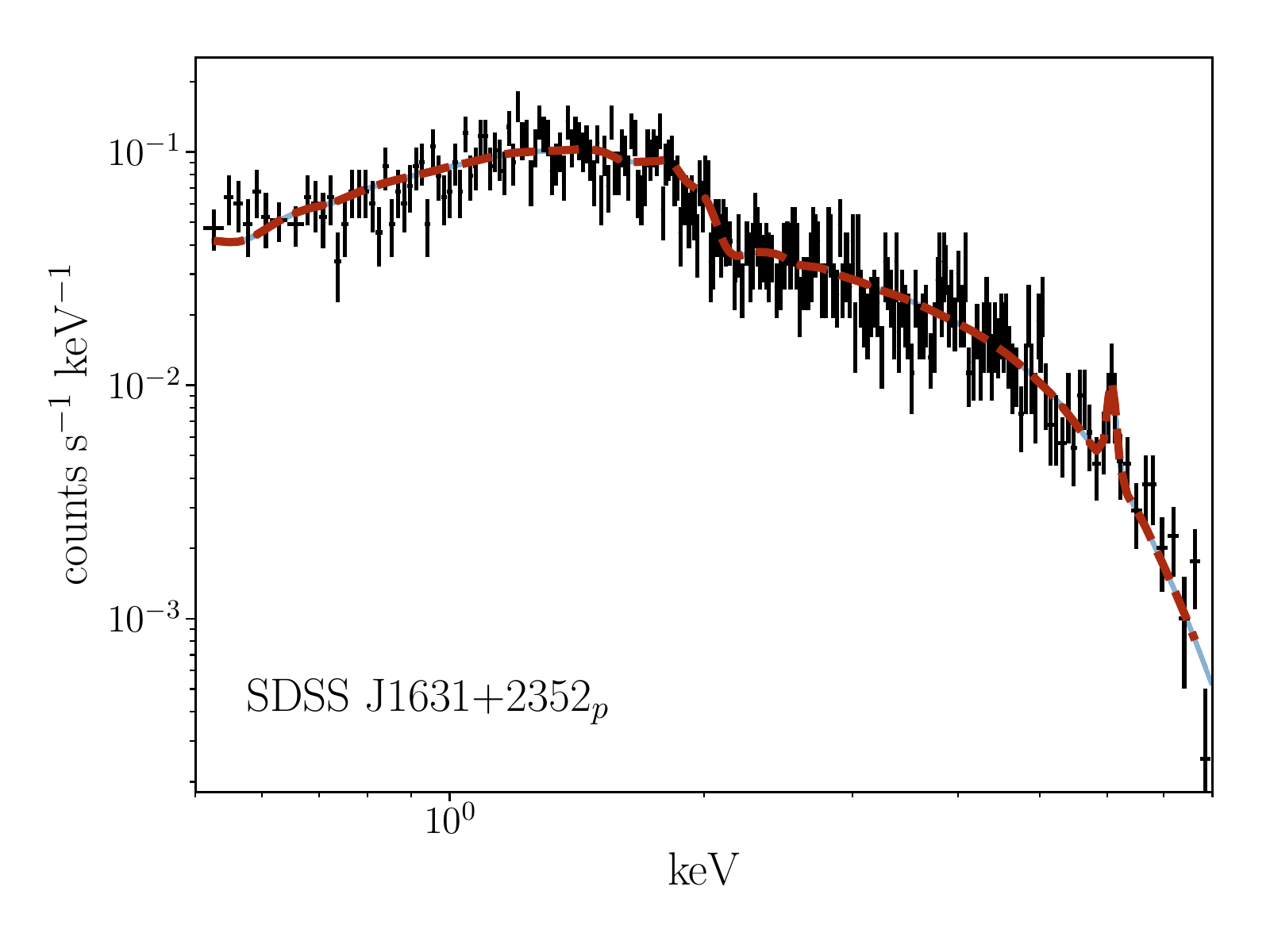}{\hskip 0pt plus 0.3fil minus 0pt}
    \includegraphics[width=0.48\linewidth]{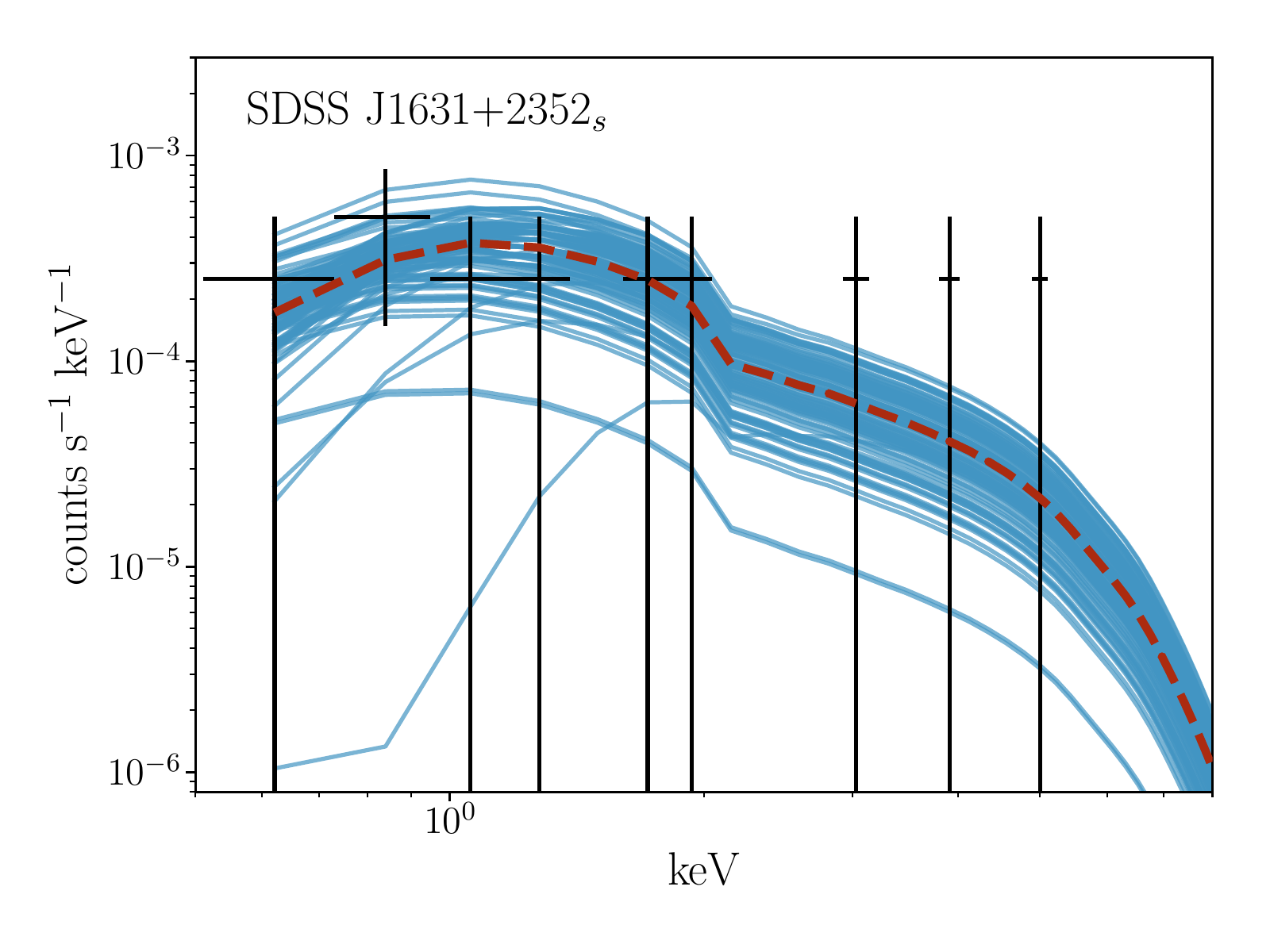}{\vskip -0.2cm plus 1fill}
    
    \caption{X-ray spectral fits of 100 realizations for the primary point source (\emph{left}) and the secondary point source (\emph{right} in SDSS J1027+1749, NGC 3341, and SDSS J1631+2352. Data have been folded through the instrument response. We overplot one of the spectral realizations in black and the median spectral fit in a red dashed line. The spectra have been rebinned for plotting purposes. We list the best-fit values for each model in Table~\ref{chap6:tabspectra}, defined as the median of the distribution of the best-fit values from the 100 realizations.}

\label{chap6:spec1}
\end{figure*}

\begin{figure*}
\centering
    \includegraphics[width=0.48\linewidth]{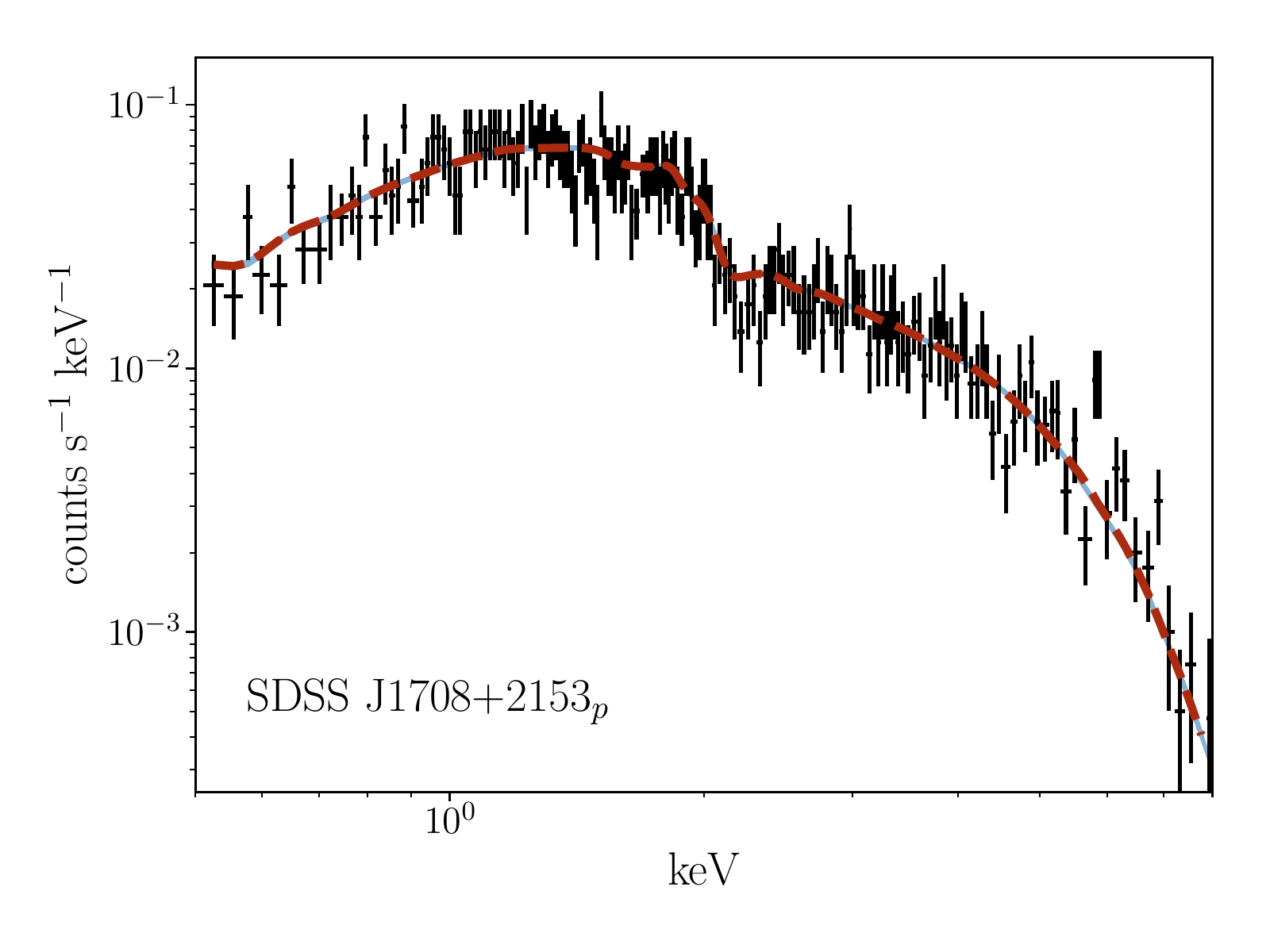}{\hskip 0pt plus 0.3fil minus 0pt}
    \includegraphics[width=0.48\linewidth]{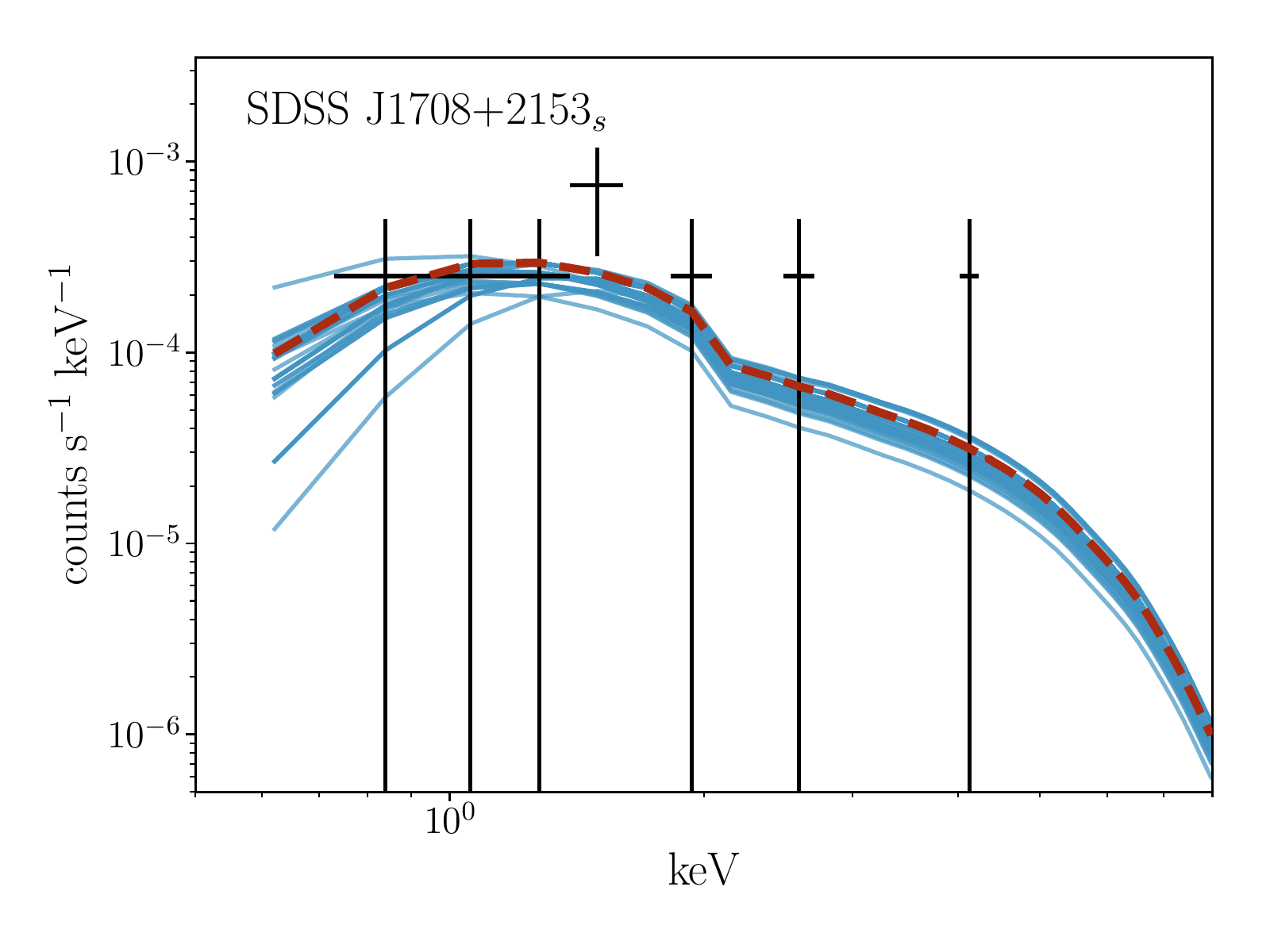}{\vskip -0.2cm plus 1fill}
     \includegraphics[width=0.48\linewidth]{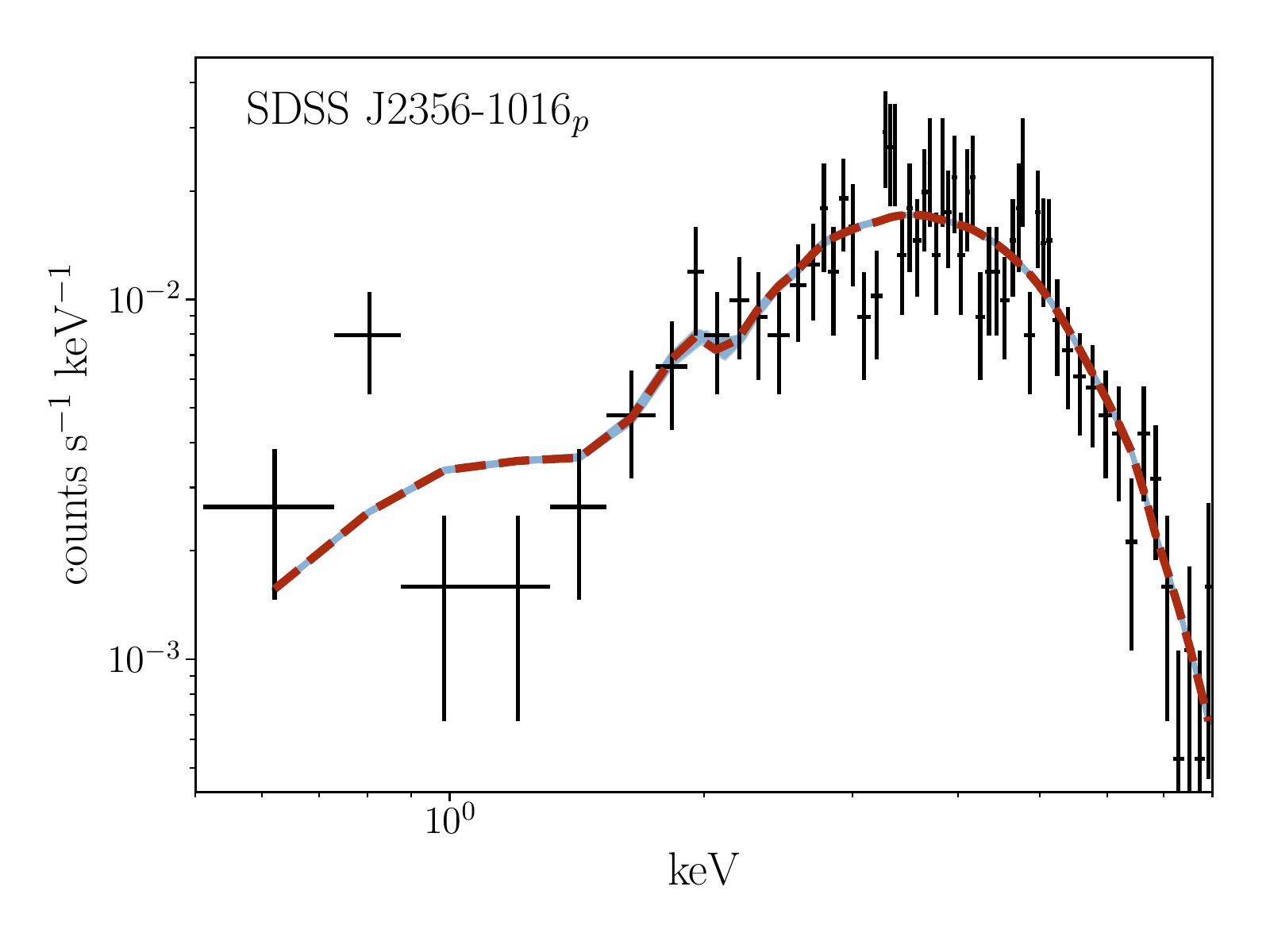}{\hskip 0pt plus 0.3fil minus 0pt}
    \includegraphics[width=0.48\linewidth]{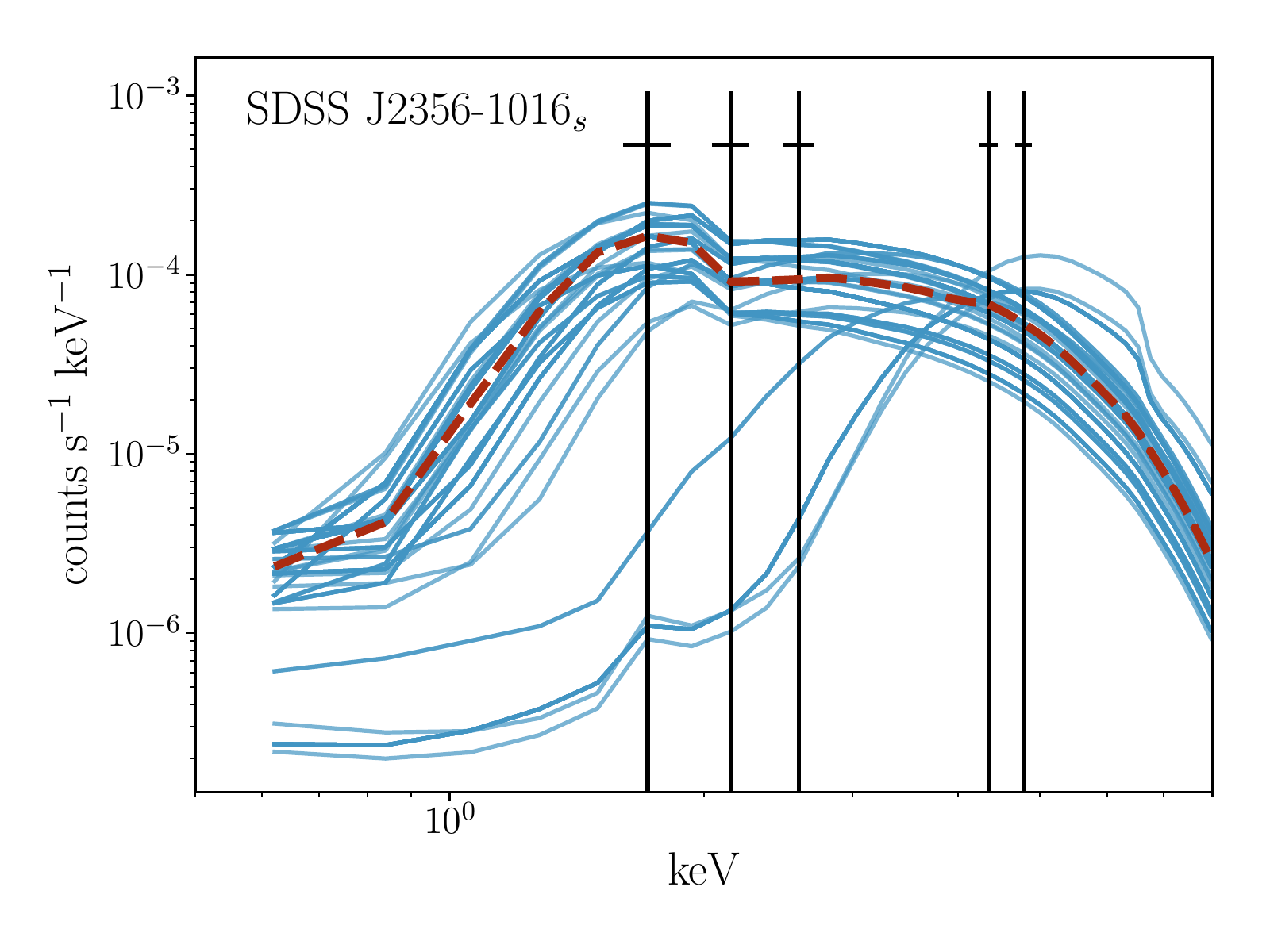}{\vskip -0.2cm plus 1fill}

    \caption{\emph{Chandra} spectral fits of 100 realizations for the primary (\emph{left}) and secondary (\emph{left}) X-ray point sources in SDSS J1708+2153 and SDSS J2356$-$1016. Plotting symbols and lines follow the the same guidelines as Fig.~\ref{chap6:spec1}.}

\label{chap6:spec2}
\end{figure*}

\begin{figure*}
\centering
    \includegraphics[width=0.48\linewidth]{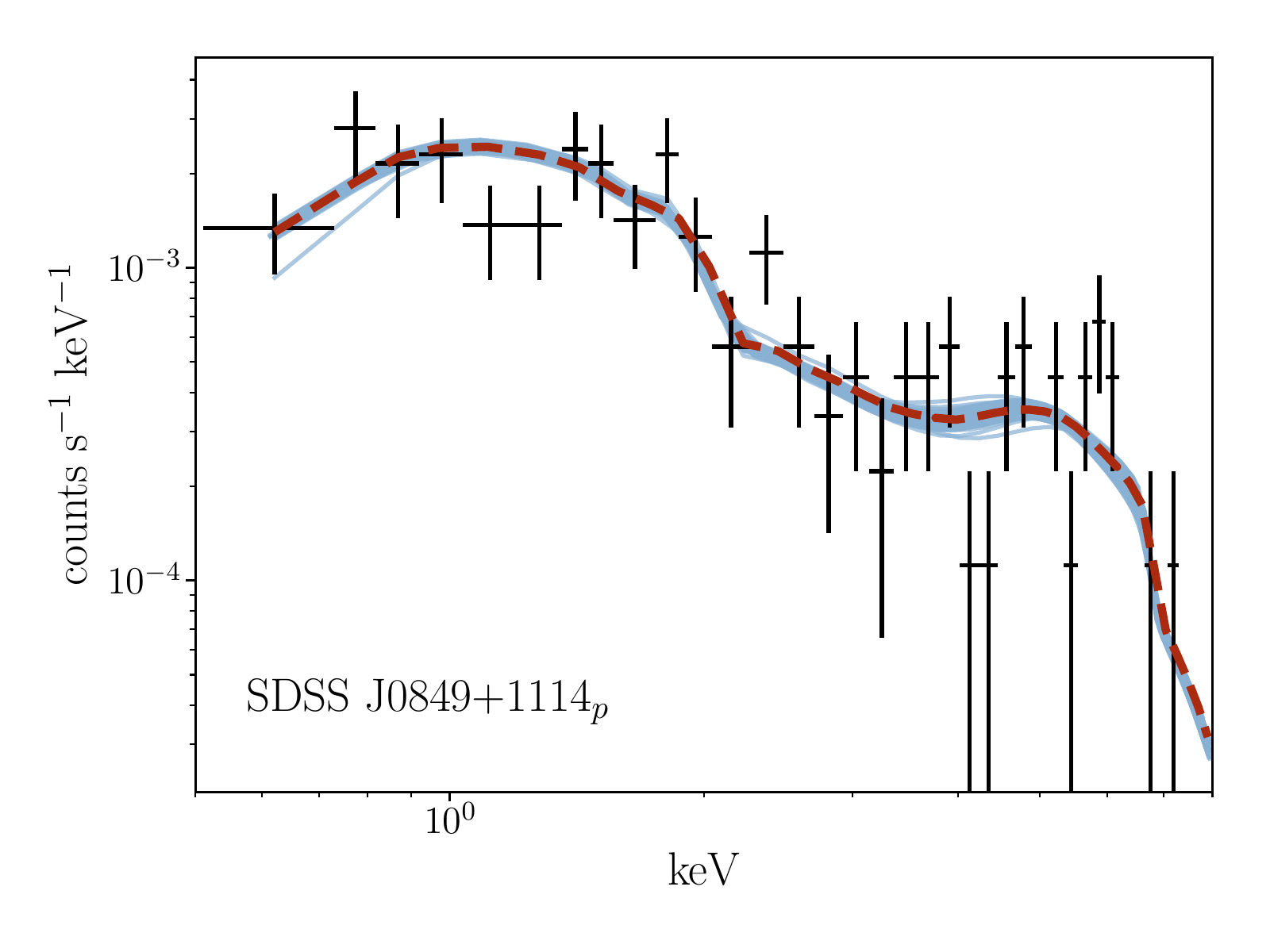}{\hskip 0pt plus 0.3fil minus 0pt}
    \includegraphics[width=0.48\linewidth]{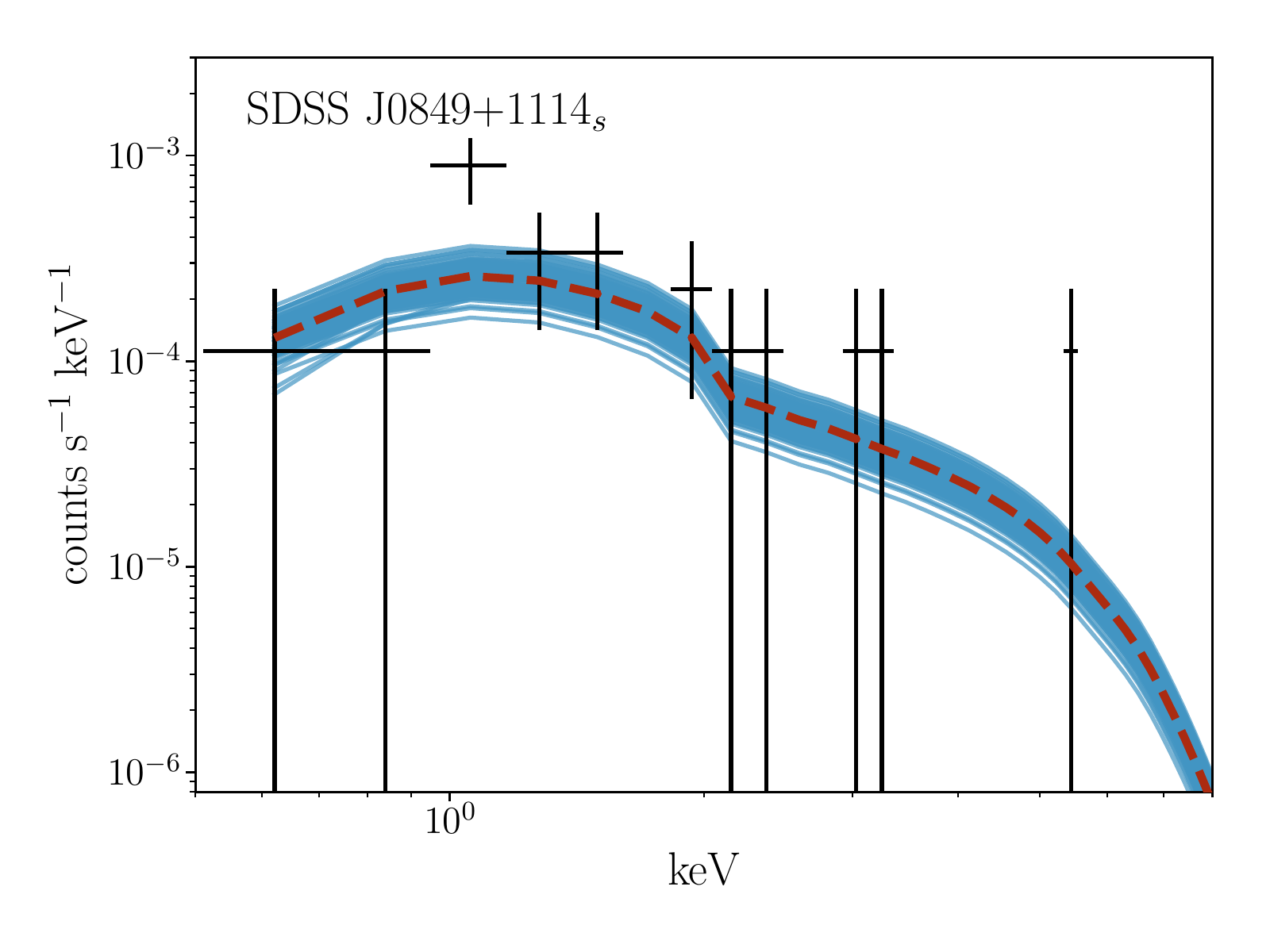}{\vskip -0.2cm plus 1fill}
    \includegraphics[width=0.48\linewidth]{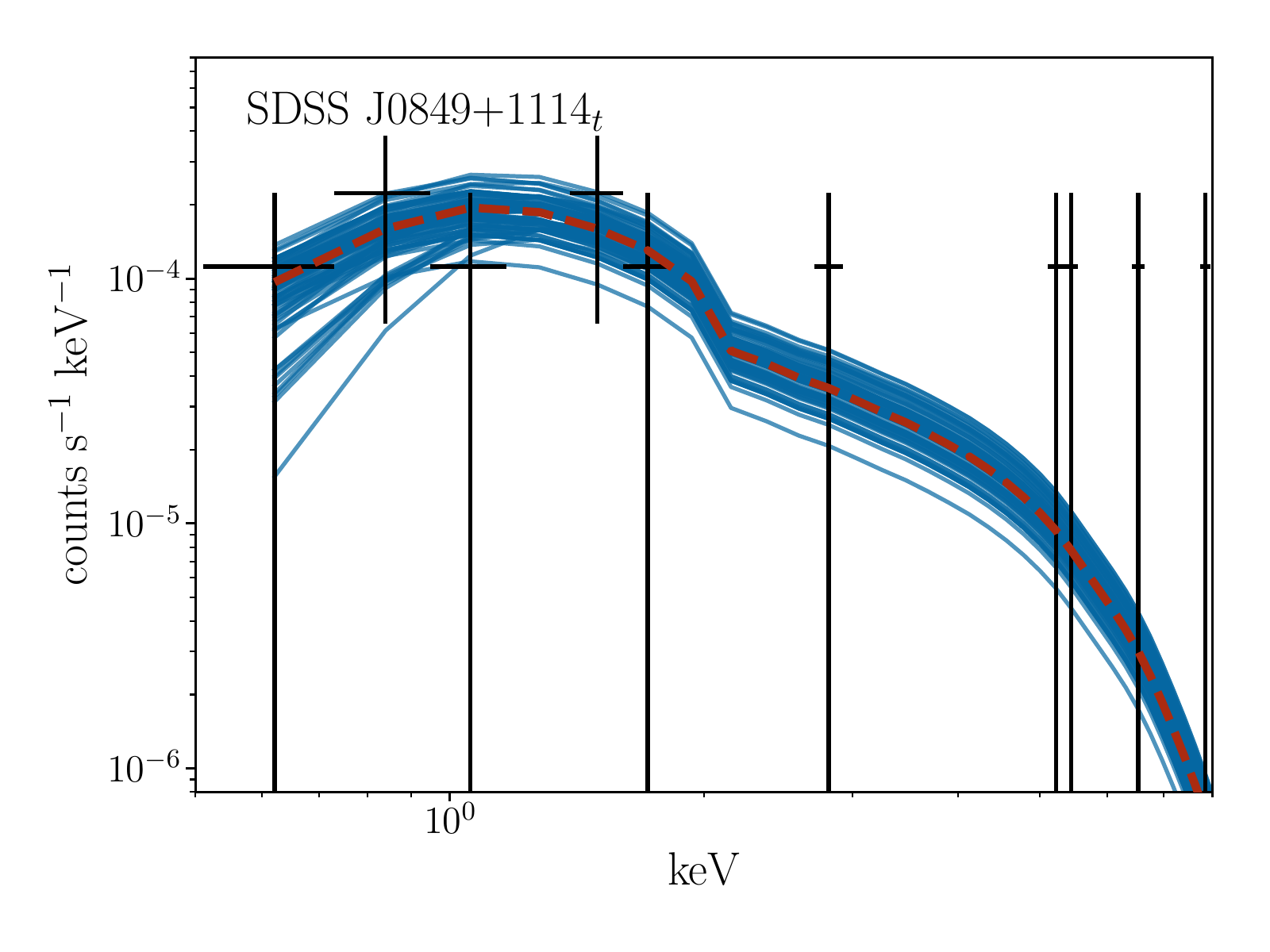}{\vskip -0.2cm plus 1fill}
    \caption{\emph{Chandra} spectral fits of 100 realizations for the primary (\emph{top left}), secondary (\emph{top right}), and tertiary (\emph{bottom}) X-ray point sources in SDSS J0849+1114. Plotting symbols and lines follow the the same guidelines as Fig.~\ref{chap6:spec1}.}

\label{chap6:spec3}
\end{figure*}

 Similar to \cite{Foord2020a}, each point source component is modeled as either a simple absorbed power-law ({\tt phabs}$\times${\tt zphabs}$\times${\tt zpow}; hereafter $m_{\mathrm{phen,1}}$) or an absorbed power-law with Compton scattering ({\tt phabs}$\times$({\tt pow} + {\tt zphabs}$\times${\tt zpow}); $m_{\mathrm{phen,2}}$), where the power-law indices are tied to one-another. We implement the Cash statistic ({\tt cstat}; \citealt{Cash1979}) in order to best assess the quality of our model fits.  Specifically, the latter model is used if it results in a statistically significant improvement in the fit, such that $\Delta C_{\mathrm{stat}}>$ $n \times$2.71 (where $n$ represents the difference in number of free parameters between the models; \citealt{Tozzi2006, Brightman2012}), corresponding to a fit improvement with 90\% confidence (\citealt{Brightman2014}). Because we are evaluating distributions of spectral parameters, we require $\Delta C_{\mathrm{stat}}>$ $n\times$2.71 at the 99.7\% confidence level. For both $m_{\mathrm{phen,1}}$ and $m_{\mathrm{phen,2}}$, if the best-fitting model where $\Gamma$ is free is a significantly better fit than the best-fitting model where $\Gamma$ is fixed (fixed to a value of 1.8; \citealt{Corral2011,Yan2015}) we choose the model with $\Gamma$ as free as the best-fitting model.  Unsurprisingly, for point sources with a low number of average counts ($<20$), the models where $\Gamma$ is fixed tends to be a significantly better fit \citep{Brightman2012}. For each model, we fix the column density to the Galactic value \citep{Kalberla2005} as well as the redshift to that of the host galaxy.  We list the best-fit values for each spectral parameter, $F_{0.5-8}$, $L_{2-7~\mathrm{keV,~unabs}}$, and hardness ratio (\emph{HR}) in Table~\ref{chap6:tabspectra}. The \emph{HR} is defined as $(H - S)/(H + S)$ where $H$ and $S$ are the number of hard (2$-$8 keV) and soft (0.5$-$2 keV) X-ray counts. In Figures~\ref{chap6:spec1},~\ref{chap6:spec2}, and \ref{chap6:spec3} we show our various spectral fits to the spectral realizations of each X-ray point source.
\subsubsection{Fitting a Physical Model}
\par In addition to the aforementioned phenomenological model, we fit the spectral realizations that have, on average, over 100 counts between 0.5$-$8 keV (with the exception of the secondary in SDSS J0849+1114, which we also include and, on average, has 99 counts between 0.5$-$8 keV) using the physically motivated model {\tt BNTorus} \citep{Brightman2011}. For a Compton thick AGN, a physical model may be able to better constrain parameters if the spectrum has a large-enough number of counts ($\sim$100 counts). We implement the {\tt BNTorus} model in XSPEC, which considers an X-ray source surrounded by a toroidal structure (see \citealt{Brightman2011} for more detail). Mirroring our analysis with the phenomenological model, we compare a simpler model ($m_{\mathrm{phys,1}}$, {\tt phabs}$\times$atable[torus1006.fits]) to a more complicated model that includes an additional scattered power-law component ($m_{\mathrm{phys,2}}$, {\tt phabs}$\times$({\tt s}$\times${\tt pow}+atable[torus1006.fits])). Here, $s$ is a constant and represents the scattering fraction of the additional power-law.
\par Similar to the {\tt powerlaw} component in XSPEC, the toroidal component in {\tt BNTorus} has parameters extragalactic $N_{H}$, photon index $\Gamma$, and redshift. Two additional parameters include the opening angle and the edge-on inclination of the torus; however we were unable to constrain either parameter during the spectral fits, and find that our results do not depend on their values. Thus, we fix them to default values of 60$^{\circ}$ and 87$^{\circ}$. When using $m_{\mathrm{phys,2}}$, we tie the photon indices and normalizations between the power-law and toroidal component. For both models, we fix the column density to the Galactic value \citep{Kalberla2005}, the redshift to that of the host galaxy, and allow each model to have both a fixed photon index ($\Gamma=1.8$) as well as a photon index that is free to vary.  We chose the most statistically significant model by analyzing the $C_{stat}$ distributions between the fits.
\par For all sources, the best-fit model to describe the spectra are consistent between the phenomenological and physical XSPEC models; furthermore, the best-fit values for the parameters are consistent with one another. On average, however, the best-fit intrinsic absorption values found via the {\tt BNTorus} toroidal model are slightly lower than those found via the phenomenological model. The exception to all of these results is the primary X-ray point source in NGC 3341, where a different model is favored when fitting with {\tt BNTorus} ($m_{\mathrm{phys,2}}$, where the photon-index is fixed) than when using a phenomenological description ($m_{\mathrm{phen,2}}$, but the photon-index is free to vary). Furthermore, the best-fit intrinsic absorption value calculated by {\tt BNTorus} is slightly larger than that calculated by the phenomenological model. Given that the primary X-ray point-source in NGC 3341 has one of the highest levels of obscuration, it is possible that fitting the spectral realizations for this source using a physically-motivated model will results in different, and better-constrained, model parameters. 
\par Lastly, we measure whether the spectral fits via {\tt BNTorus} over-fit data, which may occur when fitting lower count spectra. In this scenario, we may expect that the value of intrinsic absorption has little effect on the statistical significance of the fit. We analyze the $\chi_{\nu}$ values for each system, using the best-fit model as described above, as well as the $\chi_{\nu}$ values as a function $N_{H}$. In this latter case, we freeze the extragalactic $N_{H}$ component, and step through parameter space ($-2 < \log{N_{H}} < 2$, where $N_{H}$ is in units of $10^{22}$ cm$^{-2}$). For all spectra, we find best-fit $\chi_{\nu}$ values between 0.85$-$1.05, where the two spectra with the lowest number of counts (SDSS J0849+114$_{p}$ and SDSS J0849+114$_{s}$) have the two lowest $\chi_{\nu}$ values, both below 1. However, when stepping through $N_{H}$ parameter space, we find that all spectra have $\chi_{\nu}$ values that strongly depend on $N_{H}$, where large deviations from the best-fit $N_{H}$ value lead to large ($\gg$1) $\chi_{\nu}$ values. Thus, in the event that the physically motivated model is over-fitting the lower-count spectra, we don't expect that our estimates of $N_{H}$ are affected.
We list the best-fit values for each spectral parameter, $F_{0.5-8}$, and $L_{2-7~\mathrm{keV,~unabs}}$ in Table~\ref{chap6:tabspectraBN}.
\subsubsection{Testing for Effects of Pileup}
\par Lastly, for both SDSS J1708+2153 and SDSS J1631+2352,  we investigate how pile-up possibly affects spectrum of the primary X-ray point source. We convolve the spectral fits with the pile-up model implemented in XSPEC \citep{Davis2001}, which accounts for the energy shifts due to photon pile-up as well as flux decrements due to grade migration. The pile-up model has six parameters: the CCD frame time, the maximum number of photons to pileup, the grade correction for a single photon, the grade morphing parameter (which represents the probability, per photon count greater than one, that the piled event is not rejected by the spacecraft software as a ``bad event"; thus the larger this parameter, the larger the effects of pile-up), the PSF fraction considered, and the number of event regions. Default values were assumed for the maximum number of photons to pile-up (5), the grade correction for a single photon (1), and the number of event regions (1). The frame time was set to that of the observation and the PSF fraction was set to 0.95. However, we find that the best-fit spectral parameter values are consistent with the values presented below, and thus conclude that any levels of pile-up have negligible effect on our spectral results.
\par Additionally, we note that SDSS J1708+2153 and SDSS J1631+2352 were both observed as part of the 70-month Swift/BAT all-sky survey, observations that are sensitive to the 14–195 keV energy range. This analysis combined available X-ray observations from a range of instruments, such as \emph{XMM-Newton}, \emph{Swift/XRT}, \emph{ASCA}, \emph{Chandra}, and \emph{Suzaku}, to analyze the broadband X-ray (0.3$–$150 keV) characteristics of the AGN. We find that our X-ray spectral results agree with those presented in \cite{Ricci2017b}, where the \emph{Swift/XRT} observations of both AGN were not piled-up (see Sections~\ref{chap6:subsecJ1631} and \ref{chap6:subsecJ1708}). Thus, we conclude that any levels of pile-up have negligible effect on our spectral results.

\subsubsection{Luminosity Estimates for Tertiary AGN}
\par It is possible that additional X-ray point sources that either have low-luminosities within the \emph{Chandra} energy band, and/or are highly-obscured will be undetected by \BAYMAX{}. Thus, for systems with Bayes factors that favor the dual point source model, we estimate an upper-limit of the 2$-$7 keV luminosity of a possible tertiary X-ray point source.  For a given set of parameter vectors $\theta_{1}$, $\theta_{2}$, and $\theta_{3}$, \BAYMAX{} assigns each count to a specific model component (i.e, the primary, secondary, tertiary, or background), based on the relative probabilities of being associated with each component. Taking the best-fit values for $\theta_{2}$ as determined by \BAYMAX{}, we mask the counts associated with the primary and secondary point source, and analyze the counts within a 2\arcsec~radius extraction region centered on the optical coordinates of the galactic nucleus found to host no X-ray point source. We estimate the number of background-subtracted 0.5$-$8 keV counts associated with a possible tertiary, where the background contribution is estimated by evaluating the count-rate in 10 random and source-free regions, see section~\ref{chap6:methods}. These count rates are then converted into 2$-$7 keV luminosities via the HEASARC tool WebPIMMS (v4.11), assuming a power-law spectrum with photon index $\Gamma=1.8$. We estimate the intrinsic luminosities assuming both a relatively unobscured system (with intrinsic $N_{H} = 1\times10^{20}$ cm$^{-2}$), as well as a system with a higher-level of absorption along the line-of-sight (with intrinsic $N_{H} = 1\times10^{22}$ cm$^{-2}$; such values are expected in merging environments, e.g., \citealt{Ricci2017, DeRosa2018, Koss2012, Hou2020}).

\subsection{SDSS J1027+1749}
\label{chap6:subsecJ1027}
SDSS J1027+1749 was first identified as triple AGN candidate in \cite{Liu2011b}. Optical [\ion{O}{3}] $\lambda$5007 luminosities were individually measured for each galactic nucleus using the Dual Imaging Spectrograph (DIS) on the Apache Point Observatory 3.5 m telescope. Assuming each galaxy hosts an AGN, X-ray luminosities were estimated from the $\frac{L_{X,2-10\mathrm{keV}}}{L_{[OIII]}}$ relation for obscured AGN \citep{Panessa2007}. The estimated 2$-$10 keV luminosities for each AGN were all estimated to be greater than 10$^{42}$ erg s$^{-1}$, representing a robust classification as a triple AGN system. However, the \emph{Chandra} observation presents a more complicated scenario --- a high-count, diffuse, emission is coincident with the primary galactic nucleus, and very little X-ray emission appears to coincide with the location of the western galaxy. Our analysis with \BAYMAX{}, where we can include more complicated models that include various background regions, is necessary in order to understand whether the X-ray emission is consistent with three AGN.
\par
\begin{figure*}
\centering
\includegraphics[width=0.4\linewidth]{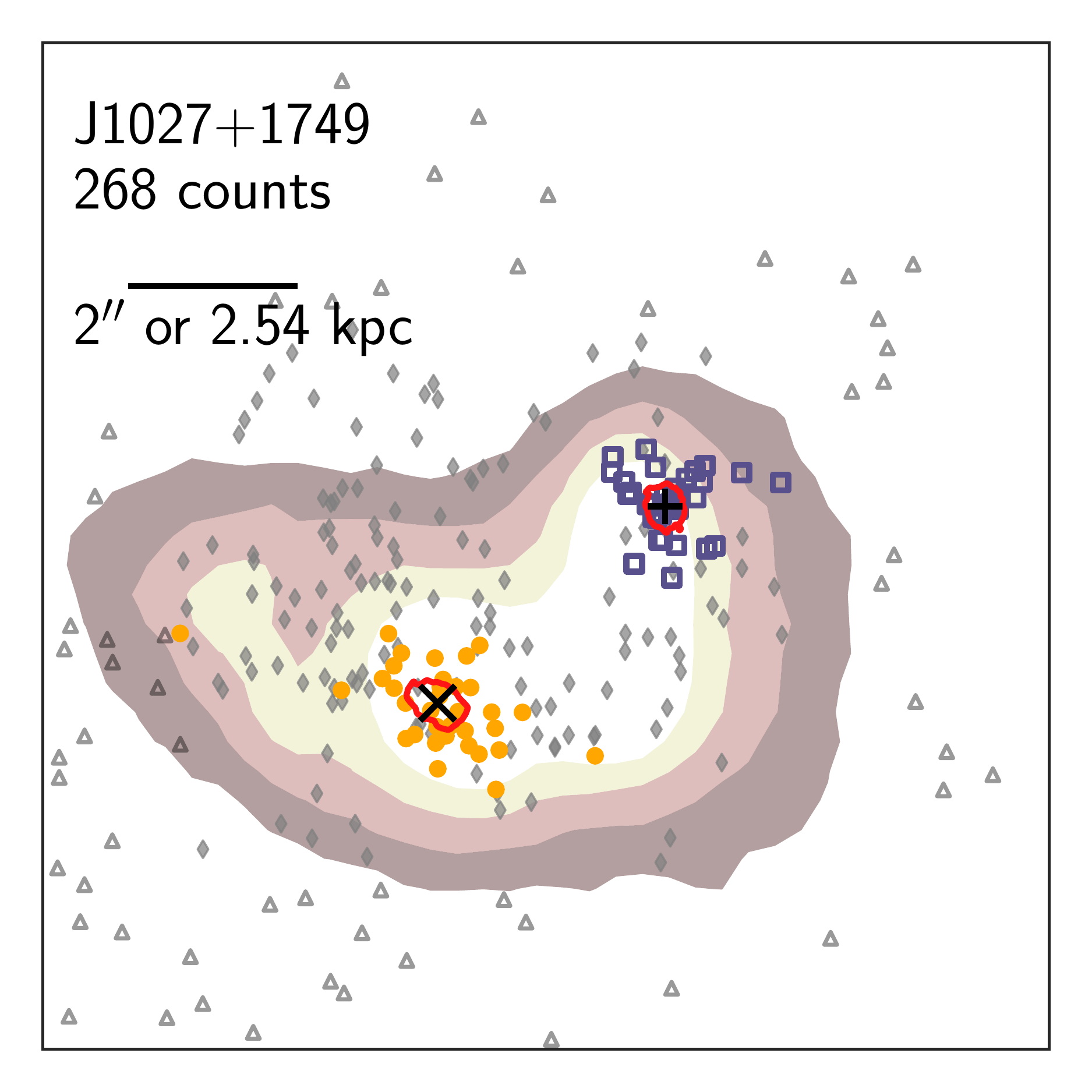}{\hskip 0pt plus 0.3fil minus 0pt}
\includegraphics[width=0.45\linewidth]{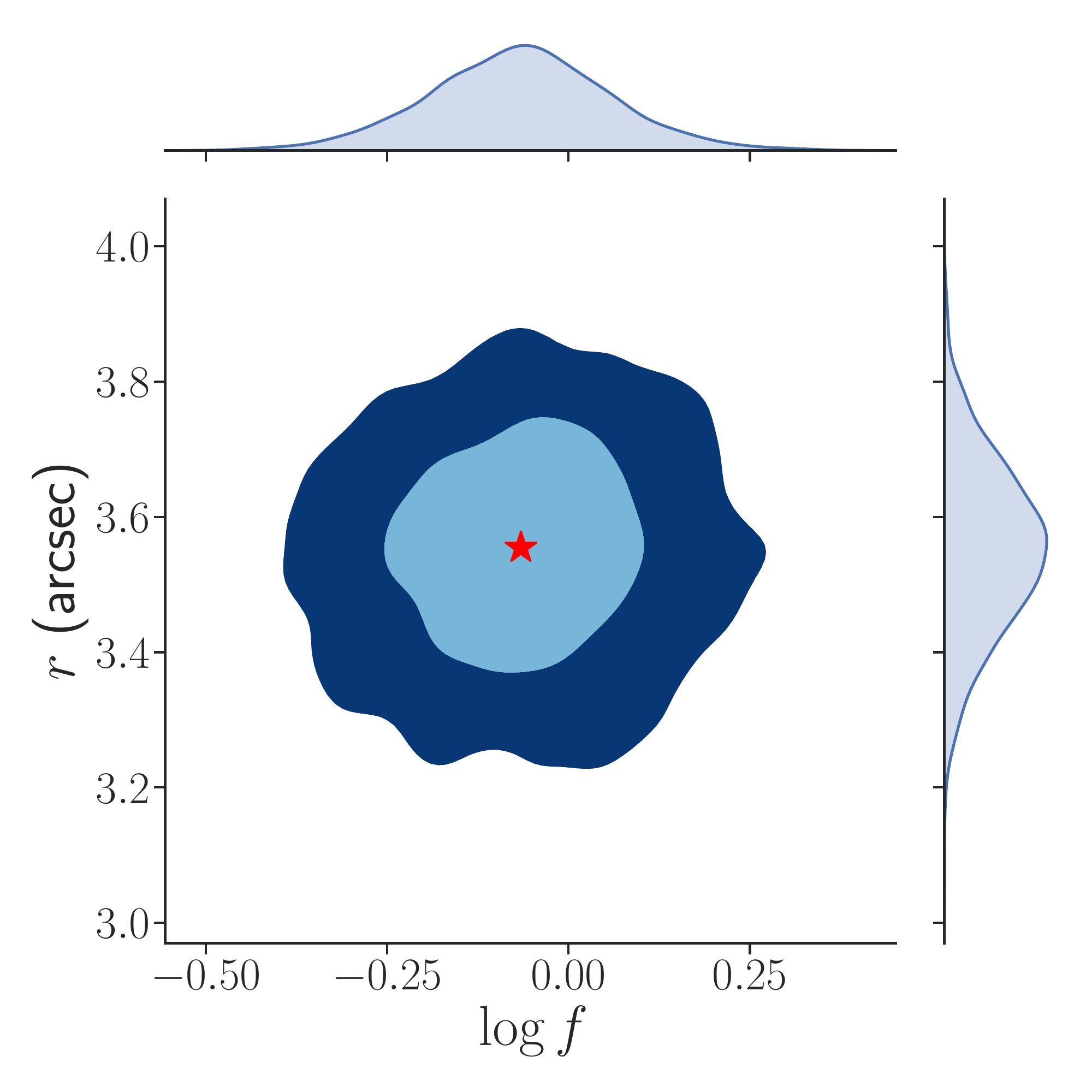}{\vskip -0.2cm plus 1fill}
    
\includegraphics[width=0.4\linewidth]{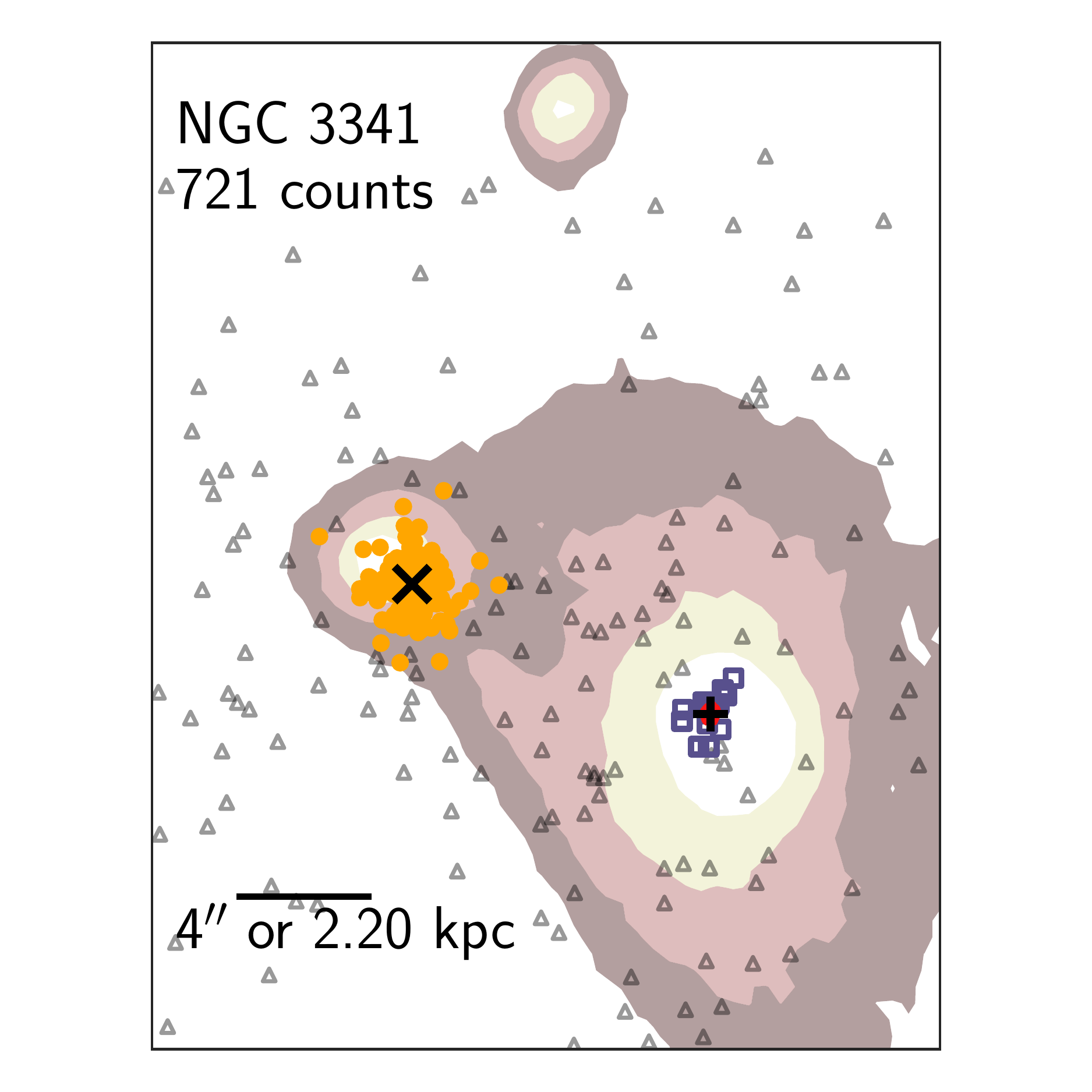}{\hskip 0pt plus 0.3fil minus 0pt}
\includegraphics[width=0.45\linewidth]{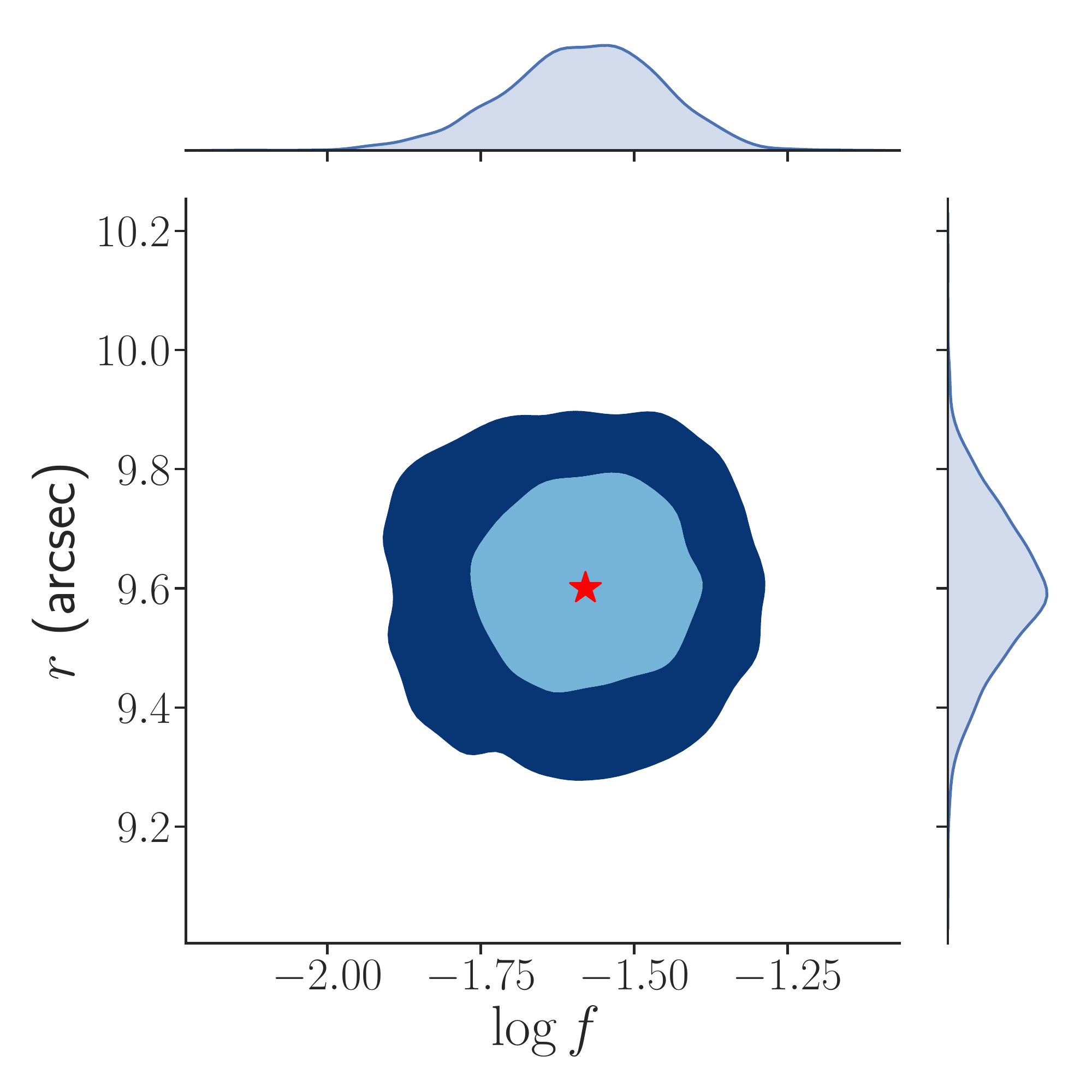}{\vskip 0.09cm plus 1fill}
    
\caption{The $0.5$--$8$ keV datasets for the two dual AGN candidates SDSS J1027+1749 and NGC 3341 (\emph{left}) and their joint posterior distributions for $r$ and $\log{f}$ (\emph{right}). In the left panels, we plot the 68\% confidence intervals (red lines) for the best-fit (as determined by 
\BAYMAX{}) sky $x$ and sky $y$ positions for a primary and secondary (which are smaller than the symbol in most instances).  Here, counts most likely associated with the primary are denoted by yellow circles, counts most likely associated with the secondary are denoted by open-faced purple squares, and counts most likely associated with background are shown as open-faced gray triangles.  In order to more clearly see the results, we do not bin the data. Contours of the SDSS i-band observations of the host galaxies are overplotted in brown. In the right panels, we show joint posterior distribution for the separation $r$ (in arcseconds) and the count ratio (in units of $\log{f}$), with the marginal distributions shown along the border. The 68\%, and 95\% confidence intervals are shown in light and dark blue contours, respectively. We denote the location of the median of the posterior distributions with a red star.}
\label{chap6:figpymc3}
\end{figure*}

 \BAYMAX{} finds the observation to have a Bayes factor that strongly favors the dual point source model, both with informative ($\ln{\mathcal{B}_{2/1, \mathrm{inform}}}=34.2\pm1.8$) and non-informative ($\ln{\mathcal{B}_{2/1}}=28.5\pm4.4$) priors. Furthermore, the locations of the primary and secondary point source are consistent between the informative and non-informative runs, and spatially coincide with the optical nuclei of the primary and northern-most galaxy. We find the best-fit values of separation and count ratio to be $r=3.4^{\arcsec+0.4}_{~-0.3}$ and $\log{f}=-0.1^{+0.5}_{-0.4}$. Thus, the separation is inconsistent with 0 at the 99.7\% confidence level. In Figure~\ref{chap6:figpymc3} we show the best-fit locations of each point source and the joint posterior distribution for $r$ and $\log{f}$.
\par Running our spectral analysis on the primary and secondary point source, we find that the primary and secondary have, on average, 37 and 35 counts. Both X-ray point sources are best-fit with $m_{\mathrm{phen,1}}$ where $\Gamma$ is fixed to a value of 1.8. For the primary, we calculate a total observed $0.5$--$8$ keV flux of $5.7^{+1.3}_{-1.6} \times 10^{-15}$ erg s$^{-1}$ cm$^{-2}$, while for the secondary we calculate a total observed $0.5$--$8$ keV flux of $4.8^{+1.1}_{-2.2} \times 10^{-15}$ erg s$^{-1}$ cm$^{-2}$ s$^{-1}$.  This corresponds to a rest-frame $2$--$7$ keV luminosity of $3.2^{+0.8}_{-0.9} \times 10^{40}$ erg s$^{-1}$ and $2.6^{+0.6}_{-1.2} \times 10^{40}$ erg s$^{-1}$ at $z=0.066$. Since we have fixed both point sources to have the same spectral shape, the count ratio that we calculate with \BAYMAX{}, should represent the flux ratio between the two sources. We find that the log of the flux ratio calculated via {\tt XSPEC} ($\approx -0.07$) is consistent with the posterior distribution of $\log{f}$ (where the median value corresponds to $\log{f}\approx-0.1$). 
\par Lastly, we estimate an upper-limit of the 2$-$7 keV luminosity of a possible tertiary X-ray point source at the location of the western galaxy. We find $0^{+1}_{-0}$ background-subtracted 0.5$-$8 keV counts within the nucleus of the SW galaxy, (where errors bars represent the 99\% C.L. as determined by Poisson statistics, see \citealt{Gehrels1986}). Assuming a power-law spectrum with photon index $\Gamma=1.8$, we estimate the upper-limit on $L_{X}$ to be between $1.50\times10^{39}$ erg s$^{-1}$ (if intrinsic $N_{H} = 1\times10^{20}$ cm$^{-2}$) and $2.36\times10^{39}$ erg s$^{-1}$ (if intrinsic $N_{H} = 1\times10^{22}$ cm$^{-2}$).

\begin{figure*}
\centering
\vspace{-0.3cm}
    \includegraphics[width=0.4\linewidth]{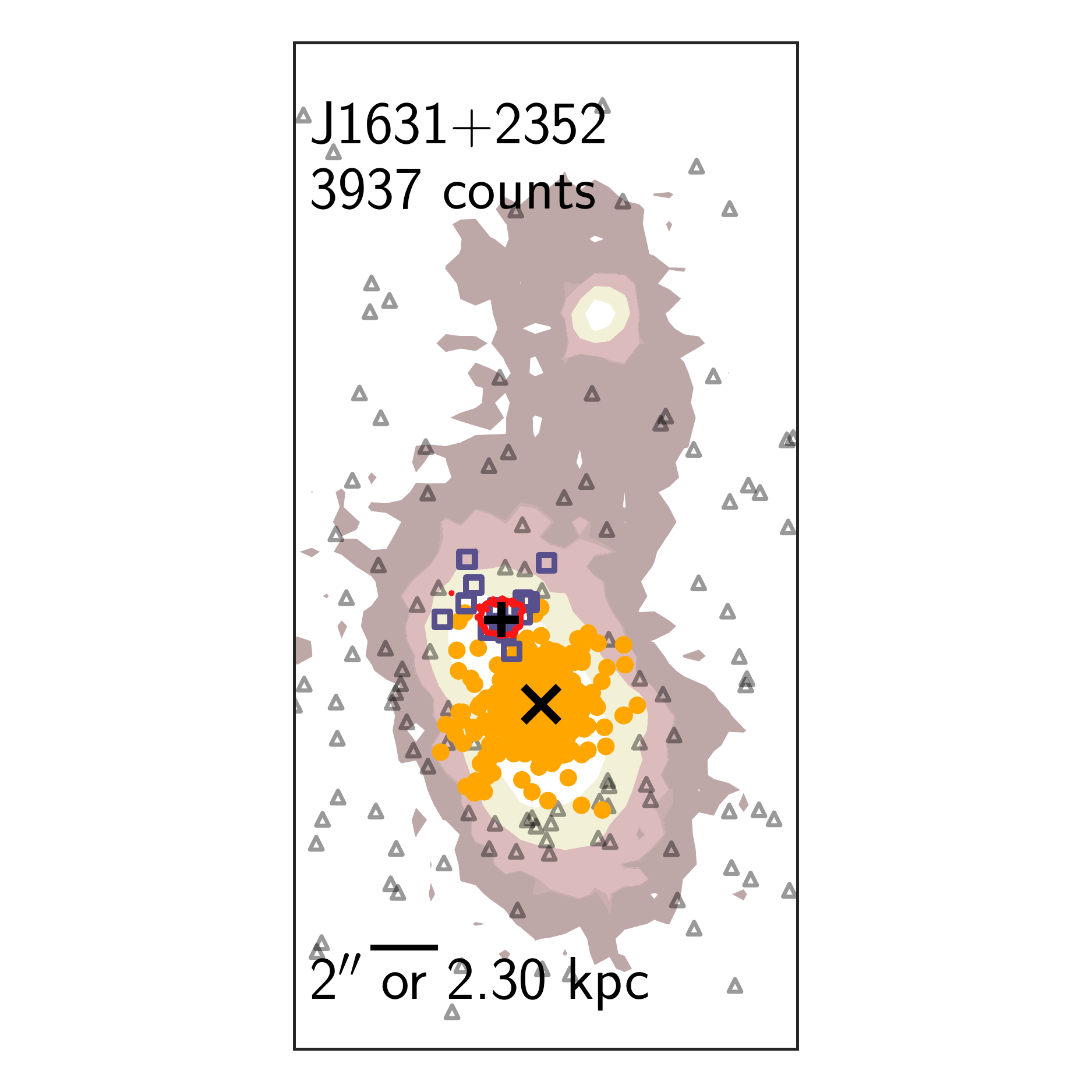}{\hskip 0pt plus 0.3fil minus 0pt}
    \includegraphics[width=0.45\linewidth]{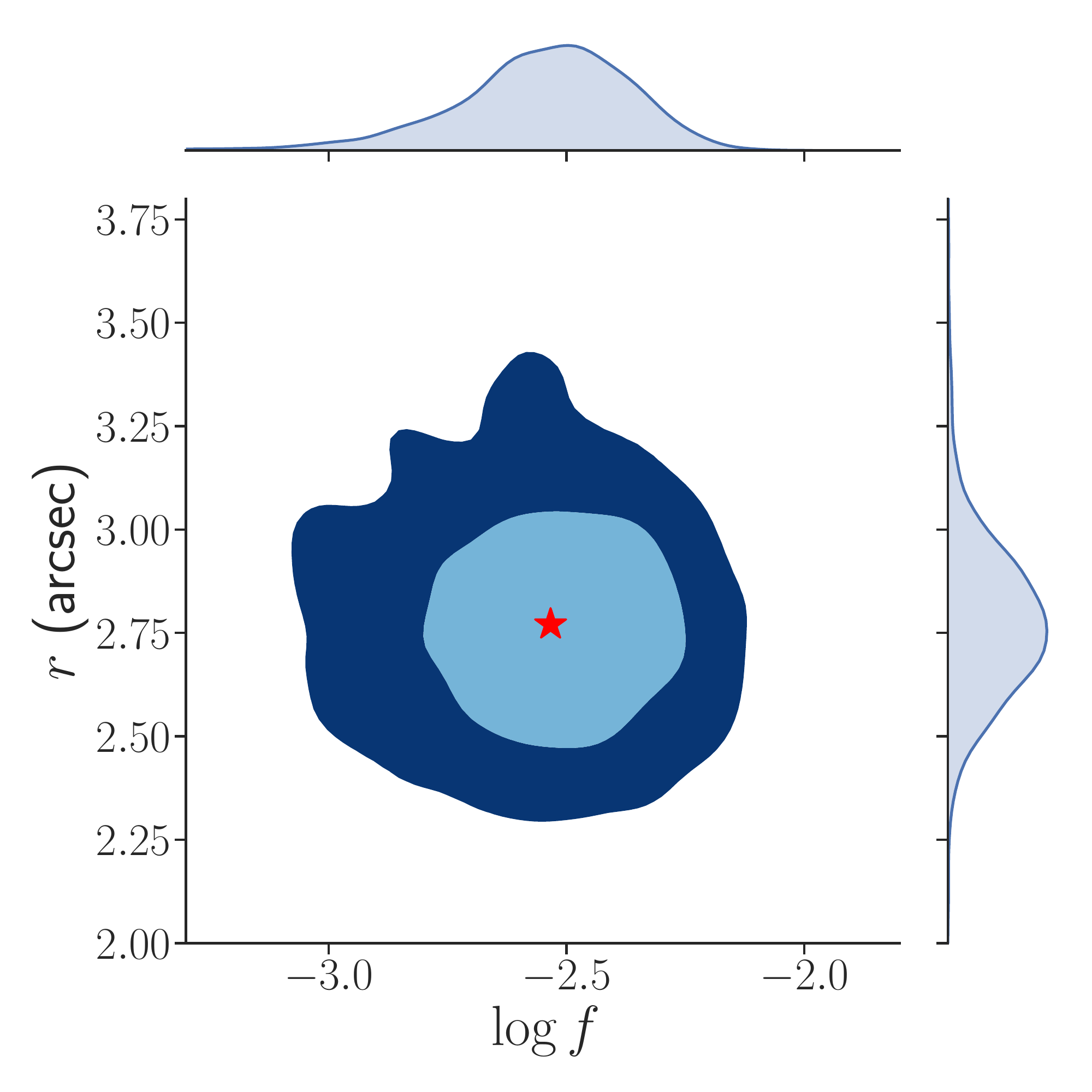}{\vskip -.28cm plus 1fill}

    \includegraphics[width=0.4\linewidth]{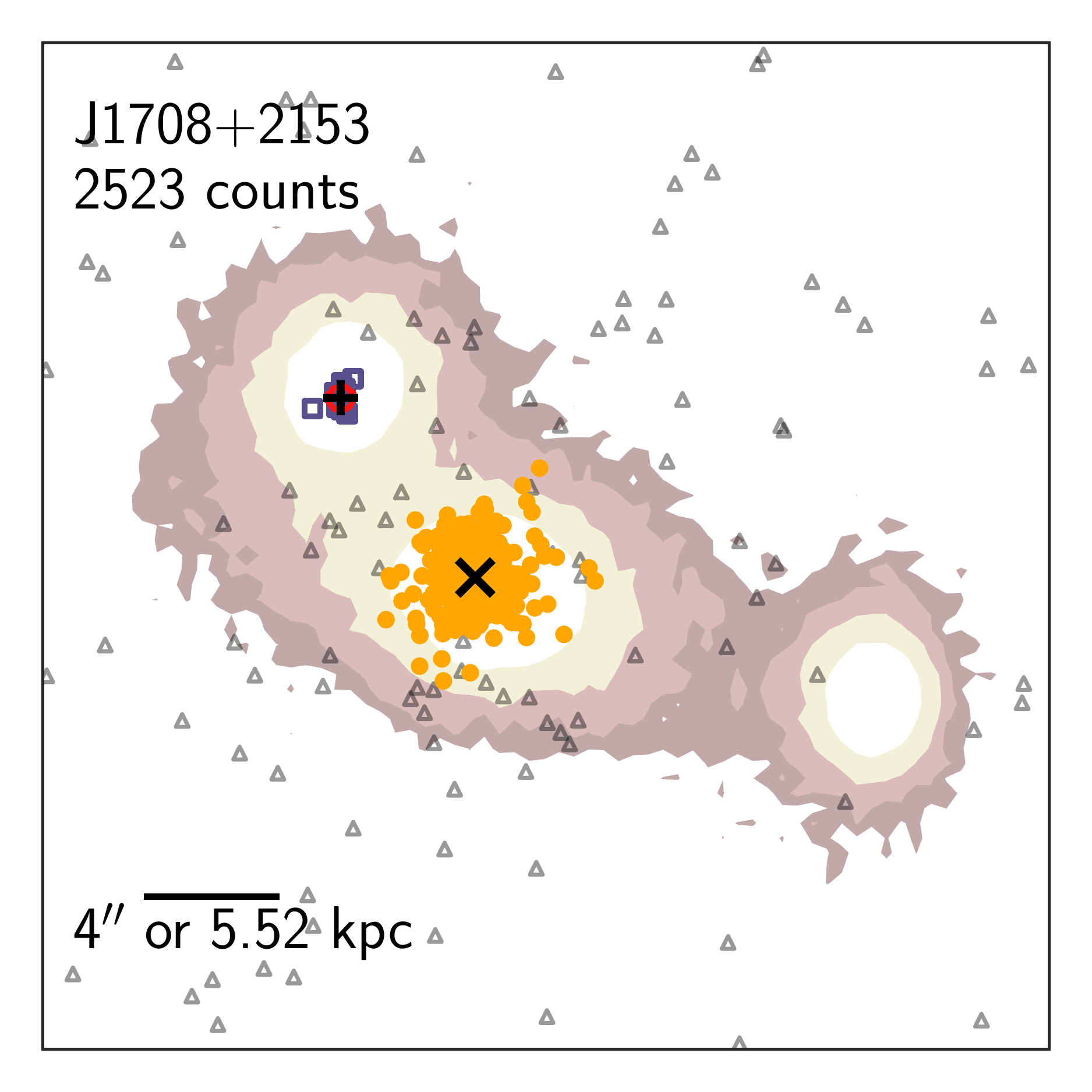}{\hskip 0pt plus 0.3fil minus 0pt}
    \includegraphics[width=0.45\linewidth]{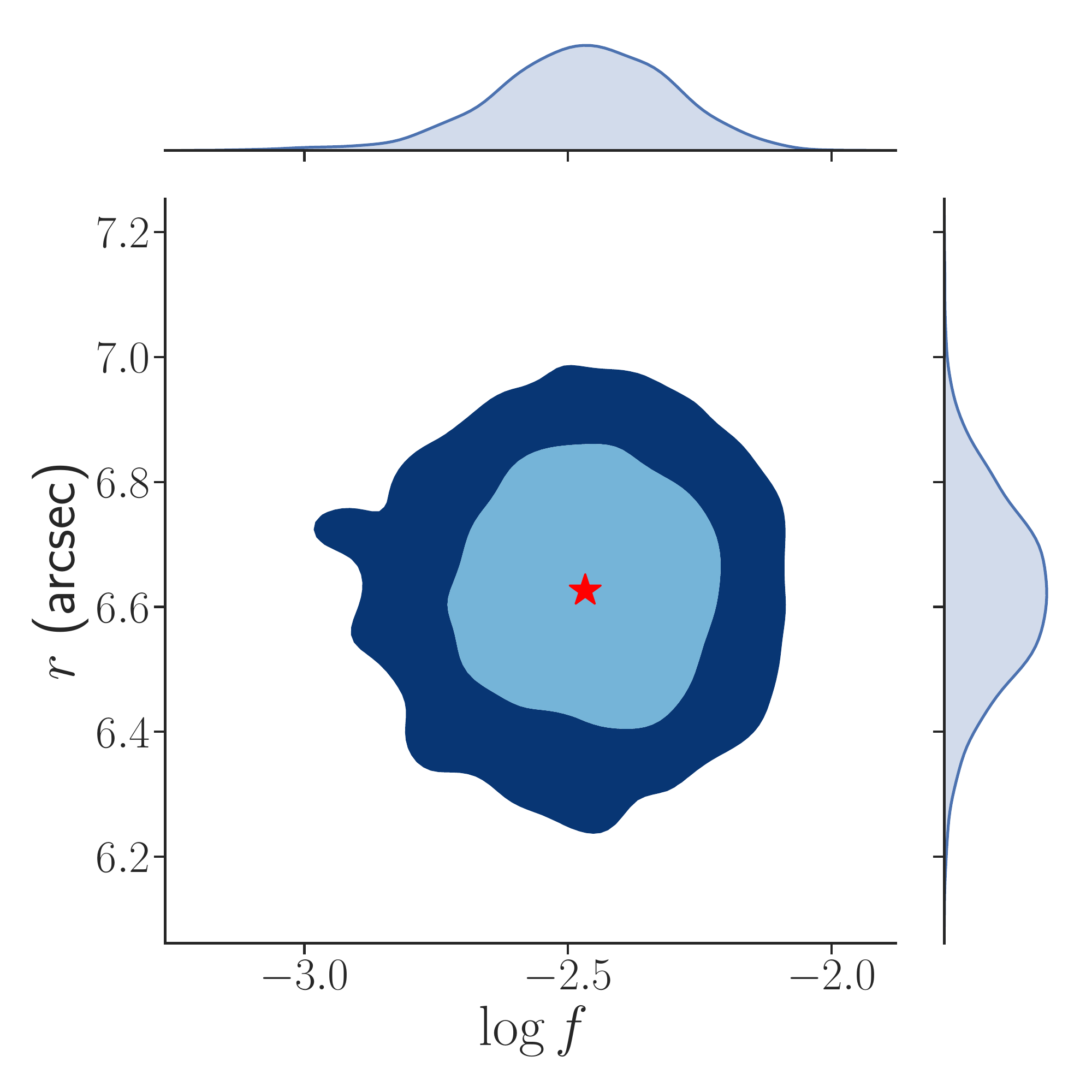}{\vskip -0.28cm plus 1fill}

    \includegraphics[width=0.4\linewidth]{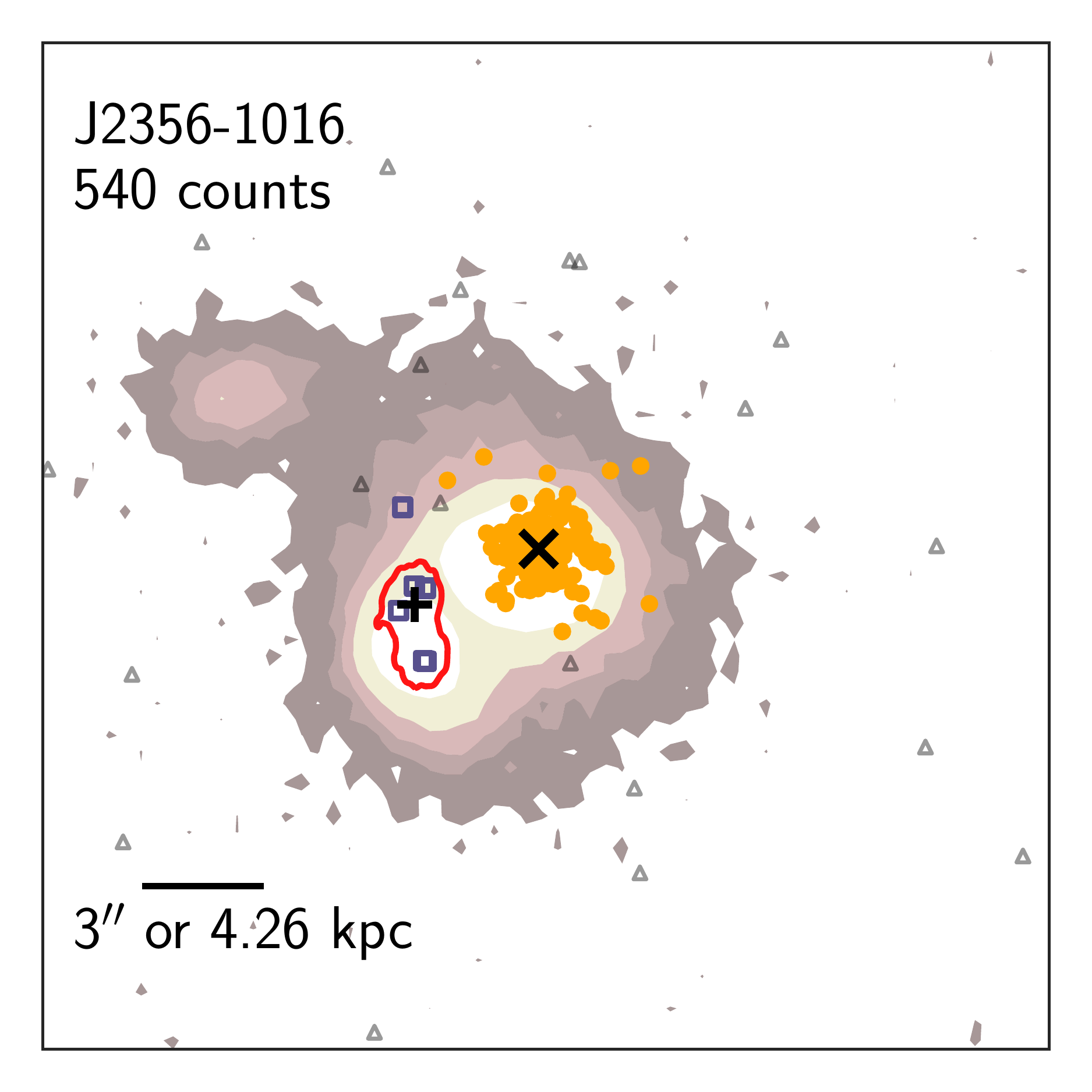}{\hskip 0pt plus 0.3fil minus 0pt}
    \includegraphics[width=0.45\linewidth]{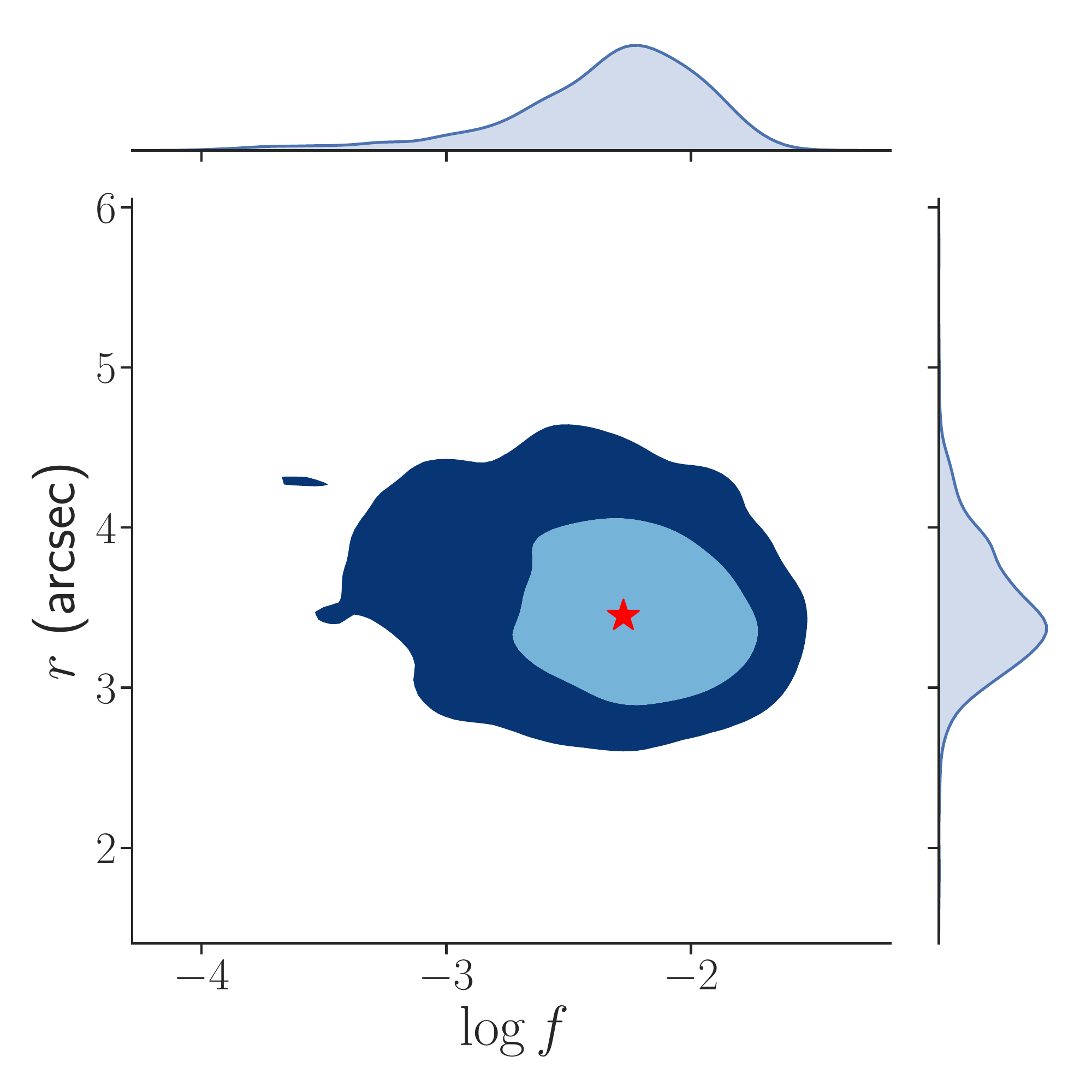}
    
    \caption{The $0.5$--$8$ keV datasets for the three dual AGN candidates J1631+2352, SDSS J1708+2153 and SDSS J2356$-$1016  (\emph{left}) and their joint posterior distributions for $r$ and $\log{f}$ (\emph{right}). Symbols and contours follow the same guidelines as Fig.~\ref{chap6:figpymc3}}.
\label{chap6:figpymc3two}
\end{figure*}

\begin{figure*}
\centering
    \includegraphics[width=0.4\linewidth]{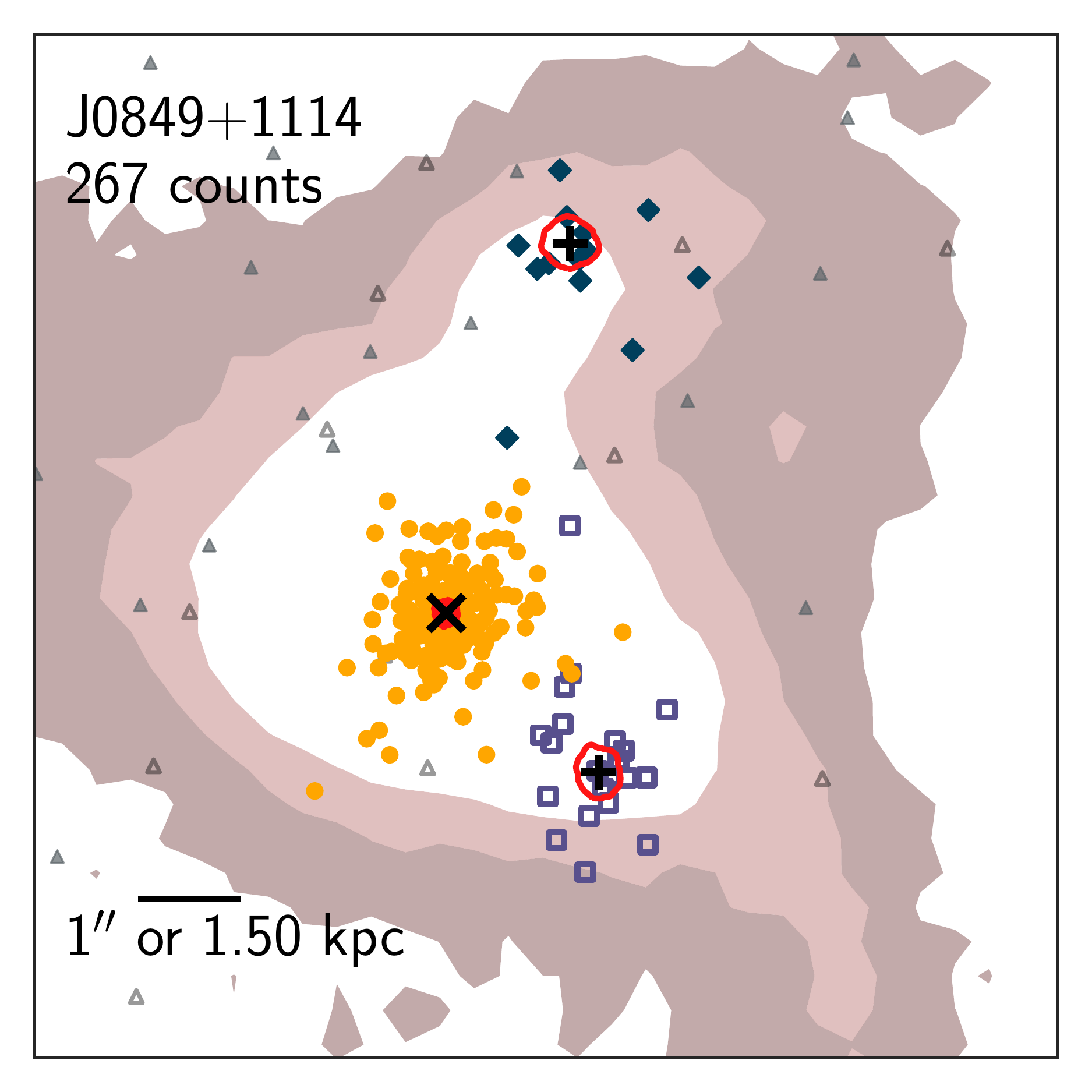}\\
    \includegraphics[width=0.42\linewidth]{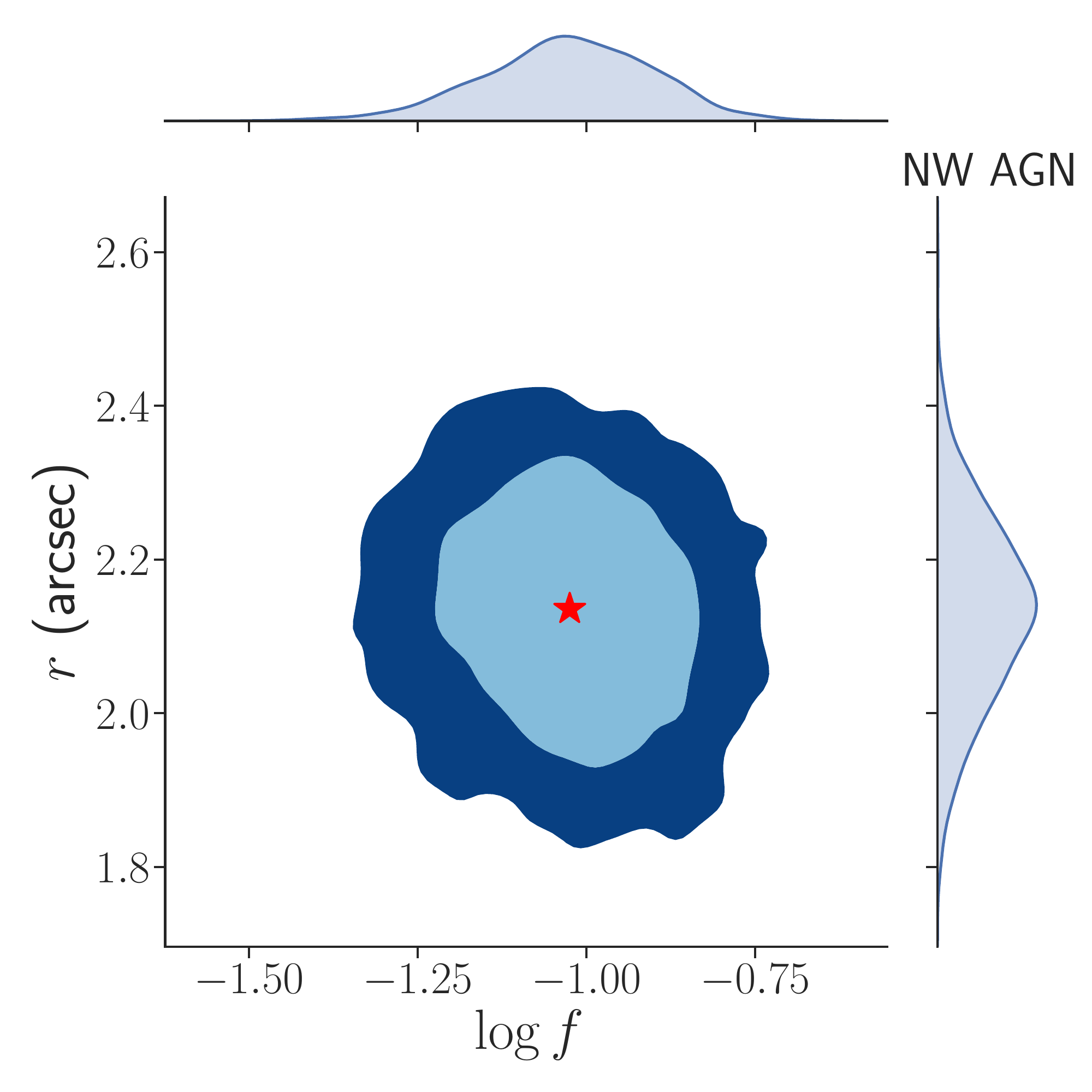}
    \includegraphics[width=0.42\linewidth]{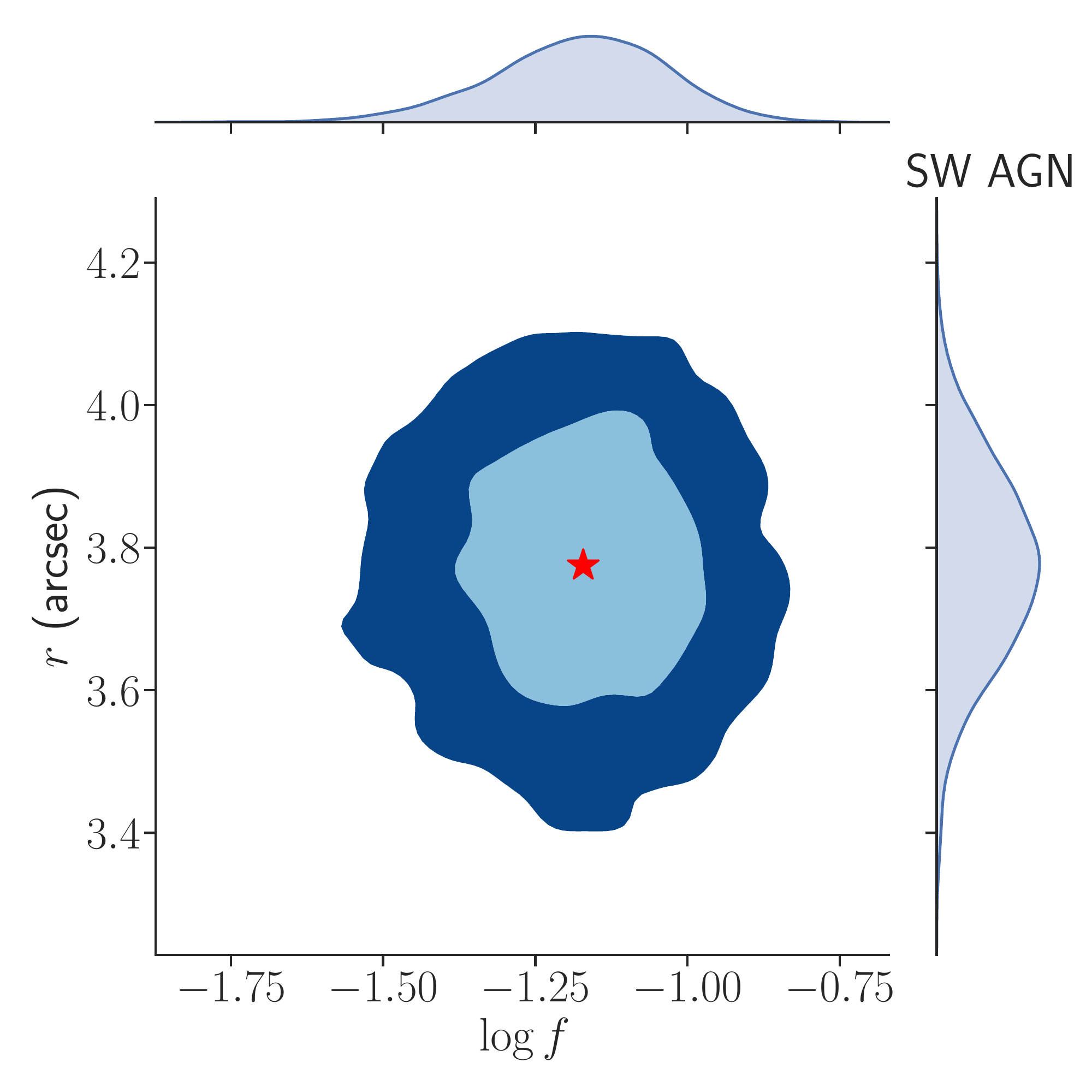}{\vskip -0.2cm plus 1fill}

    \caption{The $0.5$--$8$ keV datasets for the triple AGN SDSS J0849+1114 (\emph{top}) and the joint posterior distributions for $r$ and $\log{f}$ for the secondary and tertiary X-ray point sources (\emph{bottom}). Symbols and contours follow the same guidelines as Fig.~\ref{chap6:figpymc3}, while we denote the counts most likely associated with the tertiary point source with green filled diamonds.}

\label{chap6:figpymc3_three}
\end{figure*}

\subsection{NGC 3341}
\label{chap6:subsecN3341}
The first analysis of NGC 3341 was presented in \cite{Barth2008}, where SDSS data and new observations from the Keck Observatory were analyzed. Results from optical diagnostics were deemed too ambiguous for a proper classification of the AGN-nature of each galaxy. Expanding on this study via a multi-wavelength analysis that combined optical, X-ray, and radio data, \cite{Bianchi2013} concluded that the primary galaxy showed evidence for AGN activity. Although the SW nucleus had emission detected in almost every wave-band, the emission was consistent with star-formation ($L_{\mathrm{2-7, keV}} \approx 4\times10^{39}$ erg s$^{-1}$, while there was no sign of any compact source at 5 GHz).
\par Our analysis with \BAYMAX{} results in a Bayes factor that strongly favors the dual point source model, both with informative ($\ln{\mathcal{B}_{2/1, \mathrm{inform}}}=2766.9\pm1.6$) and non-informative priors ($\ln{\mathcal{B}_{2/1}}=22.5\pm1.8$). The locations of the primary and secondary point source are consistent between the informative and non-informative runs, and spatially coincide with the optical nuclei of the primary and SW galaxy. We find the best-fit values of separation and count ratio to be $r=9.6_{~-0.4}^{\arcsec +0.3}$ and $\log{f}=-1.6_{-0.4}^{+0.4}$. Although we find that the system weakly (given the large error bars) favors the triple point source model, over the dual point source model, when using non-informative priors ($\ln{\mathcal{B}_{3/2}}=5.8\pm4.7$), this is interpreted to be due to the presence of hot gas in the vicinity of the primary and secondary galactic nuclei; this emission is seen both in the SDSS and \emph{Chandra} observations. \BAYMAX{} finds the best-fit location of the ``tertiary" in a clumpy X-ray emission region south east of the secondary point source, spatially inconsistent with any galactic nucleus. In Figure~\ref{chap6:figpymc3} we show the best-fit locations of each point source and the joint posterior distribution for $r$ and $\log{f}$.
\par Running our spectral analysis on the primary and secondary point source, we find that the primary and secondary have, on average, 566 and 16 counts. The primary X-ray point source is best-fit with $m_{\mathrm{phen,2}}$, where $\Gamma$ is free to vary; while the secondary X-ray point source is best-fit with $m_{\mathrm{phen,1}}$ where $\Gamma$ is fixed to a value of 1.8. The primary in NGC 3341 is the only source that shows inconsistencies between our phenomenological and physically-motivated spectral fits. In particular, we find that when using {\tt BNTorus} the primary X-ray point sources is best-fit with $m_{\mathrm{phys, 2}}$ where $\Gamma$ is fixed to a value of 1.8. As such, we list the parameters for both models below (and refer the reader to Tables~\ref{chap6:tabspectra} and \ref{chap6:tabspectraBN}). Additionally, we note that the main difference between our X-ray spectral analysis and that presented in \citealt{Bianchi2013}, is that \citealt{Bianchi2013} fit the spectrum of the primary in NGC 3341 with a single power-law (analogous to $m_{\mathrm{phen,1}}$ in our analysis). They find the best-fit model one where the power-law photon index is fixed to a value of $\Gamma=1.7$ (due to not being able to constrain the value in the fit).
\par For the primary, when using the phenomenological model, we calculate a total observed $0.5$--$8$ keV flux of $3.63^{+0.07}_{-0.05} \times 10^{-13}$ erg s$^{-1}$ cm$^{-2}$, corresponding to a rest-frame $2$--$7$ keV luminosity of $8.54^{+0.41}_{-0.33} \times 10^{41}$ erg s$^{-1}$ at $z=0.027$. The spectrum of the primary is measured to have extragalactic $N_{H} = 1.0 \pm 0.06\times 10^{23}$ cm$^{-2}$, with a relatively flat photon index of $\Gamma=1.1_{-0.1}^{+0.1}$. The high level of $N_{H}$ and flat spectral shape is consistent with an obscured environment. When using the physically motivated model, where $\Gamma$ is kept fixed at a value of 1.8, we calculate a total observed $0.5$--$8$ keV flux of $1.39^{+0.01}_{-0.01} \times 10^{-12}$ erg s$^{-1}$ cm$^{-2}$, corresponding to a rest-frame $2$--$7$ keV luminosity of $1.18^{+0.02}_{-0.01} \times 10^{42}$ erg s$^{-1}$ at $z=0.027$. The intrinsic absorption is measured to be slightly higher ($N_{H} = 1.2_{-0.02}^{+0.01}\times 10^{23}$ cm$^{-2}$) than the value determined using $m_{\mathrm{phen,2}}$.
\par For the secondary, we calculate a total observed $0.5$--$8$ keV flux of $2.7^{+0.6}_{-0.8} \times 10^{-15}$ erg s$^{-1}$ cm$^{-2}$ s$^{-1}$.  This corresponds to a rest-frame $2$--$7$ keV luminosity of $2.3^{+0.9}_{-0.7} \times 10^{39}$ erg s$^{-1}$ at $z=0.027$. Although the primary has an X-ray luminosity consistent with accretion onto an AGN (regardless of spectral model used, $L_{X}>1\times10^{41}$ erg s$^{-1}$), the SW galaxy does not ($L_{X}<1\times10^{40}$ erg s$^{-1}$). These results agree with what was previously found in \cite{Bianchi2013}.
\par We estimate an upper-limit on the 2$-$7 keV luminosity of possible tertiary X-ray point source at the location of the NW galaxy. We find $0^{+1}_{-0}$ 0.5$-$8 keV background-subtracted counts within the nucleus of the NW galaxy. Assuming a power-law spectrum with photon index $\Gamma=1.8$, we estimate the upper-limit on $L_{X}$ to be between $1.80\times10^{39}$ erg s$^{-1}$ (if intrinsic $N_{H} = 1\times10^{20}$ cm$^{-2}$) and $2.86\times10^{39}$ (if intrinsic $N_{H} = 1\times10^{22}$ cm$^{-2}$).

\subsection{SDSS J1631+2352}
\label{chap6:subsecJ1631}
\par SDSS J1631+2352 is a triple merger system at $z=0.059$, with no previous analysis on the \emph{Chandra} data beyond fitting the nuclear X-ray spectra associated with the primary galaxy \citep{Ricci2017}.  It was observed as a part of a study on mergers in the \emph{Swift} BAT AGN sample \citep{Koss2010, Ricci2017}. Analyzing the archival \emph{Chandra} observation with \BAYMAX{}, we find that the data favor the dual point source model when using both informative ($\ln{\mathcal{B}_{2/1, \mathrm{inform}}}=2.7\pm1.9$) and non-informative priors ($\ln{\mathcal{B}_{2/1}}=6.2\pm 1.6$). The best-fit value for the separation between the two point sources is $r=2.8_{~-0.6}^{\arcsec+1.3}$, inconsistent with 0 at the 99.7\% confidence level; furthermore the best-fit location of the secondary point source is consistent with the location of the NE galaxy using both informative and non-informative priors. The best-fit value for count ratio is $\log{f}=-2.5_{-0.5}^{+0.4}$. We show the best-fit locations of each point source, and the joint posterior distributions for $r$ and $\log{f}$ in Figure~\ref{chap6:figpymc3two}.

\par The spectral realizations of the primary (with an average of 3810 counts) and secondary (with an average of 11 counts) point source are best-fit with $m_{\mathrm{phen,2}}$ and $m_{\mathrm{phen,1}}$, with $\Gamma$ fixed to a value of 1.8. When fitting the spectral realizations of the primary point source, we identify a statistically significant Fe K$\alpha$ fluorescent emission line, modeled by a Gaussian component ({\tt zgauss}) fixed at 6.4 keV. We measure an equivalent width of $0.23_{-0.02}^{+0.01}$ keV for the Fe K$\alpha$ line. The equivalent width of the Fe K$\alpha$ line is strongly dependent on line-of-sight absorption, as well as other parameters such as geometry of the accretion disk, inclination angle at which the reflecting surface is viewed, and the elemental abundances of the reflecting matter (see, e.g., \citealt{Brightman2011}). Although we measure a relatively mild line-of-sight hydrogen column density when fitting the 0.5$-$8 keV spectrum for the primary ($N_{H}=1.10_{-0.02}^{+0.01} \times 10^{22}$ cm$^{-2}$), analysis of the available \emph{NuSTAR} observations (which are above 10 keV), may better constrain the spectral parameters (e.g., \citealt{Marchesi2018}). However, past analyses of the \emph{Swift} BAT observations (sensitive to the 15-150 keV energy range) of the merger conclude similar levels of obscuration, with $N_{H}=5.0_{-3.0}^{+1.6}$, using more complicated spectral models \citep{Ricci2017b}. We calculate a total observed $0.5$--$8$ keV flux of $2.29^{+0.01}_{-0.01} \times 10^{-12}$ erg s$^{-1}$ cm$^{-2}$, and $5.3^{+4.8}_{-3.9} \times 10^{-15}$ erg s$^{-1}$ cm$^{-2}$ for the primary and secondary, respectively. These flux values are consistent with the extrapolated $2-7$ keV flux measurements from the \emph{Swift} BAT observations \cite{Ricci2017b}. They correspond to rest-frame $2$--$7$ keV luminosities, at $z=0.059$, of $1.33^{+0.01}_{-0.01} \times 10^{43}$ erg s$^{-1}$ and $2.5^{+2.3}_{-1.8} \times 10^{40}$ erg s$^{-1}$.
\par We estimate the upper-limit of the 2$-$7 keV luminosities of a possible tertiary X-ray point source at the location of NW host-galaxy nuclei. We find $0^{+1}_{-0}$ 0.5$-$8 keV counts associated with a possible tertiary (NW galaxy) in SDSS J1631+2352. Assuming a power-law spectrum with photon index $\Gamma=1.8$, we estimate the upper-limit on $L_{X}$ to be between $3.19\times10^{39}$ erg s$^{-1}$ (if intrinsic $N_{H} = 1\times10^{20}$ cm$^{-2}$) and $4.89\times10^{39}$ (if intrinsic $N_{H} = 1\times10^{22}$ cm$^{-2}$).

\subsection{SDSS J1708+2153}
\label{chap6:subsecJ1708}
SDSS J1708+2153 is a triple merger system at $z=0.072$.  Similar to SDSS J1631+2352, it was observed as a part of a study on mergers in the \emph{Swift} BAT AGN sample \citep{Koss2010, Ricci2017}, and the \emph{Chandra} observations have no previous analysis beyond fitting the nuclear X-ray spectra associated with the primary galaxy \citep{Ricci2017}. Analyzing the archival \emph{Chandra} observations with \BAYMAX{}, we find that the data strongly favor the dual point source model when using both informative ($\ln{\mathcal{B}_{2/1, \mathrm{inform}}}=18.4\pm 1.6$) and non-informative priors ($\ln{\mathcal{B}_{2/1}}=16.6\pm1.9$). The best-fit value for the separation is $r=6.6_{~-0.4}^{\arcsec+0.4}$, inconsistent with 0 at the 99.7\% confidence level, while the best-fit value for count ratio is $\log{f}=-2.5_{-0.6}^{+0.4}$. We show the best-fit locations of each point source, and the joint posterior distributions for $r$ and $\log{f}$ in Figure~\ref{chap6:figpymc3two}. The best-fit location for the secondary is coincident with the location of the NE galaxy, using both informative and non-informative priors.
\par Analyzing the spectral realizations of each point source, we find that the primary and secondary have, on average, 2406 and 9 counts. Both the primary and secondary point sources are best-fit with $m_{\mathrm{phen,1}}$, where $\Gamma$ is allowed to vary for the primary. For the primary, we calculate a total observed $0.5$--$8$ keV flux of $1.46^{+0.01}_{-0.01} \times 10^{-12}$ erg s$^{-1}$ cm$^{-2}$, while for the secondary we calculate a total observed $0.5$--$8$ keV flux of $4.6^{+1.6}_{-0.2} \times 10^{-15}$ erg s$^{-1}$ cm$^{-2}$ s$^{-1}$. Once again, these flux measurements are consistent with what was found in \cite{Ricci2017}, when analyzing the \emph{Swift} BAT spectrum. Our measured flux values correspond to rest-frame $2$--$7$ keV luminosities of $1.17^{+0.01}_{-0.01} \times 10^{43}$ erg s$^{-1}$ and $3.5^{+0.6}_{-1.2} \times 10^{40}$ erg s$^{-1}$ at $z=0.072$. The spectral fit of the primary point source shows low-levels of absorption, with $N_{H}=0.10_{-0.01}^{+0.01}\times10^{22}$ cm$^{-2}$ and best-fit photon index $\Gamma=1.37_{-0.01}^{+0.01}$. 
\par Lastly, we estimate an upper-limit of the 2$-$7 keV luminosity of a possible tertiary X-ray point sources at the location of the SW galaxy. Subtracting the expected background contribution, we find $0^{+1}_{-0}$ 0.5$-$8 keV counts associated with the nucleus of the SW galaxy. Assuming a power-law spectrum with photon index $\Gamma=1.8$, we estimate the upper-limit on $L_{X}$ to be between $4.80\times10^{39}$ erg s$^{-1}$ (if intrinsic $N_{H} = 1\times10^{20}$ cm$^{-2}$) and $7.15\times10^{39}$ (if intrinsic $N_{H} = 1\times10^{22}$ cm$^{-2}$).

\subsection{SDSS J2356$-$1016}
\label{chap6:subsecJ2356}
SDSS J2356$-$1016 was analyzed in \cite{Pfeifle2019a}, where they concluded that the X-ray emission was consistent with a single AGN. They find an X-ray detection at the location of the primary galaxy (at the 22.8$\sigma$ C.L.), but they find no X-ray detection at the location of the SE galaxy, which is separated from the primary by 3.5\arcsec. However, source detection was determined using the {\tt CIAO} package {\tt wavdetect}, which is not always sensitive enough to detect both low-count and closely-separated multiple point source systems (see, e.g., \citealt{Foord2020a}). Given that \BAYMAX{} is a powerful tool for low count systems, we re-analyze this triple merger to identify any previously missed detections. 
\par Analyzing the archival \emph{Chandra} observations with \BAYMAX{}, we find that the data strongly favor the dual point source model when using informative priors ($\ln{\mathcal{B}_{2/1, \mathrm{inform}}}=3.1\pm1.3$). The Bayes factor does not strongly favor the dual point source model when using non-informative priors, likely due to the low number of counts associated with the secondary point source; however the best-fit locations of the primary and secondary point source are consistent between non-informative and informative runs, and spatially coincide with the primary and SE galaxy. During our false positive tests, only 2 out of 100 simulations of single point sources analyzed by \BAYMAX{} have $\ln{\mathcal{B}_{2/1, \mathrm{inform}}}>3$, and thus we classify the merger as a dual X-ray point source system. The best-fit values of separation and count ratio are $r=3.5_{~-1.2}^{\arcsec+1.6}$ and $\log{f}=-2.4_{-1.5}^{+0.7}$. This separation is inconsistent with 0 at the 99.7\% confidence level. We show the best-fit locations of each point source, and the joint posterior distributions for $r$ and $\log{f}$ in Figure~\ref{chap6:figpymc3two}.
\par Analyzing the spectral realizations of each point source, we find that the primary and secondary have, on average, 516 and 4 counts. The primary point source source is best-fit with $m_{\mathrm{phen,2}}$ while the secondary point source is best-fit with $m_{\mathrm{phen,1}}$ where $\Gamma$ is fixed for both models. We calculate a total observed $0.5$--$8$ keV flux of $1.60^{+0.01}_{-0.01} \times 10^{-12}$ erg s$^{-1}$ cm$^{-2}$ for the primary, while for the secondary we calculate a total observed $0.5$--$8$ keV flux of $7.3^{+7.0}_{-3.5} \times 10^{-15}$ erg s$^{-1}$ cm$^{-2}$ s$^{-1}$. This corresponds to a rest-frame $2$--$7$ keV luminosity of $3.11^{+0.04}_{-0.04} \times 10^{43}$ erg s$^{-1}$ and $8^{+10}_{-1} \times 10^{40}$ erg s$^{-1}$ at $z=0.074$. The spectral fit of the primary point source shows relatively high (with respect to the other dual X-ray point sources in the sample) levels of absorption, with $N_{H}=7.8_{-0.1}^{+0.2}\times10^{22}$ cm$^{-2}$.
\par We estimate an upper-limit of the 2$-$7 keV luminosity of possible tertiary X-ray point sources at the location of the NE galaxy. We find $0^{+1}_{-0}$ 0.5$-$8 keV background-subtracted counts associated with the nucleus of the NE galaxy. Assuming a power-law spectrum with photon index $\Gamma=1.8$, we estimate the upper-limit on $L_{X}$ to be between $1.02\times10^{40}$ erg s$^{-1}$ (if intrinsic $N_{H} = 1\times10^{20}$ cm$^{-2}$) and $1.50\times10^{40}$ (if intrinsic $N_{H} = 1\times10^{22}$ cm$^{-2}$).
\par Although our analysis and false positive tests for SDSS J2356$-$1016 show the weakest evidence for a dual X-ray source, the detected secondary has the hardest X-ray spectrum of all our the systems in our sample. Follow-up observations with either optical and/or IR IFU will help reveal the true origin of X-ray emission.

\subsection{SDSS J0849+1114}
SDSS J0849+1114 is a triple AGN candidate first published by \cite{Pfeifle2019a}, and in-depth analyses followed in \cite{Pfeifle2019b} and \cite{Liu2019}. \cite{Pfeifle2019a} conclude that the primary is detected at a 10.2$\sigma$ level, while the SW and NE galaxy are detected at 2.2$\sigma$ and 1.4$\sigma$, respectively. Given the lower significance of the SW and NE detections, we re-evaluate the system with \BAYMAX{} to determine the likelihood that the merger is composed of three X-ray point sources. We find that \BAYMAX{} strongly favors the triple point source system using both informative ($\ln{\mathcal{B}_{3/2, \mathrm{inform}}}=19.4\pm1.9$) and non-informative priors ($\ln{\mathcal{B}_{3/2}}=22.4\pm2.1$); furthermore the locations of each point-source are consistent between the informative and non-informative runs, and spatially coincide with the optical nuclei of each galaxy in the triple merger. The best-fit values for the separation and count ratio of the secondary (associated with the SW galaxy) X-ray point source are $r=2.1_{~-0.3}^{\arcsec+0.3}$ and $\log{f}=-1.0_{-0.4}^{+0.4}$; while the best-fit values for separation and count ratio of the tertiary (associated with the NW galaxy) X-ray point source are $r=3.8_{~-0.4}^{\arcsec+0.4}$ and $\log{f}=-1.2_{-0.5}^{+0.4}$. Here, the separation and count-ratio are defined relative to the position and number of counts associated with the primary point source. These separations are inconsistent with 0 at the 99.7\% confidence level. We show the best-fit locations of each point source, and the joint posterior distributions for $r$ and $\log{f}$ in Figure~\ref{chap6:figpymc3_three}.
\par Creating 100 spectral realizations of each point source, we find that the primary, secondary, and tertiary have, on average, 189, 99, and 13 counts. The primary point source source is best-fit with $m_{\mathrm{phen,2}}$ where $\Gamma$ is fixed to a value of 1.8, while the secondary and tertiary point sources are best-fit with $m_{\mathrm{phen,1}}$ where $\Gamma$ is fixed to value of 1.8. Although \cite{Pfeifle2019a} find that adding an Fe K$\alpha$ emission component into the primary's spectral model results in statistically significant better fit (using only one of the two \emph{Chandra} observations), we find that the combined \emph{Chandra} dataset results in spectral realizations that do not favor the addition of an Fe K$\alpha$ emission component (and a similar conclusion was reached in \citealt{Pfeifle2019b} when combining all available observations form \emph{Chandra} and \emph{NuSTAR}). We calculate a total observed $0.5$--$8$ keV fluxes of: $6.54^{+0.28}_{-0.22} \times 10^{-14}$ erg s$^{-1}$ cm$^{-2}$ for the primary; $4.0^{+1.6}_{-1.3} \times 10^{-15}$ erg s$^{-1}$ cm$^{-2}$ for the secondary; and $3.2^{+1.0}_{-1.3} \times 10^{-15}$ erg s$^{-1}$ cm$^{-2}$ for the tertiary. This corresponds to rest-frame $2$--$7$ keV luminosities (at $z=0.059$) of: $1.60^{+.80}_{-.40} \times 10^{42}$ erg s$^{-1}$ for the primary, $1.9^{+0.6}_{-0.7} \times 10^{40}$ erg s$^{-1}$ for the secondary, and $1.3^{+0.6}_{-0.5} \times 10^{40}$ erg s$^{-1}$ for the tertiary. The spectral fit of the primary point source shows one of the highest-levels of absorption in our sample, with $N_{H}=5.2_{-1.1}^{+2.3}\times10^{23}$ cm$^{-2}$. 

\subsection{SDSS J0858+1822: Clumpy Diffuse Emission or Compton Thick AGN}
SDSS J0858+1822 is a unique system within our sample, where there appears to be an excess of X-ray emission south of the primary galaxy (see ~Fig.~\ref{chap6:figTripleGalaxyImages}), which does not spatially coincide with a galactic nucleus. Interestingly, when using non-informative priors, \BAYMAX{} favors the single point source model, but with the best-fit location for the point source corresponding to the location of the southern region of X-ray emission. This is consistent with our results when running \BAYMAX{} with informative priors -- the single point source model is favored, but the location of the point source is shifted to the southern-most edge of the spatial constraints placed on $\mu$. These results are complementary to the results of the optical analysis presented in \cite{Husemann2020}, where the [\ion{O}{3}] emission-line peak was found to be offset by 1\arcsec~from the position of the primary galactic nucleus. We aim to better understand the accretion nature of the southern emission region (which, for example, may be explained by a background AGN or a recoiling SMBH), and do so by analyzing the spectrum.
\par We use the results from the non-informative run to create 100 spectral realizations of the point source south of the primary galaxy; this is done by sampling from the posterior distributions of $\theta_{1}$ and probabilistically assigning counts to either the point source or background component. The spectral realizations of the point source are best-fit with $m_{\mathrm{phen,1}}$ where $\Gamma$ is free to vary. We calculate a total observed $0.5$--$8$ keV flux of $3.0^{+2.6}_{-1.3} \times 10^{-15}$ erg s$^{-1}$ cm$^{-2}$, corresponding to a rest-frame $2$--$7$ keV luminosity of $2.1^{+0.1}_{-0.1} \times 10^{39}$ erg s$^{-1}$ at $z=0.077$.  The region is extremely soft, with best-fit $\Gamma=8.5_{-5.4}^{+0.7}$ and $\emph{HR}=-0.9_{-0.1}^{+0.4}$. These results can be well-explained by a concentrated region (or, clump) of the surrounding diffuse emission, that is better modeled as a point-source sitting among a uniform background. 
\par The very soft emission can possibly originate from an AGN if the system is Compton thick; here, all emission above 10 keV is absorbed, and the observed $2-8$ keV X-ray emission is a result of the reflected and reprocessed (such as from narrow-line regions; see, e.g., \citealt{Liu2013, Comerford2015}) emission.  In such a scenario, the intrinsic X-ray luminosity can be $\sim$60$-$70 times higher with respect to what is observed (corresponding to an X-ray luminosity on the order of $10^{41}$ erg s$^{-1}$, compatible with an AGN; e.g., \citealt{Panessa2006, Lamastra2009, Marinucci2012}). We test this hypothesis by fitting the spectrum of the southern source in SDSS J0858+1822 with $m_{\mathrm{phen,1}}$, fixing the photon index of the power-law to 1.8, and calculating the expected value of intrinsic $N_{H}$. Doing so, we find that best-fit $N_{H}<1\times10^{-20}$ cm$^{-2}$. However, an additional complication to this analysis is that Compton-thick sources may appear to have low-levels of absorption if using simple models and/or low-count data sets \citep{Risaliti1999}. Thus, deeper \emph{Chandra} X-ray observations, or observations with instruments that are more sensitive to harder X-rays (such as XMM and / or NuSTAR), and/or multi-wavelength follow-up (i.e, optical/IR spectroscopy and/or radio observations) can potentially help constrain the origin of X-ray emission.

\subsection{Classifications of X-ray Point Sources}
\par In general, bona fide AGN can usually be classified as point sources with unabsorbed 2$-$7 keV luminosities $L_{2-7~\mathrm{keV, unabs}}>10^{41}$ erg s$^{-1}$, while likely AGN are classified as point sources with unabsorbed 2$-$7 keV luminosities $L_{2-7~\mathrm{keV, unabs}}>10^{40}$ erg s$^{-1}$.  Any point source with $L_{2-7~\mathrm{keV, unabs}}<10^{40}$ erg s$^{-1}$ is conservatively not classified as an AGN.
\par The main source of contamination are X-ray binaries (XRBs) or ultraluminous X-ray sources (ULXs); however, most high-mass X-ray binary (HMXBs) have  $2$--$7$ keV X-ray luminosities between 10$^{38}$--10$^{39}$ erg s$^{-1}$, while the ULX population dominates at the highest luminosities, with $L_{2-7~\mathrm{keV}}>$ 10$^{39}$ erg s$^{-1}$ (e.g., \citealt{Swartz2011, Walton2011}). The overall X-ray luminosity function (XLF) of HMXBs and ULXs indicates a general cutoff at $L_{2-7~\mathrm{keV}}=$10$^{40}$ erg s$^{-1}$ (e.g., \citealt{Mineo2012, Sazonov2017, Lehmer2019}), and previous studies on XRB contamination in both late- and early-type galaxies have concluded that the majority of nuclear (within $2\arcsec$ of the galactic nucleus) X-ray point sources with $L_{2-7~\mathrm{keV}}>$ 10$^{40}$ erg s$^{-1}$ are highly unlikely to be emission associated with accretion onto XRBs \citep{Foord2017a, Lehmer2019}. However, these studies have yet to focus on a sample of merging galaxies, where amplified star formation rates can increase the surrounding X-ray emission. Thus, from our X-ray data alone, we can safely classify any system with unabsorbed 2$-$7 keV luminosities $L_{2-7~\mathrm{keV, unabs}}>10^{41}$ erg s$^{-1}$ as an AGN. However, each point source with $10^{40} < L_{2-7~\mathrm{keV, unabs}} < 10^{41}$ erg s$^{-1}$, will need a separate and more detailed analysis using complementary IR observations in order to estimate the expected X-ray contribution from high-mass X-ray binaries. We defer this facet of the analysis to a separate paper analysing the multi-wavelength emission of each triple galaxy merger.
\par Thus, from our spectral analysis alone (see Table~\ref{chap6:tabspectra}) we can classify NGC 3341 as a single AGN system; while SDSS J161+2352, SDSS J1708+2153, SDSS J2356$-$1016, and SDSS J0849+1114 have at least one AGN (the primary point source in all cases). We note that these other dual X-ray point source systems all have primary and secondary point sources with bright X-ray luminosities ($L_{2-7\mathrm{kev}}>10^{40}$erg s$^{-1}$), possibly associated with 4 new undetected dual AGN. A multi-wavelength analysis is required for a better understanding of the true duality, or triality, of all these systems.
%
\begin{table*}
\begin{center}
\caption{Posterior Results for Multiple X-ray Point Sources}
\label{chap6:tabpymc3}
\setlength\tabcolsep{1.7pt}
\begin{tabular*}{\textwidth
}{lcccccccc}
	\hline
	\hline
	\multicolumn{1}{c}{Galaxy Name} &
	\multicolumn{1}{c}{$\alpha$} &
	\multicolumn{1}{c}{$\delta$} &
	\multicolumn{1}{c}{$\alpha_{s}$} &
	\multicolumn{1}{c}{$\delta_{s}$} &
	\multicolumn{1}{c}{$r$ (arcsec)} & 
	\multicolumn{1}{c}{$\theta_{\mathrm{PA}}$ (degrees)} &
	\multicolumn{1}{c}{$\log{f}$} & \multicolumn{1}{c}{$\log{f_{bkg}}$} \\
	\multicolumn{1}{c}{(1)} & \multicolumn{1}{c}{(2)} & \multicolumn{1}{c}{(3)} & \multicolumn{1}{c}{(4)}  & \multicolumn{1}{c}{(5)}  & \multicolumn{1}{c}{(6)} &
	 \multicolumn{1}{c}{(7)}  & \multicolumn{1}{c}{(8)} & \multicolumn{1}{c}{(9)}\\
	\hline \\ [-1.7ex]
	SDSS J1027+1749 & 10:27:00.39 & +17:49:02.94 & 10:27:00.57 & +17:49:00.63 & $3.4^{+0.4}_{-0.3}$ & $-130_{-8}^{+8}$ & $-0.1^{+0.5}_{-0.4}$ & $-0.2^{+0.1}_{-0.1}$ \\ [0.4ex]
    NGC 3341 & 10:42:31.46 & +05:02:37.94 & 10:42:32.05 & +05:02:41.75 & $9.6_{-0.4}^{+0.3}$ & $114_{-4}^{+3}$ & $-1.6_{-0.4}^{+0.4}$ & $-0.7_{-0.1}^{+0.1}$ \\ [0.4ex]
    SDSS J1631+2352 & 16:31:15.52 & +23:52:57.62 & 16:31:15.60 & +23:53:00.10 & $2.8_{-0.6}^{+1.3}$ & $-23_{-25}^{+20}$ & $-2.5_{-0.5}^{+0.4}$ & $-1.5_{-0.1}^{+0.1}$ \\ [0.4ex]
    SDSS J1708+2153 & 17:08:59.12 & +21:53:08.06 & 17:08:59.41 & +21:53:13.33 & $6.6_{-0.4}^{+0.4}$ & $-37_{-5}^{+4}$ & $-2.5_{-0.6}^{+0.4}$ & $-1.4_{-0.1}^{+0.1}$ \\ [0.4ex]
    SDSS J2356$-$1016 & 23:56:54.36 & $-$10:16:05.45 & 23:56:54.56 & $-$10:16:06.77 & $3.5_{-1.2}^{+1.6}$ & $-113_{-41}^{+14}$ & $-2.4_{-1.5}^{+0.7}$ & $-1.4_{-0.4}^{+0.3
    }$ \\ [0.4ex]
    SDSS J0849$+$1114 & 08:49:05.54 & +11:14:47.94 & 08:49:05.44 & +11:14:46.42 & $2.1_{-0.3}^{+0.3}$ & $137_{-10}^{+8}$ & $-1.0_{-0.4}^{+0.4}$ & $-0.8_{-0.2}^{+0.2}$ \\ [0.4ex]
    \dots & \dots & \dots & 08:49:05.46$^{\mathlarger{\dagger}}$ & +11:14:51.49$^{\mathlarger{\dagger}}$ & $3.8_{-0.4}^{+0.4 \mathlarger{\dagger}}$ & $18.8_{-7.7}^{+6.1}$ & $-1.2_{-0.5}^{+0.4\mathlarger{\dagger}}$ & -- \\ [0.4ex]
	\hline 

\end{tabular*}
\end{center}
Note. -- Columns: (1) SDSS galaxy designation; (2) the central R.A. of the primary X-ray source; (3) the central declination of the primary X-ray source; (4) the central R.A. of the secondary X-ray source; (5) the central declination of the secondary X-ray source; (6) the separation between the two point sources in arcseconds; (7) the position angle between the primary and secondary, measured East of North; (8) the log of the count ratio between the secondary and primary; (9) the log of the count ratio between the background and point source contribution. The dagger represents posterior results for the tertiary point source in SDSS J0854+1114. For SDSS J1027+1749, the background component is defined as the diffuse emission component, $f_{diff}$.  Each value is the best-fit value from the posterior distributions, defined as the median of the distribution. All posterior distributions are unimodal, and thus the median is a good representation of the value with the highest likelihood. Error bars represent the 99.7\% confidence level of each distribution.
\end{table*}

\begin{table*}
\begin{center}
\caption{Best-fit Spectral Parameters for Phenomenological Model}
\label{chap6:tabspectra}
\setlength\tabcolsep{2pt}
\begin{tabular*}{0.8\textwidth
}{lcccccr}
	\hline
	\hline
	\multicolumn{1}{c}{Galaxy Name} &
	\multicolumn{1}{c}{$m_{\mathrm{phen, x}}$} & \multicolumn{1}{c}{$N_{H}$ (10$^{22}$ cm$^{-2}$)} & \multicolumn{1}{c}{$\Gamma$} &
	\multicolumn{1}{c}{$F_{0.5-8~\mathrm{keV}}$} & 
	\multicolumn{1}{c}{$L_{2-7~\mathrm{keV,~unabs}}$} &
	\multicolumn{1}{c}{\emph{HR}}\\
	\multicolumn{1}{c}{(1)} & \multicolumn{1}{c}{(2)} & \multicolumn{1}{c}{(3)} & \multicolumn{1}{c}{(4)} & \multicolumn{1}{c}{(5)} & \multicolumn{1}{c}{(6)} & \multicolumn{1}{c}{(7)} \\
	\hline \\ [-2.5ex]
    SDSS J1027+1749$_{p}$ & 1 & $<10^{-2}$ & 1.8 & $5.7_{-1.6}^{+1.3}$ & $3.2_{-0.9}^{+0.8}$ & $0.1_{-0.4}^{+0.2}$ \\ [0.4ex]
	SDSS J1027+1749$_{s}$ & 1 & $<10^{-2}$ & 1.8 & $4.8_{-2.2}^{+1.1}$ & $2.6_{-1.2}^{+0.6}$ & $-0.1^{+0.2}_{-0.4}$\\ [0.4ex]
	\hline \\ [-2.5ex]
	NGC 3341$_{p}$ & 2 &  $10.1_{-0.6}^{+0.5}$ & $1.1_{-0.1}^{+0.1}$ & $363.0_{-5.6}^{+7.0}$ & $85.4_{-3.3}^{+4.1}$ & $0.8_{-0.1}^{+0.1}$ \\ [0.4ex]
    NGC 3341$_{s}$ & 1 & $<10^{-2}$ & 1.8 & $2.7_{-0.8}^{+0.6}$ & $0.2_{-0.1}^{+0.1}$ & $-0.5_{-0.5}^{+0.2}$ \\ [0.4ex]
    \hline \\ [-2.5ex]
    SDSS J1631+2352$_{p}$ & 2 & $1.10_{-0.02}^{+0.01}$ & 1.8 & $2.29^{+0.01}_{-0.01}\times10^{3}$ & $1.33^{+0.01}_{-0.01}\times10^{3}$ & $0.15_{-0.01}^{+0.01}$ \\ [0.4ex]
    SDSS J1631+2352$_{s}$ & 1 & $0.13_{-0.01}^{+2.26}$ & 1.8 & $5.3^{+4.8}_{-3.9}$ & $2.5^{+2.3}_{-1.8}$ & $0.0_{-0.7}^{+0.3}$ \\ [0.4ex]
    \hline \\ [-2.5ex]
    SDSS J1708+2153$_{p}$ & 1 & $0.10_{-0.01}^{+0.01}$ & $1.4_{-0.1}^{+0.1}$ & $1.46^{+0.01}_{-0.01}\times10^{3}$ & $1.17_{-0.01}^{+0.01}\times10^{3}$ & $-0.02_{-0.01}^{+0.01}$ \\ [0.4ex]
    SDSS J1708+2153$_{s}$ & 1 & $0.2_{-0.1}^{+0.5}$ & 1.8 & $4.6_{-1.6}^{+0.2}$ & $3.5_{-1.2}^{+0.6}$ & $-0.4_{-0.2}^{+0.2}$ \\ [0.4ex]
    \hline \\ [-2.5ex]
    SDSS J2356$-$1016$_{p}$ & 2 & $7.8_{-0.1}^{+0.2}$ & 1.8 & $1.60^{+0.01}_{-0.01}\times10^{3}$ & $3.11^{+0.04}_{-0.04}\times10^{3}$ & $0.73_{-0.01}^{+0.01}$ \\ [0.4ex]
    SDSS J2356$-$1016$_{s}$ & 1 & $3.2_{-0.1}^{+1.1}$ & 1.8 & $7.3^{+7.0}_{-3.5}$ & $8^{+10}_{-1}$ & $0.6_{-0.5}^{+0.4}$ \\ [0.4ex]
    \hline \\ [-2.5ex]
    SDSS J0849+1114$_{p}$ & 2 & $52_{-11}^{+23}$ & 1.8 & $65.4^{+2.8}_{-2.2}$ & $160^{+80}_{-40}$ & $0.34_{-0.04}^{+0.05}$ \\ [0.4ex]
    SDSS J0849+1114$_{s}$ & 1 & $<10^{-2}$ & 1.8 & $4.0^{+1.6}_{-1.3}$ & $1.9^{+0.6}_{-0.7}$ & $-0.2_{-0.2}^{+0.3}$ \\ [0.4ex]
    SDSS J0849+1114$_{t}$ & 1 & $<10^{-2}$ & 1.8 & $3.2^{+1.0}_{-1.3}$ & $1.3^{+0.6}_{-0.5}$ & $0.0_{-0.2}^{+0.3}$ \\ [0.4ex]
	\hline 
	\hline
\end{tabular*}

\end{center}
Note. -- Columns: (1) SDSS galaxy designation, we denote the primary, secondary, and tertiary with subscripts $p$, $s$, and $t$; (2) the spectral model used; (3) the best-fit extragalactic column density; (4) the assumed or best-fit spectral index; (5); the measured $0.5$--$8$ keV flux, in units of 10$^{-15}$ erg s$^{-1}$ cm$^{-2}$; (6) the rest-frame, unabsorbed, $2$--$7$ keV luminosity in units of 10$^{40}$ erg s$^{-1}$; (7) the hardness ratio, defined as $= (H-S) / (H+S)$.  Each best-fit value is defined as the median of the full distribution. We identify a statistically significant Fe K$\alpha$ fluorescent emission line in the spectrum of SDSS J1631$-$1016$_{p}$, modeled by a Gaussian component ({\tt zgauss}) fixed at 6.4 keV, with an equivalent width of $0.23_{-0.02}^{+0.01}$ keV (see Fig.~\ref{chap6:spec1}). Error bars represent the 99.7\% confidence level of each distribution.
\end{table*}

\begin{table*}
\begin{center}
\caption{Best-fit Spectral Parameters for {\tt BNTorus}}
\label{chap6:tabspectraBN}
\setlength\tabcolsep{4pt}
\begin{tabular*}{0.9\textwidth
}{lcccccc}
	\hline
	\hline
	\multicolumn{1}{c}{Galaxy Name} & 
	\multicolumn{1}{c}{$m_{\mathrm{phys, x}}$} & \multicolumn{1}{c}{$s$ (\%)} & \multicolumn{1}{c}{$N_{H}$ (10$^{22}$ cm$^{-2}$)} & \multicolumn{1}{c}{$\Gamma$} &
	\multicolumn{1}{c}{$F_{0.5-8~\mathrm{keV}}$} & 
	\multicolumn{1}{c}{$L_{2-7~\mathrm{keV,~unabs}}$} \\
	\multicolumn{1}{c}{(1)} & \multicolumn{1}{c}{(2)} & \multicolumn{1}{c}{(3)} & \multicolumn{1}{c}{(4)} & \multicolumn{1}{c}{(5)} & \multicolumn{1}{c}{(6)} & \multicolumn{1}{c}{(7)} \\
	\hline \\ [-2.5ex]
	NGC 3341$_{p}$ & 2 & $0.19_{-0.06}^{+0.07}$ & $11.8_{-0.2}^{+0.2}$ & $1.8$ & $1.39_{-0.01}^{+0.01}\times10^{3}$ & $118.2_{-1.3}^{+1.6}$ \\ [0.4ex]
    \hline \\ [-2.5ex]
    SDSS J1631+2352$_{p}$ & 2 & $35.72_{-0.40}^{+0.50}$ & $0.99_{-0.02}^{+0.01}$ & 1.8 & $2.29^{+0.01}_{-0.01}\times10^{3}$ & $1.38^{+0.01}_{-0.01}\times10^{3}$ \\ [0.4ex] 
    \hline \\ [-2.5ex]
    SDSS J1708+2153$_{p}$ & 1 & -- & $0.09_{-0.01}^{+0.1}$ & $1.38_{-0.01}^{+0.01}$ & $1.46^{+0.01}_{-0.01}\times10^{3}$ & $1.17_{-0.01}^{+0.01}\times10^{3}$ \\ [0.4ex]
    \hline \\ [-2.5ex]
    SDSS J2356$-$1016$_{p}$ & 2 & $1.13_{-0.02}^{+0.01}$ & $7.0_{-0.1}^{+0.1}$ & 1.8 & $1.61^{+0.01}_{-0.01}\times10^{3}$ & $3.18^{+0.03}_{-0.04}\times10^{3}$ \\ [0.4ex]
    \hline \\ [-2.5ex]
    SDSS J0849+1114$_{p}$ & 2 & $10.6_{-2.8}^{+3.5}$ & $46_{-11}^{+14}$ & 1.8 & $66^{+8}_{-6}$ & $160.6^{+21.1}_{-10.0}$ \\ [0.4ex]
    SDSS J0849+1114$_{s}$ & 1 & $-$ & $<10^{-2}$ & 1.8 & $4.0^{+1.6}_{-1.2}$ & $1.8^{+0.7}_{-0.6}$ \\ [0.4ex]
	\hline 
	\hline
\end{tabular*}

\end{center}
Note. -- Columns: (1) SDSS galaxy designation, we denote the primary, secondary, and tertiary with subscripts $p$, $s$, and $t$; (2) the spectral model used; (3) the scattering fraction of the additional power-law; (4) the best-fit extragalactic column density; (5) the assumed or best-fit spectral index; (6); the measured $0.5$--$8$ keV flux, in units of 10$^{-15}$ erg s$^{-1}$ cm$^{-2}$; (7) the rest-frame, unabsorbed, $2$--$7$ keV luminosity in units of 10$^{40}$ erg s$^{-1}$.  Each best-fit value is defined as the median of the full distribution. In agreement with our phenomenological fits, we identify a statistically significant Fe K$\alpha$ fluorescent emission line in the spectrum of SDSS J1631$-$1016$_{p}$, modeled by an additional Gaussian component ({\tt zgauss}) fixed at 6.4 keV, with an equivalent width of $0.23_{-0.01}^{+0.02}$ keV. Error bars represent the 99.7\% confidence level of each distribution. $\emph{HR}$ values for each system are independent of model, and are listed in Table~\ref{chap6:tabspectra} for each X-ray point source.
\end{table*}

\section{Conclusions}
\label{chap6:conclusions}
In this study, we present the first X-ray analysis of AGN activity in triple galaxy mergers. We analyze 7 nearby ( $0.059 < z < 0.077$) triple galaxy mergers with existing archival \emph{Chandra} and SDSS DR16 observations. Each of these systems are confirmed as triple mergers with available spectroscopic and/or photometric redshift measurements. Running \BAYMAX{} on these observations, we aim to detect low-count and closely-separated multiple AGN systems. Archival SDSS DR16 and/or \emph{HST} observations allow for informative priors on the locations of each AGN, while \BAYMAX{} allows for a statistical analysis on the presence of an X-ray point source at each galactic nucleus. The main results of this study are summarized below: \\
\begin{enumerate}
    \item We find that 1 triple merger system favors the single point source model (SDSS J0858+1822); 5 triple merger systems favor the dual point source model (SDSS J1027+1749, NGC 3341, SDSS J1631+2352, SDSS J1708+2153, and SDSS J2356$-$1016); and one triple merger system favors the triple point source model (SDSS J0849+1114). All of the multiple point source systems have Bayes factors that favor the same model when using both informative and non-informative priors, with the exception of SDSS J2356$-$1016, which has a Bayes factor that favors the dual point source model only when using informative priors.
    \item We quantify the strength of the Bayes factor, by running false positive tests. We find that there is less than a 1\% chance (or, $\le$2\% for SDSS J1631+2352) that the X-ray emission of each system is actually described by a single point source.
    \item The posterior distributions returned by \BAYMAX{} show that the best-fit locations of each multiple point source system coincide with the optical nuclei of galaxies within the merger (as determined by SDSS DR16) and all separations are inconsistent with 0 at the 99.7\% C.L. 
    \item Running our spectral analysis on the multiple point source systems, we find that all point sources have unabsorbed $2-7$ keV luminosities greater than 10$^{40}$ erg s$^{-1}$, with the exception of the secondary point source in NGC 3341. 
    \item For the one system with a Bayes factors in favor of the single point source model, further investigation reflects that SDSS J0858+1822 likely hosts no AGN. Classifying X-ray point sources with $L_{2-7\mathrm{keV}} > 10^{41}$ erg s$^{-1}$ as bona fida AGN, we conclude that NGC 3341 is a single AGN system, while SDSS J161+2352, SDSS J1708+2153, SDSS J2356$-$1016, and SDSS J0849+1114 have at least one AGN (the primary point source in all cases). A multi-wavelength analysis is required for a better understanding of the true duality, or triality, of all these systems.
\end{enumerate}

\noindent \begin{center}{\MakeUppercase{\small Acknowledgements}}\end{center}
\vspace{-0.3cm}
\noindent We thank our referee for thorough and thoughtful feedback, which strengthened our analysis and results. A.F. and K.G. acknowledge support provided by the National Aeronautics and Space Administration through Chandra Award Numbers TM8-19007X, GO7-18087X, and GO8-19078X, issued by the Chandra X-ray Observatory Center, which is operated by the Smithsonian Astrophysical Observatory for and on behalf of the National Aeronautics Space Administration under contract NAS8-03060. A.F. also acknowledges support provided by the National Aeronautics and Space Administration through Chandra proposal ID 21700319. MK acknowledges support from NASA through ADAP award 80NSSC19K0749. The scientific results reported in this article are based on data obtained from the \emph{Chandra Data} Archive. This research has made use of NASA’s Astrophysics Data System.

\noindent \software{ {\tt CIAO} (v4.8; \citealt{Fruscione2006}), \newline XSPEC (v12.9.0; \citealt{Arnaud1996}), \newline {\tt nestle}~(https://github.com/kbarbary/nestle), \newline {\tt PyMC3} \citep{Salvatier2016}, \newline {\tt MARX} (v5.3.3; \citealt{Davis2012})}

\bibliographystyle{aasjournal}
\bibliography{foord.bib}

\end{document}